\documentclass[iop, revtex4]{emulateapj}
\usepackage{amsmath}
\usepackage{natbib}
\usepackage{booktabs}
\usepackage{xcolor}
\usepackage{color}
\usepackage{epsfig}
\usepackage{rotating}

\shorttitle{Strong Clustering of Lyman Break Galaxies around Luminous Quasars at $z \sim 4$}
\shortauthors{Garc\'ia-Vergara, C. et al.}

\begin{document}

\title{Strong Clustering of Lyman Break Galaxies around Luminous Quasars at $z \sim 4$\footnotemark[1,2]}
\footnotetext[1]{Based on observations collected at the European Organization for Astronomical Research in the Southern Hemisphere, Chile. Data obtained from the ESO Archive, Normal program, visitor mode. Program ID: 079.A-0644.}
\footnotetext[2]{We dedicate this work to the memory of Josef Fried, who originally obtained and analyzed the data on which this work is based.}

\author{Cristina Garc\'ia-Vergara \altaffilmark{3, 4}}
\author{Joseph F. Hennawi \altaffilmark{4,5}}
\author{L. Felipe Barrientos \altaffilmark{3}}
\author{Hans-Walter Rix \altaffilmark{4}}

\altaffiltext{3}{Instituto de Astrof\'isica, Pontificia Universidad Cat\'olica de Chile, Avenida Vicu\~na Mackenna 4860, Santiago, Chile.}
\altaffiltext{4}{Max-Planck Institut f\"ur Astronomie (MPIA), K\"onigstuhl 17, D-69117 Heidelberg, Germany.}
\altaffiltext{5}{Department of Physics, University of California, Santa Barbara, CA 93106, USA}
\email{cjgarci1@uc.cl}

\begin{abstract}
In the standard picture of structure formation, the first massive
galaxies are expected to form at the highest peaks of the density field, which constitute
the cores of massive proto-clusters.
Luminous quasars (QSOs) at $z\sim4$ are the most strongly clustered
population known, and should thus reside in massive dark matter halos
surrounded by large overdensities of galaxies, implying a strong
QSO-galaxy cross-correlation function. We observed six $z\sim4$ QSO fields
with VLT/FORS exploiting a novel set of narrow band filters
custom designed to select Lyman Break Galaxies (LBGs) in a thin redshift slice of
$\Delta z\sim0.3$, mitigating the projection effects that have limited the
sensitivity of previous searches for galaxies around $z\gtrsim4$ QSOs.  
We find that LBGs are strongly clustered around QSOs, and present 
the first measurement of the QSO-LBG
cross-correlation function at $z\sim4$, on scales of $0.1\lesssim R\lesssim9\,h^{-1}\,{\rm Mpc}$
 (comoving). Assuming a power law form for
the cross-correlation function $\xi=(r/r^{QG}_0)^\gamma$,
we measure $r^{QG}_{0}=8.83^{+1.39}_{-1.51}\,h^{-1}\,{\rm Mpc}$ for a fixed
slope of $\gamma=2.0$.
This result is in agreement with the expected cross-correlation length
deduced from measurements
of the QSO and LBG auto-correlation function, and assuming a linear bias model.
We also measure a strong auto-correlation of LBGs in our QSO fields finding
$r_{0}^{GG}=21.59^{+1.72}_{-1.69}\,h^{-1}\,{\rm Mpc}$ for a fixed
slope of $\gamma=1.5$, which is $\sim4$ times larger than the LBG
auto-correlation length in random fields, providing further evidence that
QSOs reside in overdensities of LBGs. Our results qualitatively
support a picture where luminous QSOs inhabit exceptionally massive ($M_{\rm
  halo}>10^{12}\,\rm M_{\odot}$) dark matter halos at $z\sim4$.
\end{abstract}


\section{Introduction}
\label{sec:intro}

Our understanding of structure formation suggests that
small inhomogeneities in the density field shortly after the Big
Bang grew over cosmic time via gravitational instability
\citep[e.g.][]{Dodelson03, Padmanabhan06, Schneider15} into
massive dark matter halos at $z=0$. As clusters of galaxies are the
most massive, gravitationally bound structures in the Universe, they
must have formed from the highest density peaks at early times.
This make them
ideal laboratories for studying the formation and evolution of cosmic structure.

Because of the small areas of sky surveyed at high-redshift, and the
low comoving number density $\sim10^{-7}$\,Mpc$^{-3}$ of local
clusters \citep{Gioia01, Vikhlinin09}, the evolutionary link between
these low-redshift clusters and high-redshift galaxies has been
challenging to make. The progenitors of clusters are extremely
difficult to identify when the density contrast between the forming
cluster and its surroundings is small. For this
reason, a commonly adopted approach is to search for these so-called
proto-clusters around known massive galaxies at high redshift.

One very fruitful technique to find high-redshift proto-clusters has been to use the presence of an active super massive black hole (BH) as a signpost for a massive galaxy and hence massive
dark matter halo in the distant Universe
\citep[e.g.][]{Venemans07, Kashikawa07, Overzier08, Morselli14}. This
technique is motivated by several considerations. First, the masses of
supermassive BHs ($M_{\rm BH}$) are known to tightly correlate with the
bulge mass of their host galaxy \citep{Magorrian98, Ferrarese00, Gebhardt00}, 
and possibly 
with the
masses of their host dark halos ($M_{\rm halo}$) \citep[but see \citealt{Kormendy11}]{Ferrarese02}.
Intriguingly, the most luminous quasars (QSOs) at $z>3$ have $M_{\rm
  BH} \sim 1-6 \times 10^{9}\,\rm M_{\odot}$ \citep{Shen11},
comparable to the most massive known local BHs. If the
present day $M_{\rm BH} - M_{\rm halo}$ relation holds at early times,
such BHs should reside in exceptionally massive halos.  Second, some
studies have suggested that the nuclear activity in active galactic
nuclei (AGN) is triggered by processes related to the environment
where they reside. For example, galaxy mergers could trigger the AGN
activity \citep{Bahcall97, Wyithe02, Hennawi15}, and galaxy mergers occur
preferentially in dense environments \citep{Lacey93}. This would imply
that the existence of an AGN requires a dense environment around it.
Finally, another line of evidence that QSOs trace the rarest
environments at high redshift arises from their extremely strong
clustering. Indeed, \citet{Shen07} determined that QSOs at $z>3.5$
have a comoving auto-correlation length of $r_0=24.3\,h^{-1}\,{\rm Mpc}$
(for a fixed correlation function slope of $\gamma=2.0$),
making them the most strongly clustered population in the universe,
and demanding that they reside in the most massive $M_{\rm
  halo}>10^{12}\,\rm M_{\odot}$ dark matter halos at this epoch. Additionally,
the \citet{Shen07} correlation function, agrees with that required to
explain the abundance of binary QSOs at $z>3.5$
\citep{Hennawi10,Shen10}, indicating that overdense structures around
QSOs extend down to scales as small as $100\,h^{-1}\rm kpc$.
Since in hierarchical clustering models, QSOs and galaxies trace
the same underlying dark matter density distribution, the generic
prediction is that galaxies should be very strongly clustered around
QSOs at $z\gtrsim 3.5$. Observationally this should be reflected as a 
strong QSO-galaxy cross-correlation function.

The QSO-galaxy cross-correlation function has been measured at $z<4$
in the past. At $z\lesssim1$ it is found to be in good agreement with
the auto-correlation of galaxies and QSOs, and it has been shown to be
independent of the
QSO luminosity, and weakly dependent on redshift
\citep[e.g.][]{Padmanabhan09, Coil07}.
\citet{Adelberger05} measured
the AGN-galaxy cross-correlation function at higher redshifts
($2\lesssim z \lesssim3$) finding a cross-correlation length of
$r_{0}\sim 5\,h^{-1}\,{\rm Mpc}$ for a slope of $\gamma=1.6$ which is
similar to the auto-correlation of Lyman Break Galaxies (LBGs) at $z\sim3$
\citep{Adelberger03}. They also claim an independence of the
cross-correlation length with the AGN luminosity, implying that both
faint and bright AGNs should be found in halos with similar
masses.
The highest redshift measurement of QSO environments
is the work of \citet{Trainor12}, who quantified the clustering of LBGs
around 15 hyper-luminous QSOs at $z=2.7$. They find a QSO-LBG
cross-correlation length of $r_{0}= 7.3\pm1.3 \,h^{-1}\,{\rm Mpc}$ for
a fixed slope of $\gamma=1.5$ and claim that this measurement is in
agreement with the \citet{Adelberger05} results. Additionally, they
compute a halo mass for those QSOs of log($M_{\rm halo}/\rm
M_{\odot})=12.3 \pm0.5$, which is in agreement with the halos masses
inferred for
fainter QSOs at the same redshift \citep{White12}.

Theoretical considerations suggest that high-redshift QSOs live in
massive dark matter halos, but not necessarily the most massive ones
\citep{Fanidakis13}. 
However, a high
signal to noise clustering analysis is necessary to confirm this
hypothesis.

In addition to these statistical clustering analyses, many
studies of individual AGN environments have been conducted. 
The population of AGNs whose environments have been
studied most intensively are the high-redshift radio galaxies (HzRGs)
at $z\sim2-4$, which have been shown to often reside in proto-cluster
environments \citep[e.g.][]{Venemans07, Intema06, Overzier08,
  Hennawi15}.
At higher redshifts the environments of other classes of AGN, such as
optically-selected QSOs, are currently less well constrained. Most
previous work focuses on searching for galaxies around the most
distant $z\gtrsim5$ QSOs, and these results paint a diverse and rather
confusing picture: \citet{Stiavelli05}, \citet{Zheng06},
\citet{Kashikawa07}, \citet{Utsumi10}, and \citet{Morselli14} find a
quite strong enhancement of galaxies compared to control fields around
$z\sim5-6$ QSOs, whereas \citet{Willott05}
\citet{Banados13},\citet{Simpson14} and \citet{Mazzucchelli16}, find
no significant excess of galaxies around QSOs at
$z\sim6-7$.
\citet{Kim09} studied five QSO fields at $z\sim6$ and reported a mix
of overdensities and underdensities, and \citet{Husband13} find galaxy
overdensities in $z\sim5$ QSOs environments, but they note that even
some randomly chosen patches of sky without AGN signposts (`blank
fields') at the same redshift contain similar galaxy overdensities.
Indeed, surveys of a few deg$^{2}$ for $z\sim6$ LBGs or LAEs
have identified comparable overdensities in blank field pointings
\citep[e.g.][]{Ouchi05, Ota08, Toshikawa12}.
These mixed results at
$z\gtrsim5$ do not yet provide compelling
evidence QSOs inhabit
massive dark matter halos at the highest redshifts, and more work
is clearly required.

One complication of these studies is that the majority of them are
focused on dropout selection, which selects galaxies over a broad
redshift range of $\Delta z\sim1$ \citep[e.g.][]{Ouchi04a}, 
corresponding to $\sim520\,h^{-1}$\,cMpc at $z=4$. 
A large part of such a volume is unassociated with the QSO, which
introduces projection effects that dilute the overdensity around the
QSO making it much more difficult to detect.  Furthermore, most work
at the highest redshifts have focused their searches around
a handful of individual QSOs, and given the poor statistics and
large cosmic variance (which is typically not taken into account),
this could preclude the detection of an overdensity.

In this paper we study the environs of QSOs at
$z\sim4$. There are several advantages to working at this
redshift. First, it is the highest redshift at which auto-correlation
measurements exist for QSOs \citep{Shen07}, establishing that
they reside in massive dark matter halos. Second, the
luminosity function and clustering properties of
$z\sim4$ galaxies are also well known \citep[e.g.][]{Shen07, Ouchi04a,
  Ouchi08}. The well-measured luminosity function
allows us to accurately determine the background 
number density, essential for a robust clustering analysis. Furthermore,
the fact that the auto-correlation of QSOs and galaxies are both known,
gives us an idea of what the cross-correlation should be on large scales
where linear bias models apply. In practical terms, redshift $z\sim 4$ also
represents a compromise since the dark matter halos hosting QSOs are still expected
to be massive \citep{Shen07}, while at the same time 
the characteristic galaxy luminosity $\rm L_{*}$ can be imaged with
much shorter exposure times than galaxies at $z\gtrsim5$, allowing us
to observe a larger statistical sample of QSO fields.  Note that at
$z\sim4$ the universe was only $\sim1.5$\,Gyr old, and only 0.5\,Gyr
has elapsed since the end of reionization. Thus, our QSO targets are
definitely  young objects residing in large scale structures that are
still forming.

Here we present VLT/FORS imaging of six $z\sim4$ luminous QSOs
fields. Using a novel narrow band (NB) filter technique designed to select
LBGs
in a narrow redshift range ($\Delta
z\sim0.3$) around QSOs. This minimizes the line-of-sight
contamination, dramatically reducing the projection effects which are
inherent in broad-band selection.
We measure the QSO-LBG cross-correlation function at $z\sim4$ for the
first time,
to determine whether luminous QSOs at $z\sim4$ are surrounded by
overdensities of galaxies. The large sample of QSOs studied allows
us to beat down the noise from limited numbers of galaxies and cosmic
variance. 

The outline of this paper is as follows. In section \S~\ref{sec:data}
we describe the QSO target selection, we explain the novel NB
imaging technique used to select LBGs, and we give details of the
imaging observations, data reduction, and photometry. We present the
color criteria used to select LBGs and compute the redshift selection
function of the sample in section \S~\ref{sec:LBG_sel}. The
measurement of the QSO-LBG cross-correlation function and LBG
auto-correlation in QSO fields are presented in section
\S~\ref{sec:clustering}, where we also estimate the power law
correlation function $\xi(r) = (r\slash r_0)^{-\gamma}$ parameters
$r_{0}$ and $\gamma$. We test the robustness of our results in section
\S~\ref{sec:testing}, and summarize and conclude in
section
\S~\ref{sec:summ}.

Throughout this paper magnitudes are given in the AB system
\citep{Oke74,Fukugita95} and we adopt a cosmology with $H_{0}=70\,{\rm
  km\,s}^{-1}\,{\rm Mpc}^{-1}$, $\Omega_{m}=0.26$ and
$\Omega_{\Lambda}=0.74$ which is consistent with the nine-year
Wilkinson Microwave Anisotropy Probe (WMAP) Observations
\citep{Hinshaw13}. Comoving and proper Mpc are denoted as ``cMpc" and
``pMpc", respectively.

\section{Observations and Data Reduction} 
\label{sec:data}

The dataset presented in this section was obtained from the ESO
Archive (Program ID: 079.A-0644, P.I: Rix).  This program was designed
to search for LBGs in $z\sim4$ QSOs environments
using a novel NB filter technique. The aim was to test whether QSOs
with the most massive BHs at $z\sim4$
live in the most massive dark matter halos. In this section we
summarize the strategy used to select the targeted QSOs, we explain
the NB technique used to select LBGs, and we provide details of
the imaging observations, reduction process, and photometry.

\subsection{QSO Target Selection} 
\label{sec:qso_selec}

The PI of this program designed a custom set of filters (see
\S~\ref{sec:novel} for details) to search for LBGs in QSO
environments. Using experiments with mock-catalogs, they showed that
this filter set allowed one to select galaxies with $z=3.78\pm0.08$. Given
this small redshift interval,  and with the goal of stacking the
galaxy number counts from several QSO fields, the QSO targets were
selected to span a narrow redshift range of $\Delta z=0.04$, centered
at $z=3.78$.

Taking advantage of the large sample of QSOs from the Sloan Digital
Sky Survey (SDSS; \citealt{York00}), they first selected all QSOs in
this redshift range. Given the goal of studying the most massive dark
matter halos at $z\sim4$, believed to be correlated with the most
massive BHs, only QSOs with M$_{\rm BH}\gtrsim10^{9}$
M$_{\odot}$ were selected. As is typical, the M$_{\rm BH}$ was
estimated from the emission line widths and continuum luminosities
using the so-called single-epoch reverberation mapping technique
\citep{Vestergaard02}. One of the targeted QSOs was not selected from SDSS, 
but it was added to the sample because it belongs to the redshift and M$_{\rm BH}$ 
range of interest. The final sample is comprised of six bright QSOs
with $i<20.2$ mag. 

We verified that none of the QSOs had a detected radio emission
counterpart at 20cm by checking the Faint Images of the Radio Sky at
Twenty-centimeters \citep[FIRST][]{Becker95} catalog, since it is
known that radio emission could strongly affect the galaxy clustering
properties in AGN environments \citep[e.g.][]{Venemans07, Shen09}. A
summary of the QSO properties are listed in Table
\ref{table:qso_prop}, where we show more recent $M_{\rm BH}$ estimates
taken from \citet{Shen11}.

\begin{deluxetable*}{lrrrrl}
\tabletypesize{\small}
\tabletypesize{\scriptsize}
\tablecaption{Targeted QSOs properties.\label{table:qso_prop}}
\tablewidth{0pt}
\tablewidth{0.8\textwidth}
\tablehead{
\colhead{Field}&
\colhead{RA (J2000)}&
\colhead{DEC (J2000)}&
\colhead{Redshift}&
\colhead{$i$}&
\colhead{log($M_{\rm BH}/\rm M_{\odot})$ $^a$}
}
\startdata
SDSSJ0124+0044 &  01:24:03.78 &   00:44:32.67  &   3.834  &    17.99  &    10.15  $\pm$   0.03 \\%&   47.70 $\pm$   0.01\\
SDSSJ0213--0904 &  02:13:18.98 &  -09:04:58.28  &  3.794   &     19.03 &     9.57 $\pm$     0.18 \\%&   47.26 $\pm$    0.01\\
J2003--3300$^b$ &  20:03:24.12  &  -32:51:45.02 & 3.773      & 17.04    &     9.7                  \\%&       \\ 
SDSSJ2207+0043 &  22:07:30.48   &  00:43:29.37   &   3.767 &   19.47    &   9.13  $\pm$    0.16  \\%&  47.14  $\pm$   0.01 \\
SDSSJ2311--0844 &  23:11:37.05 &  -08:44:09.56  &  3.745   &     20.18 &     9.41  $\pm$    0.24 \\%&   46.71  $\pm$   0.03\\
SDSSJ2301+0112 &  23:01:11.23 &    01:12:43.34 &    3.788 &    19.44  &    8.55  $\pm$    0.80 %&   46.93  $\pm$   0.03\\
\enddata
\tablenotetext{a}{Virial BH masses from \citet{Shen11}.}
\tablenotetext{b}{This QSO was not selected from SDSS, but it was targeted because it belongs to the redshift range of interest. The properties shown here are from \citep{McLeod09}, who do not report the error for the BH mass measurement.\\}
\end{deluxetable*}

\subsection{A Novel Method to Select LBGs} 
\label{sec:novel}

The traditional Lyman break technique used to select high-redshift
galaxies relies on the detection of the 912\AA\, flux break (the
so-called Lyman limit break) observed in galaxies due to the
absorption of photons with $\lambda<912$\AA\, by neutral hydrogen
in their interstellar and circumgalactic media.  For this selection method,
two bands are typically used bracketing the break, 
one located at $\lambda<912(1+z)$\AA, and the other at 
$\lambda>912(1+z)$\AA, where $z$ is the redshift of the galaxies
in question.  Given this configuration, a non-detection is expected
in the band blueward of the break, whereas
a clear detection is expected redward of it, such that a very red color
will be measured.  Additionally, a third band is added at longer wavelengths in
order to eliminate possible contaminants. This method was originally
explored using the \textit{UGR} filter system to detect galaxies at
$z\sim3$ \citep{Steidel95, Steidel96, Steidel03}, however, it was
subsequently generalized to higher redshift ($z\sim4-5$) by using a
filter set shifted to longer wavelengths \citep{Steidel99, Ouchi04a}.

At higher redshifts ($z\gtrsim4$), a second break in galaxy spectra
becomes important. The Ly$\alpha$ opacity of the intergalactic medium
(IGM) rapidly increases with redshift, such that a large fraction of
photons emitted by galaxies with $\lambda<1216$\AA\ are absorbed by
neutral hydrogen. This implies a break at $\lambda=1216$\AA\, (the
so-called Lyman alpha break), which can be used to select galaxies
analogous to the traditional Lyman break technique described
above. This Ly$\alpha$ break detection technique has been used to
successfully identify galaxies and QSOs at $z\gtrsim 6$
\citep{Fan00, Bouwens07, Bouwens10, Oesch10, Banados16}.

In order to achieve our goal of selecting galaxies physically
associated with high-redshift QSOs, we need to select LBGs within a narrow
redshift range centered on the QSO. However, the Lyman break method
(using either the Lyman limit or Ly$\alpha$ breaks) efficiently
selects LBGs in a broad redshift slice of $\Delta z\sim1$
\citep[e.g.][]{Ouchi04a, Bouwens07, Bouwens10}, corresponding to 
$\sim520\,h^{-1}$\,cMpc at $z=4$.
For such a broad redshift range, the overdensity
signal around the QSO will be significantly diluted by the
projection of galaxies at much larger distances, hundreds
of comoving Mpc away. 

\begin{figure}
\centering{\epsfig{file=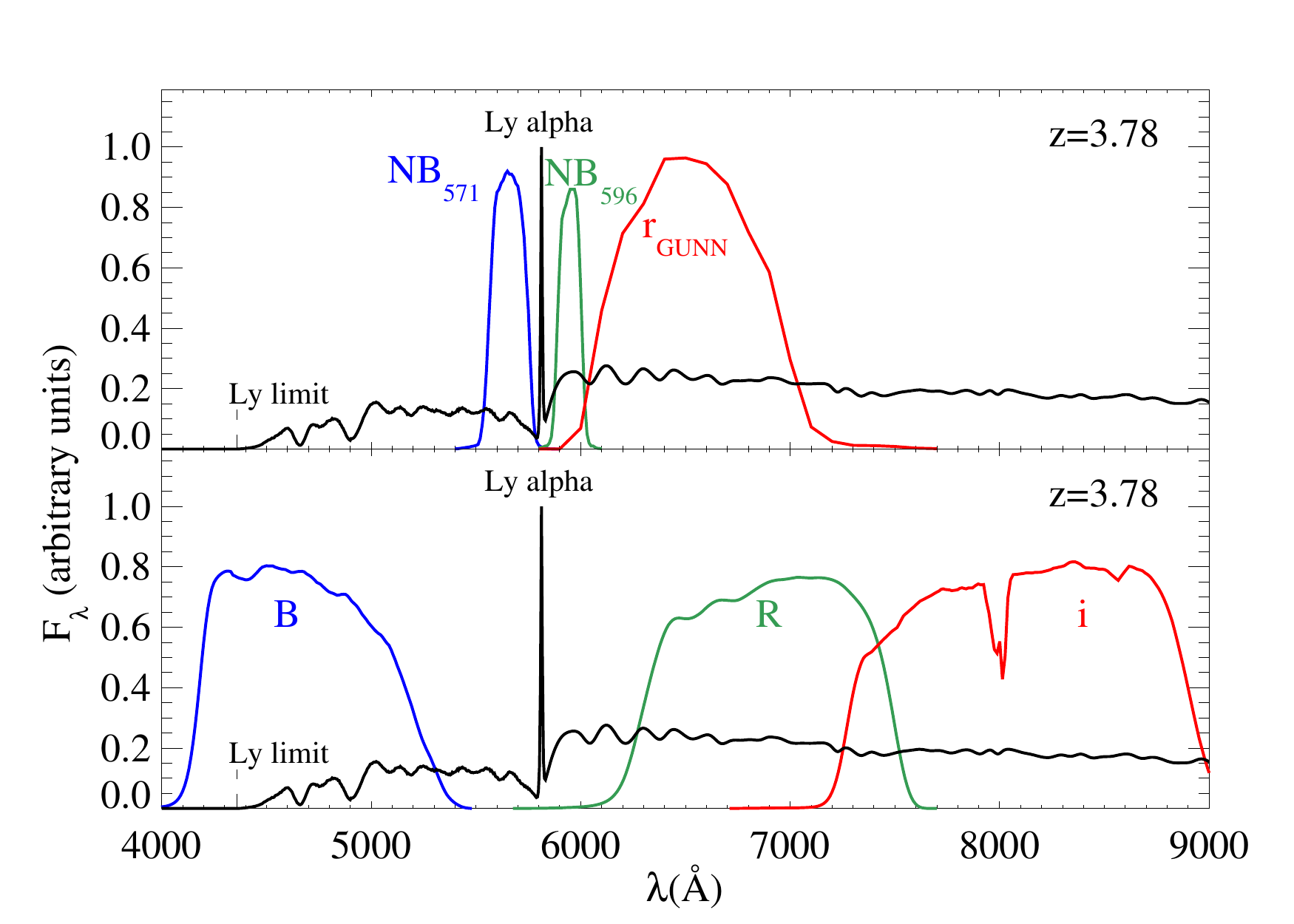, width=\columnwidth}}
\caption{\textit{Upper panel:} Filter configuration used in this study, shown on a LBG simulated spectrum at $z=3.78$ (see section \S~\ref{sec:modeling} for the simulated spectra details). The NBs were designed specially for this program, in order to identify LBGs at $z\sim3.78$ by detecting the Lyman alpha break. This filter configuration selects galaxies in a quiet narrow redshift slice of $\Delta z\sim0.3$. \textit{Lower panel:} Example of a filter set used to identify galaxies with the standard Lyman break technique which is based in the detection of the Lyman limit break. The filter curves shown are those used by \citet{Ouchi04a} to find LBGs at $z\sim4$ over a redshift slice of $\Delta z\sim1.0$\\}
\label{fig:filters}
\end{figure}

In order to address this problem, the PI proposed a novel selection
technique analogous to the Ly$\alpha$ break method, but with the
difference that the selection of LBGs is performed using two NB
filters located very close to each other, instead of using broad
bands.  These filter curves are compared to those used for traditional
LBG selection in Fig.~\ref{fig:filters}.  The advantage of using NB
filters is that they allow one to select LBGs in a much narrower
redshift range of $\Delta z\sim0.3$ ($\sim167$\,cMpc at $z=3.78$) (see
section \S~\ref{sec:comple}), which is $\sim3.3$ times smaller than
the redshift range covered when broad bands are used,
allowing one to minimize line-of-sight projections from physically unassociated
galaxies.

This method has never been used before to select LBGs, and the filters
used to perform the observations were custom designed
to select LBGs at $z\sim3.78$ centered on the redshift of our six QSO targets. 
The two NB
filters used in this study are $\rm NB_{571}~(\lambda_{eff}=5657$\AA,
$\rm FWHM=187$\AA), and $\rm NB_{596}~(\lambda_{eff}=5947$\AA, $\rm
FWHM=116$\AA), which were designed to have a gap between them to exclude the Ly$\alpha$
emission line at $z =3.78$. Then the galaxy selection is not influenced by the Ly$\alpha$ line-strength
, but rather is sensitive to the Ly$\alpha$ break. Additionally data was also collected in the broad band
filter $\rm r_{GUNN}\,(\lambda_{eff}=6490$\AA) to help remove low-redshift
interlopers.

\subsection{VLT Imaging and Data Reduction}
\label{sec:obs}

Imaging observations were acquired on three consecutive nights during 2007 September 9 - 11,
using the FOcal Reducer and low dispersion Spectrograph 1 (FORS1; \citealt{Appenzeller92}) 
instrument on the Very Large Telescope (VLT). The field-of-view (FOV) of FORS1 is $6.8\times6.8$ arcmin$^{2}$ which corresponds to $\sim 3.0\times3.0\,{\rm pMpc}^{2}$ at $z=3.78$.
The instrument pixel scale is 0.251 arcsec/pix for images binned $2\times 2$. 

Each QSO field was observed in the three filters shown in Fig.~\ref{fig:filters}. The total exposure time for the filters was 8000s,
4000s, and 1800s for $\rm NB_{571}$, $\rm NB_{596}$, and $\rm r_{GUNN}$
respectively. Observations were acquired in shorter individual 
dithered exposures, in order to fill the gap between the CCDs
and to facilitate the data reduction process (cosmic ray and bad pixel
rejection, building a superflat, etc). A spectrophotometric
standard
star was observed only in the second and third night. The typical
seeing during the three nights was 0.6 - 0.8 arcsec.

Science images were reduced using standard IRAF\footnote{Image
  Reduction and Analysis Facility} tasks and our own custom codes
written in the Interactive Data Language (IDL).
The reduction process included bias subtraction and flat
fielding. As our images exhibited illumination patterns, we performed
the flat fielding with superflat images,
created using the unregistered science frames. For that, we first
masked all the objects out and then combined the science frames with
an average sigma-clipping algorithm.

SExtractor \citep{Bertin96} was used to create a source catalog for each individual image and then SCAMP \citep{Bertin06} was used to compute an astrometric solution,  
using the SDSS-DR7 $r$-band star catalogs as the astrometric reference.
Finally, the individual images were sky-subtracted, re-sampled, and median-combined using SWarp \citep{Bertin02}, and then the noisy edges of the combined images were trimmed.

For the flux calibration, we only had observations of the
spectrophotometric standard star SA109-949 at the beginning of the
last two nights. The tabulated spectrum of this star has a coarse
sampling of 25\AA\,\citep{Stone96} which is not suitable when NB
 filters are used. For the first night, spectrophotometric
standard stars were not observed, but we took advantage of two
existing SDSS star spectra in one of the fields taken during that
night. The coordinates of the stars with available SDSS spectra are
$\rm RA_{star1}=21.014$, $\rm DEC_{star1}=0.740872$ and $\rm
RA_{star2}=21.057$, $\rm DEC_{star2}=0.686577$ and the median
signal-to-noise ratio per angstrom of their spectra at the wavelengths of interest
is $13.3$ and $8.5$ respectively.

The flux calibration process was as follows. For the first night
calibration we convolved the SDSS star spectra with the three filters
curves in order to obtain standard magnitudes. These magnitudes were
compared with the stars instrumental magnitude (obtained using the
MAG\_AUTO of SExtractor on the combined science images) to obtain the
zero-points (ZPs) for each filter. A mean final ZP was computed from
the two stars and the typical error for this ZP measurement was $\sim0.08$
mag.
For the second and third night calibration, we used the spectrum
of the observed spectrophotometric star to convolve it only with the
broad-band filter curve to obtain the $\rm r_{GUNN}$ ZP. The error in this computation was 
$\sim0.02$ mag. 
After that,
the differential ZPs from the first night were used to determine the
NB zero-points for the second and third nights for which we obtained 
a typical error of $\sim0.11$ mag.

\subsection{Photometric Catalogs}
\label{sec:phot}

Object detection and photometry were performed using SExtractor 
in dual mode, with the $\rm r_{GUNN}$ image as the detection image. We set the parameters BACK\_SIZE and BACKPHOTO\_THICK such that the background was calculated in regions of 64 pixels in size and then recomputed locally in an annulus area of 24 pixels of thickness centered around the object. The parameters DETECT\_MINAREA and DETECT\_THRESH were set such that every group of at least five contiguous pixels having a value above $1.5\sigma$ (with $\sigma$ the background RMS) was considered as an object.

In order to ensure an adequate color measurement we need to carry out
photometry in the same object area for the three different
filters. Therefore, we convolved our images with a
Gaussian kernel to degrade its PSF to match it with the worst seeing
image for each field. Then, the object magnitudes were estimated by
the MAG\_APER parameter of SExtractor using a fixed aperture of
$2\arcsec$ diameter. This magnitude is not necessarily 
the total magnitude of
the object, but is used to compute the colors of galaxies. With this
choice, if galaxies at $z\sim4$ are unresolved by the PSF,
we are
including the flux out to
$\sim3\sigma$ of the object's PSF (for a seeing of $0.8\arcsec$). This
ensures that we measure the majority of the object's flux, as well as
avoid contamination from other close sources. Magnitudes of objects not
detected or detected with a signal-to-noise ratio (${\rm S\slash N})$ $< 2$ either
in $\rm NB_{\rm 571}$ or $\rm NB_{596}$ were assigned the value of the
corresponding $2\sigma$ limiting magnitude.

Here, the ${\rm S\slash N}$ of each object is defined as the ratio of counts in the $2\arcsec$
aperture, given by SExtractor, to the rms sky noise in the aperture. This rms sky noise
 is calculated using an IDL procedure which performs $2\arcsec$ aperture
photometry in $\sim5000$ different random positions in the image
(avoiding the locations of objects) to compute a robust measurement of the
mean sky noise.
The rms sky noise is calculated as the standard deviation of the distribution of
mean values.

Magnitudes were corrected for extinction due to airmass using the
atmospheric extinction curve for Cerro Paranal \citep{Patat11}, and
by galactic extinction calculated using the \citet{Schlegel98} dust
maps and extinction laws of \citet{Cardelli89} with R$_{V} = 3.1$. 
The error in the measured magnitude was computed by error propagation, with the 
object flux error given by the rms noise $N$ in the aperture computed as we described above. 

The
mean $4\sigma$ limiting magnitude of the reduced images was of 26.06
for $\rm NB_{571}$, 25.53 for $\rm NB_{596}$ and 25.82 for $\rm
r_{GUNN}$ for $2\arcsec$ diameter apertures. These limiting magnitudes
are listed in Table \ref{table:limit_mag} for each field. 

\begin{deluxetable}{lllll}
\tabletypesize{\small}
\tabletypesize{\scriptsize}
\tablecaption{$4\sigma$ limit magnitudes per field measured in a  $2\arcsec$ diameter aperture and seeing measured on the $\rm r_{GUNN}$ images.\label{table:limit_mag}}
\tablewidth{0pt}
\tablewidth{1.0\columnwidth}
\tablehead{
\colhead{Field}&
\colhead{$\rm NB_{571}$}&
\colhead{$\rm NB_{596}$}&
\colhead{$\rm r_{GUNN}$}&
\colhead{Seeing [$\arcsec$]}
}
\startdata
SDSSJ0124+0044 & 26.04  &25.51   &25.86 & 0.83\\
SDSSJ0213--0904 & 26.18	 &25.71	 &25.92 & 0.89 \\
J2003--3300         & 26.05&25.44    &25.62 & 0.45\\
SDSSJ2207+0043 & 26.03	&25.38	 &25.78 &  0.53 \\
SDSSJ2311--0844 & 26.02	&25.60	 &25.84 & 0.76\\
SDSSJ2301+0112 & 26.04	&25.55	 &25.91 & 0.70		
\enddata
\tablenotetext{}{\\}
\end{deluxetable} 

For each field, we computed the completeness of the photometric catalogs for the image detection $\rm r_{GUNN}$. For that, we linearly fitted the logarithmic magnitude distribution in the magnitude range $21.0<\rm r_{GUNN}<24.5$ where the photometric catalogs are assumed to be 100\% complete. We extrapolated the linear fit to fainter magnitudes and we measured the completeness as a function of magnitude as the ratio of the histogram relative to that linear fit. We find that at our $4\sigma$ limiting magnitude the completeness is on average $\sim 12$\%. 

\section{LBG Selection at $z$=3.78}
\label{sec:LBG_sel}

LBG candidates at $z=3.78$ were selected using the Ly$\alpha$ break
technique adapted to our custom filters, which target the Ly$\alpha$
break at
$\lambda_{\rm rest-frame}=(1 + z)1216$\AA.
Our two NB filters were chosen to bracket this break, and thus
we expect that LBGs
at $z=3.78$ will have red colors in $\rm
NB_{571}-NB_{596}$.
But if we used only this color criteria, we could be
including some low-redshift galaxy interlopers in the sample. In order to remove them, a
third filter is used to give a measurement of the LBG continuum slope using the
$\rm NB_{596}-r_{GUNN}$ color.

Since the filters used in this study are not standard, the color
criteria to select LBGs is unknown. We also do not know what colors
low-redshift galaxy contaminants have in this filter system. For this
reason, we must explore how galaxies populate the color space in order
to select a complete LBGs sample while avoiding low-redshift
interlopers. Furthermore, in order to perform a LBGs clustering
analysis in QSO fields we need to know the number density of LBGs
expected at random locations in the universe. When a standard filter
set is used (e.g. LBG selection using broad band filters),
this number density can be computed directly from the LBG luminosity
function measured from work using similar filters.
However, in our case if we compute the number density from this LBG
luminosity function, we have to correct this quantity to take into
account the fact that our filter system is mapping a different survey
volume and does not necessarily identify all of the LBGs selected by
broad-band selection.
Specifically, we need to a) determine what fraction of LBGs
we are detecting at any redshift (i.e. the completeness) and b)
determine the redshift range over which we are selecting LBGs ($\Delta
z$). Both of these goals can be achieved by performing an accurate computation of
the redshift selection function $\phi_{z}(z)$, defined as the LBG
completeness as a function of redshift.

In order to perform the optimal LBG selection and compute $\phi_{z}(z)$, we conducted detailed
simulations to model the distribution of LBG colors in the color-space. In this section
we detail how the color modeling was performed, we study what contaminants could be affecting our LBGs selection, and we define a color criteria to select LBGs at $z=3.78$. Finally, we present the redshift selection function providing the completeness as a function of
redshift for the sample.

\subsection{LBG Color Modeling}
\label{sec:modeling}

We performed a Monte Carlo simulation of 1000 LBG spectra at each redshift, that were created to have 
different UV continuum slopes and Ly$\alpha$ equivalent widths (EW$_{\rm
  Ly\alpha}$), such that they reproduce the space of
possible LBG spectra informed by our knowledge of LBG properties.  

Each simulated rest-frame spectrum was created in the following
way. As a starting point, we considered a template galaxy spectrum
generated from \citet{Bruzual03} population synthesis
models\footnote{Obtained from http://bruzual.org/}, corresponding to
an instantaneous burst model with an age of 70Myr, a
\citet{Chabrier03} IMF, and a metallicity of 0.4Z$_{\odot}$, as
expected for LBGs at $z\sim4$ \citep{Jones12}. We assumed a power law
UV continuum for this template with amplitude $A$ and a slope
$\alpha_{\rm BC}$, such that we modeled its flux as $F_{\rm
  BC}(\lambda) = A \lambda^{\alpha_{\rm BC}}$. We fit this model to
the template spectrum over the UV continuum range (here defined as
1300\AA $<\lambda<$2000\AA) by least-squares minimization to
obtain the best fit $A$ and $\alpha_{\rm BC}$ parameters.

First, we modified the UV slope of this template by multiplying its flux by $\lambda^{\alpha - \alpha_{\rm BC}}$ in
order to obtain a spectrum with a power law UV continuum given by $A \lambda^{\alpha}$. The new slope $\alpha$ was chosen as a value taken randomly from
a Gaussian distribution with mean $\mu=-1.676$ and $\sigma=0.39$. These
values are motivated by \citet{Bouwens09}, who presented the UV
continuum slope distribution of LBGs at $z\sim4$ for samples selected
in different magnitude ranges.

\begin{figure}
\centering{\epsfig{file=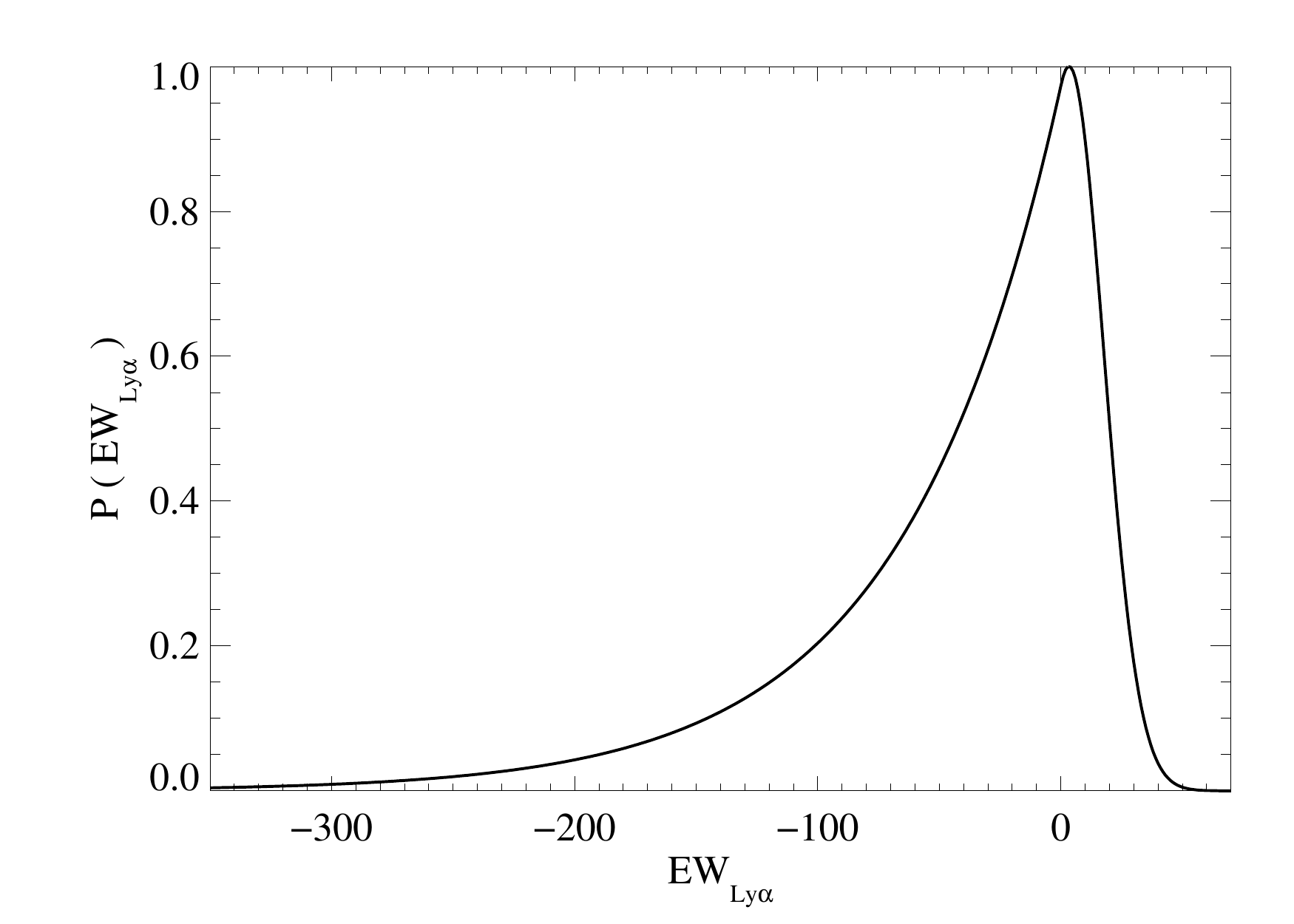, width=\columnwidth}}
\caption{Normalized probability distribution function of EW$_{\rm Ly\alpha}$ used for the simulated spectra, where negative values correspond to emission lines. EW$_{\rm Ly\alpha}$ was chosen from a Gaussian distribution with rest-frame mean $\mu=-25$\AA\, and $\sigma=40$\AA\, \citep{Shapley03} plus an exponential tail of high EW$_{\rm Ly\alpha}$ values with scale length of W$_{0}=-64$\AA\,\citep{Ciardullo12}.\\} 
\label{fig:ew_dist}
\end{figure}

Second, we added a Gaussian Ly$\alpha$ line with rest-frame central
wavelength $\lambda_{\rm Ly\alpha}=1215.7$\AA, standard deviation
$\sigma_{\rm Ly\alpha}$ and amplitude $B$ which adjusts the intensity
of the line. For all the simulated spectrum we used a fixed
$\sigma_{\rm Ly\alpha}=1$\AA\, which agrees with the $\sigma_{\rm
  Ly\alpha}$ of the composite spectrum of LBGs at $z\sim4$
\citep{Jones12}. The $B$ value was adjusted in order to model a
Ly$\alpha$ line with a EW$_{\rm Ly\alpha}$ value drawn randomly from a
distribution chosen to agree with observations of LBGs. The
EW$_{\rm Ly\alpha}$ distribution was given by a Gaussian core
plus a tail to large negative equivalent widths to represent
strong line emitters.  For the Gaussian core
we adopted a mean $\mu=-25$\AA\, and standard deviation
$\sigma=40$\AA\,(rest-frame), based on the measurements
of \citet{Shapley03}, who studied the
spectra of 811
LBGs at $z\sim3$. We thus assume that the Gaussian core of the LBG EW$_{\rm
  Ly\alpha}$ distribution does not evolve significantly from
$z\sim3$ to $z\sim4$. For the tail representing strong line-emitters, we
modified the Gaussian by adding an exponential function
with rest-frame EW$_{\rm Ly\alpha}$ scale length of W$_{0}=-64$\AA, as presented in
\citet{Ciardullo12}.
In this way our model of line emission encompasses both LBG and LAE spectra.
Fig.~\ref{fig:ew_dist} shows the
EW$_{\rm Ly\alpha}$ probability distribution function used to simulate our spectral models. The EW$_{\rm Ly\alpha}$ are defined as: 
\begin{equation}
{\rm EW}_{\rm Ly\alpha} = - \int \frac{F_{\rm Ly\alpha}}{F_{\rm cont}} d\lambda, 
\label{eq:ew}
\end{equation}
where $F_{\rm Ly\alpha}$ is the flux of the ${\rm Ly\alpha}$ line
(with the continuum subtracted), which is given by a Gaussian with amplitude $B$, as we described above, and $F_{\rm cont}$ is the flux of the
continuum given by $A \lambda^{\alpha}$.
Note that we defined negative values of
EW$_{\rm Ly\alpha}$ for emission lines and positive for absorption
lines.

Once $\alpha$ and EW$_{\rm Ly\alpha}$ were chosen for a given
simulated spectrum, we dust-attenuated it using the starburst
reddening curve from \citet{Calzetti00} and adopted a color excess
value of $\rm E(B-V)=0.16$ according to the values estimated for LBGs
at $z\sim3$ \citep{Shapley03}.

After the dust-attenuation is applied, we model the fact that only a
small fraction of Lyman limit photons escape LBGs with an escape
fraction parameter $f_{\rm esc}^{\lambda<912}$. Although this value is
observationally poorly constrained, studies suggest it is in the range
0.04-0.14 \citep{Fernandez03, Shapley06, Ouchi04a}. We assumed a fixed
value of $f_{\rm esc}^{\lambda<912} = 0.05$, and multiplied the
spectrum at $\lambda\leq 912$\AA\, by this value. We also tested our
results using different values of $f_{\rm esc}^{\lambda<912}$, finding
that the colors of simulated galaxies are relatively insensitive to
the exact value of $f_{\rm esc}^{\lambda<912}$ used, because these wavelengths
are subsequently significantly attenuated by the IGM transmission
function (see below).

\begin{figure}
\centering{\epsfig{file=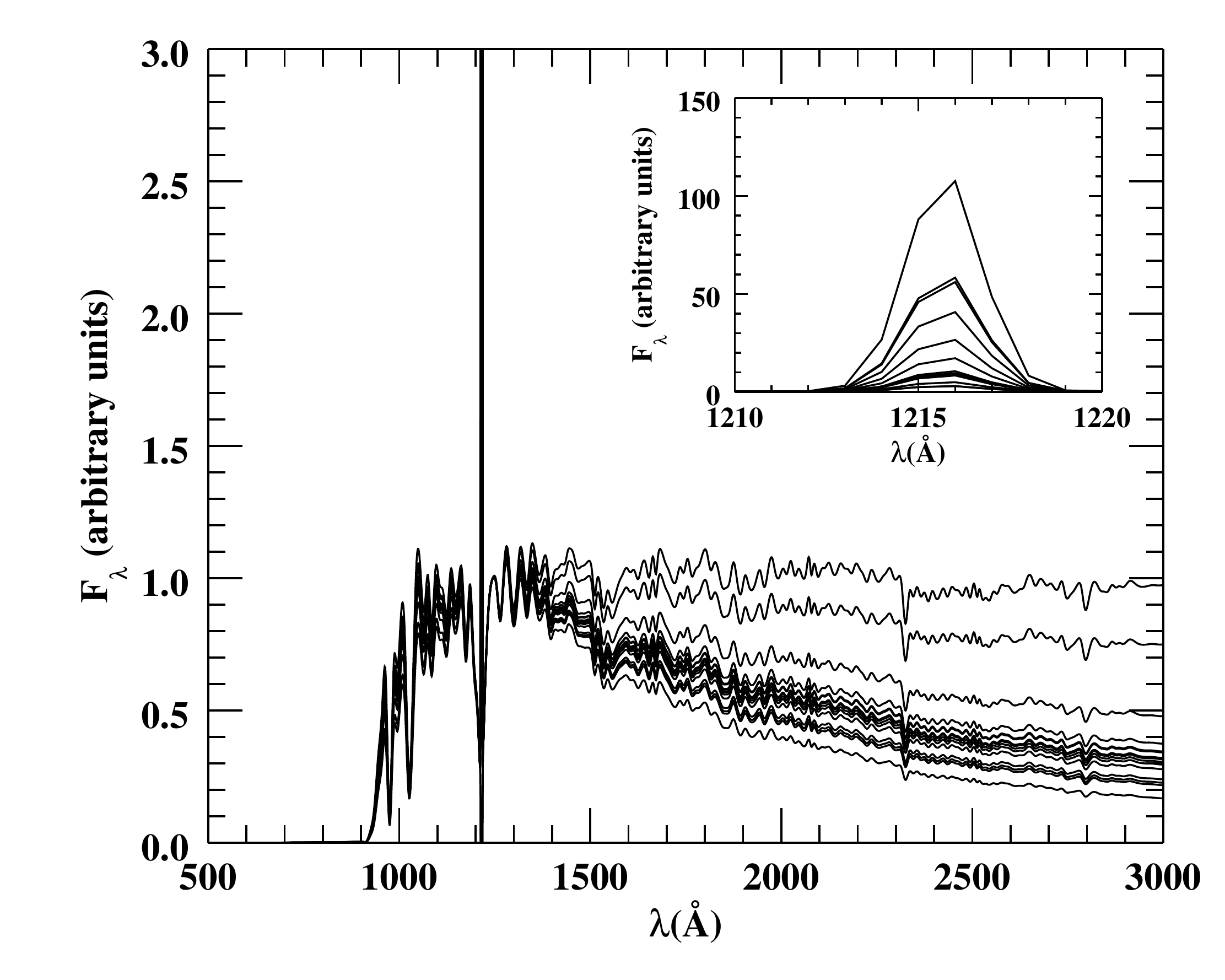, width=\columnwidth}}
\caption{Example of ten rest-frame simulated spectra using our Monte Carlo simulation. The spectra have been normalized to have the same flux value at $\lambda=1245$\AA. The subplot in the upper right corner shows a zoom-in of the region of the ${\rm Ly\alpha}$ line.\\} 
\label{fig:simulated_spec}
\end{figure}

Finally, we redshifted 
each model spectrum to different redshifts on a grid with a grid spacing
of 0.02 and ranging from $z=3.2$ to $z=4.4$. In the redshifting process we used the
IGM transmission model $T_{z}(\lambda)$ for the corresponding redshift
$z$ from \citet{Worseck11} to attenuate the flux blueward of
the ${\rm Ly\alpha}$ line\footnote{Kindly provided to us by G. Worseck.}. Note that in
principle we should attenuate both the
continuum blueward of the ${\rm Ly\alpha}$ line and the line itself, however, the
EW$_{\rm Ly\alpha}$ values used in this simulation are taken from the
literature, which are observed values that are
not corrected for IGM attenuation, such that this line emission is effectively
already attenuated. In Fig.~\ref{fig:simulated_spec} we show some
examples of our rest-frame simulated spectra, which have been
normalized to have the same flux at $\lambda=1245$\AA.

At each redshift, we integrated the spectra against our three filter transmission curves
to obtain the fluxes and then the LBG colors. In order to model the
impact of noise, we added photometric errors to the simulated LBG
photometry. To this end we first assigned an $\rm r_{GUNN}$ magnitude
to each simulated object by randomly drawing a value from the $z\sim4$
LBG luminosity function, integrated over the same magnitude range as
our LBG sample ($24.0 \leq r_{\rm GUNN} \leq 25.6$ or $0.76\leq L/\rm
L_{*}\leq3.5$; see \S~\ref{sec:sample})\footnote{Given that for each
  field we reached slightly different limiting magnitudes, we
  simulated the LBG photometry field by field according to their
  respective $r_{\rm GUNN}$ limiting magnitude. This results in a
  slightly different redshift selection function for each field.}.
We also weighted the luminosity function by the completeness of the
source detection at each apparent magnitude and for each field
(computed in \S~\ref{sec:phot}),
which takes into account the fact that the fraction of sources detected
depends on their magnitude, such that the photometric catalog is complete for bright sources but less complete at the faint end. In this way the incompleteness of our photometry is
also factored into our color modeling.

\begin{figure*}
\centering{\epsfig{file=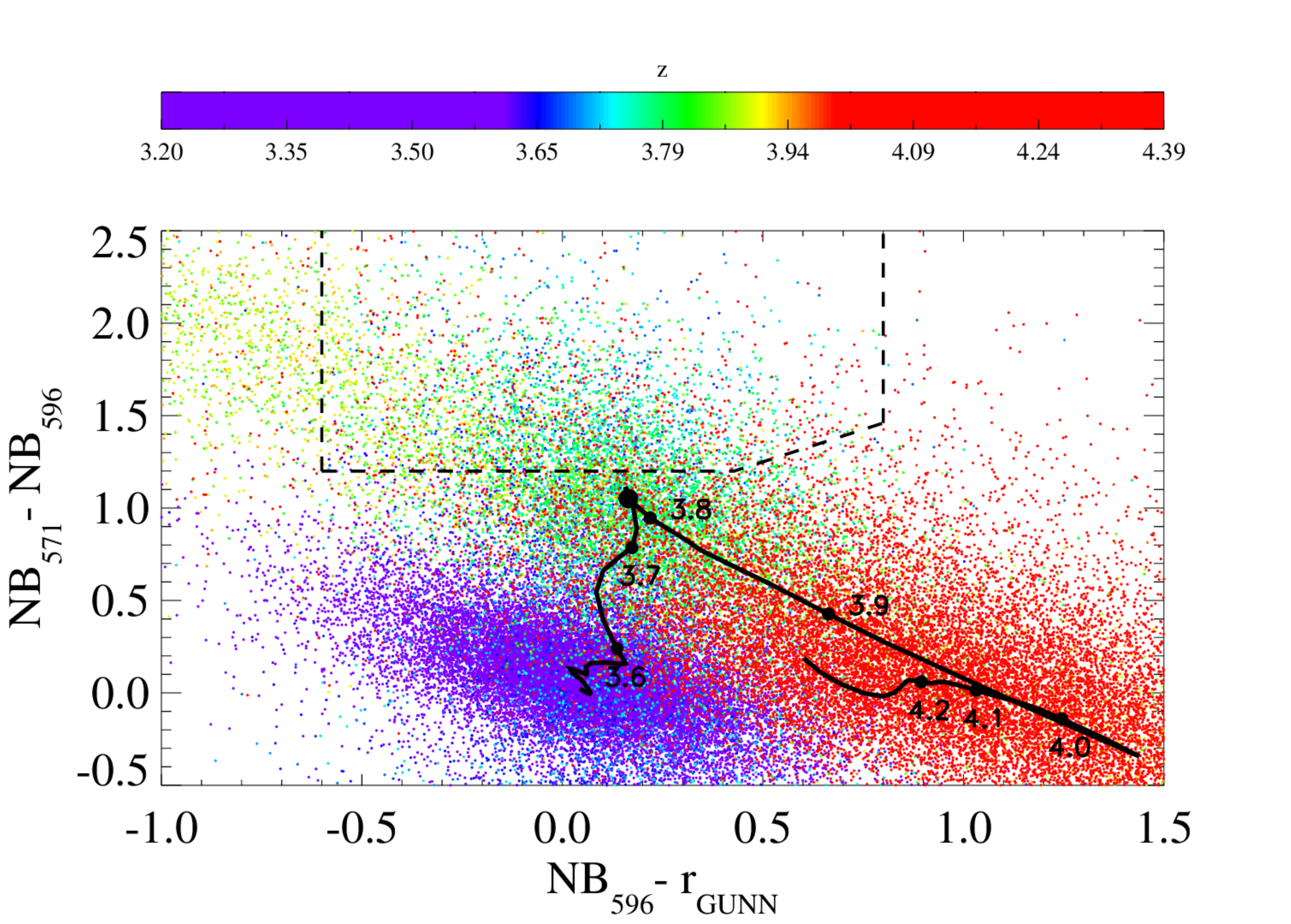, width=\textwidth}}
\caption{Color-color diagram showing the simulated colors for 1000 LBGs spectra, plotted as redshift color-coded points according to the color bar. The median LBG evolutionary track is plotted as a black curve. The filled points over this curve indicate the median LBGs colors at different redshift ranging from 3.6 to 4.2. The largest circle shows the exact position of the median $z=3.78$ LBG colors. The dashed line indicates the selection region used to select LBGs according to the eqn.~(\ref{eq:colorcut}). \\} 
\label{fig:colors_yerr}
\end{figure*}

Based on the simulated LBG colors the chosen $\rm r_{GUNN}$ value, we
then determined the magnitude in the other two filters $\rm NB_{571}$
and $\rm NB_{596}$ for each spectrum in each redshift bin. In order to
construct a noise model , we selected a galaxy sample from our
photometric catalogs, and we computed the median magnitude error as a
function of the magnitude for each filter (with the magnitude error
computed as we explained in \S~\ref{sec:phot}). Finally, we assigned
random Gaussian distributed magnitude errors using our median relations, and then added
this noise to the model photometry which defined the final photometry
of the simulated spectra.  The colors for the 1000 simulated spectra
at each redshift are shown in Fig.~\ref{fig:colors_yerr}.
We also computed the median
of our 1000 rest-frame Monte Carlo spectra, redshifted it, and obtained the colors at each redshift to compute
the median evolutionary track of LBG colors, shown as the black solid line in
Fig.~\ref{fig:colors_yerr}. 

Fig.~\ref{fig:colors_yerr} indicates that the median colors of LBGs at
$z=3.78$ are $\rm NB_{571}-NB_{596}= 1.05$ and $\rm NB_{596}-r_{GUNN}=
0.16$. However, if we consider the 
intrinsic scatter in LBG properties (continuum slope and EW$_{\rm
  Ly\alpha}$) and photometric uncertainties, the $z\simeq 3.78$ LBGs
(indicated by green points) span a wider color range 
with
$\rm NB_{571}-NB_{596} \gtrsim 0.5$ and $-0.6 \lesssim \rm
NB_{596}-r_{GUNN} \lesssim 0.8$. In principle, we should select LBGs
in this broad selection region to obtain a highly complete sample,
however, we also need to take into account the colors of low-redshift
galaxies in our filter system to define a final selection criteria. We
perform this analysis in \S~\ref{sec:interlopers}, where we also test
our LBG color modeling by reproducing the LBG evolutionary track
presented in previous work using broad band LBG selection.

\subsection{Low-Redshift Galaxy Colors}
\label{sec:interlopers}

We use template galaxy spectra to develop a basic understanding of how low-redshift galaxies
populate the color-color diagram in our new filters. We used a set of five commonly used
templates for estimating photometric redshifts, such that they span
the range of galaxy spectral energy distributions (SEDs). The templates are from the
photo-$z$ code EASY \citep{Brammer08}, which are distilled from the PEGASE
spectral synthesis models.  

\begin{figure}
\centering{\epsfig{file=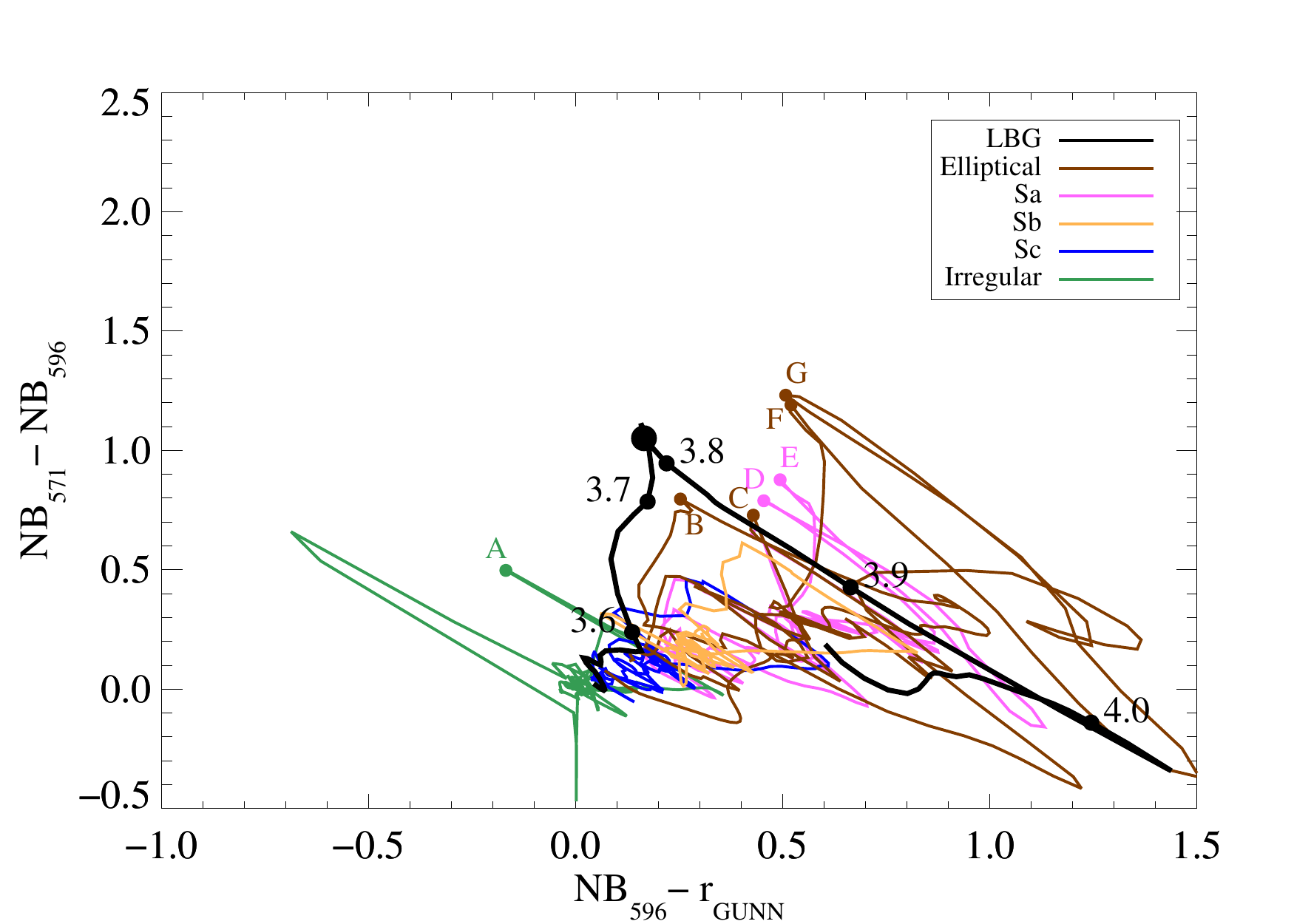, width=\columnwidth}}
\caption{Evolutionary tracks of low-redshift galaxies redshifted from $z=0$ to $z=3$. We plot as brown, magenta, orange, blue, and red curves the evolutionary track of elliptical, Sa, Sb, Sc, and Irregular galaxies, respectively. We overplotted the track of LBGs computed as was explained in section \S~\ref{sec:modeling} as a black curve. Filled circles over the black curve indicate colors of LBGs from redshift 3.6 to 4.0, and the largest black point indicates the exact position of the color of LBGs at $z=3.78$. Filled circles labeled with letters over the low-redshift galaxies evolutionary tracks are indicating the colors of some contaminants that could be affecting our selection: galaxies at $z=0.60$ (A), $z=1.83$ (B), $z=0.45$ (C), $z=1.23$ (D and F) and $z=1.04$ (E and G).\\}
\label{fig:evoltrack_all} 
\end{figure}

We redshifted these template spectra from $z=0$ to $z=3$, and
integrated them over our filter transmission curves to generate their
evolutionary track. Note that we need not attenuate these spectra by the IGM transmission function $T_{z}(\lambda)$, since our NB filters never cover rest-frame wavelengths lower than $1216$\AA\,
for the low redshifts considered.
In Fig.~\ref{fig:evoltrack_all} we show the evolutionary tracks for different galaxy types together with the median
LBG evolutionary track that we computed in \S~\ref{sec:modeling}. 

In order to test our Monte Carlo simulation as well as the
evolutionary tracks for low-redshift galaxies, we have used our 1000
simulated spectra at each redshift to compute the median LBG evolutionary track in the
standard $BRi$ filter set used to select LBGs at $z\sim 4$ (see Fig.~\ref{fig:filters})
by \citet{Ouchi04a}. We also computed the
evolutionary track of these low-redshift
galaxies in the standard LBG filters in the same way as described above. These results are shown in Fig.~\ref{fig:evoltrack_all_ouchi}, where we also overplot the selection
region used by \citet{Ouchi04a} to select $z\sim 4$ LBGs. We find that the
median  LBG evolutionary track from our Monte Carlo model lies within the \citet{Ouchi04a} selection region, and selects LBGs at $z\gtrsim 3.5$ as claimed. Note also that
our LBG evolutionary track agrees well with the \citet{Ouchi04a} evolutionary track
(see Fig.~4 of their paper) determined from a
much simpler model of LBG spectra and IGM transmission.
In addition we see that the evolutionary tracks of low-redshift galaxies lie comfortably
outside the $BRi$ LBG selection region as claimed by  \citet{Ouchi04a}.

\begin{figure}
\centering{\epsfig{file=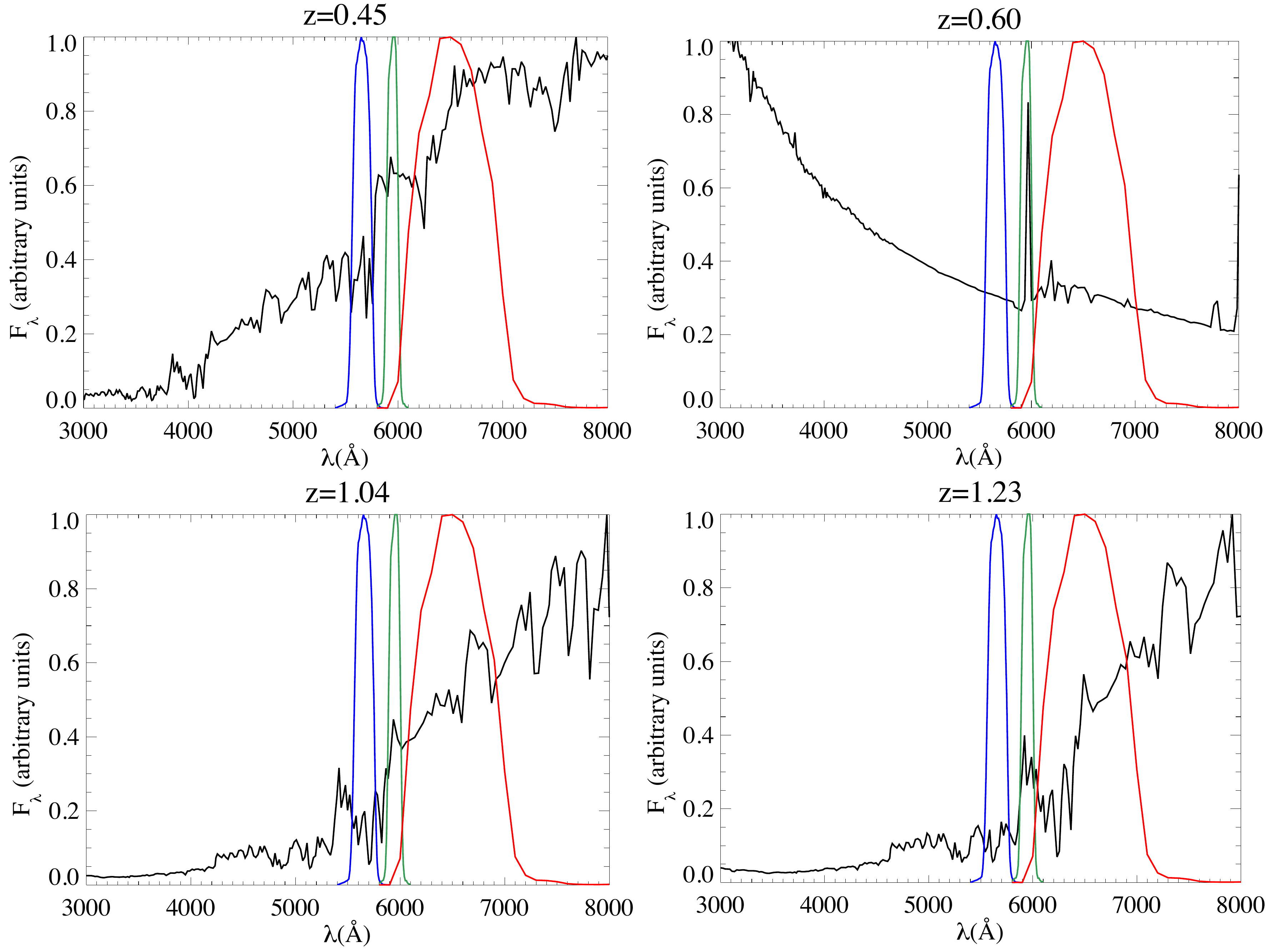, width=\columnwidth}}
\caption{Examples of interlopers that could affect our LBGs selection. We show the galaxy spectra and the position of our three filters over it. \textit{Top left panel:} The spectra of a elliptical galaxy at $z=0.45$, with strong Balmer break located at $\lambda_{obs}=5840$\AA\,  and intense Calcium H \& K absorption. \textit{Top right panel:}  The spectra of a galaxy at $z=0.60$, with intense OII emission line at $\lambda_{obs}=5925$\AA. \textit{Bottom left panel:} The spectra of a galaxy at $z=1.04$, with MgI and MgII absorption at $\lambda_{obs}\sim5650$\AA. \textit{Bottom right panel:} The spectra of a galaxy at $z=1.23$, with a large break at $\lambda_{obs}=5887$\AA. \\}
\label{fig:contaminants} 
\end{figure}

\begin{figure}
\centering{\epsfig{file=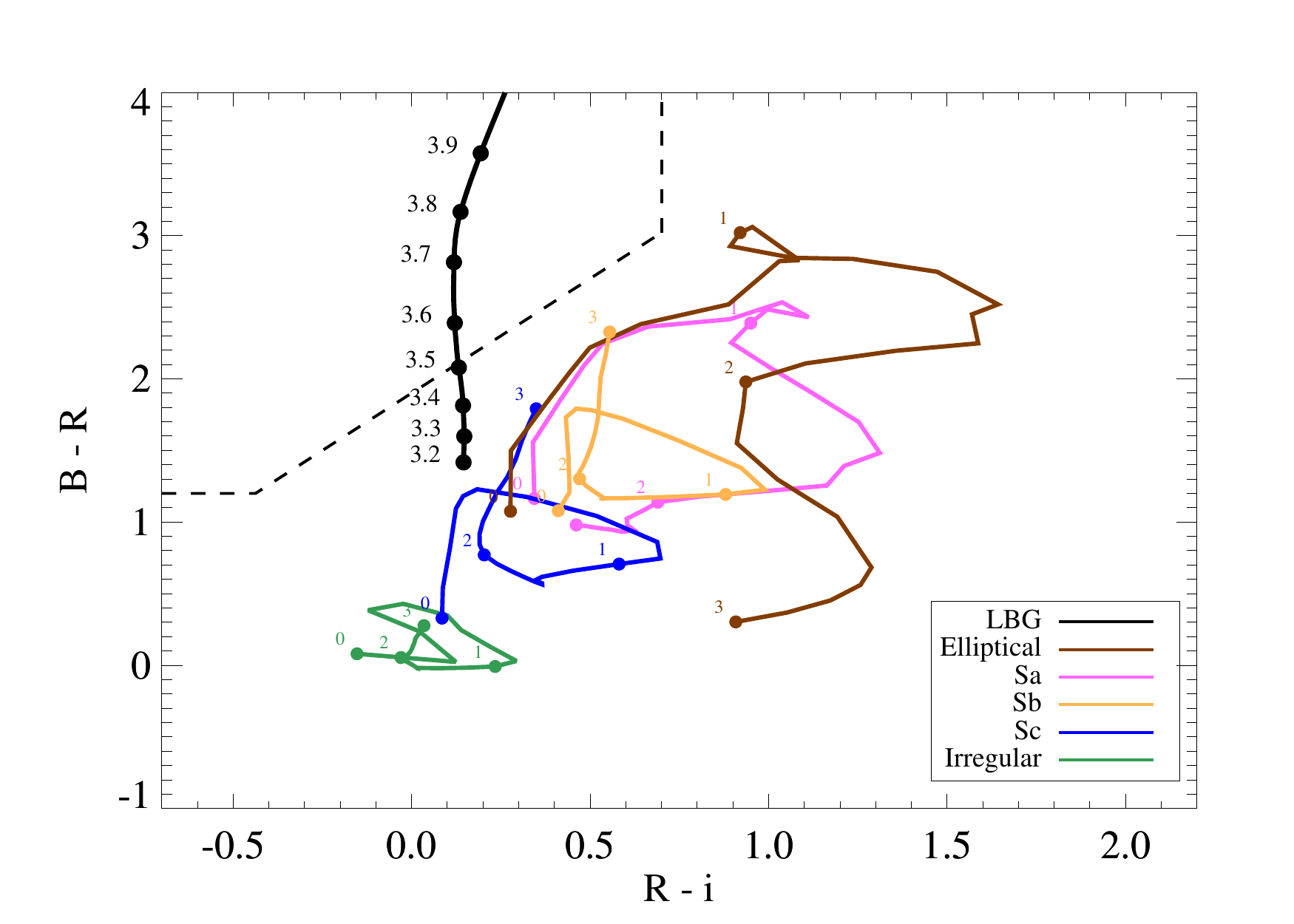, width=\columnwidth}}
\caption{Same as Fig.~\ref{fig:evoltrack_all} but using the filter system used by \citet{Ouchi04a} to select LBGs at $z\sim4$ (broad band filters B, R, i). Filled circles over the black curve indicate colors of LBGs from redshift 3.2 to 3.9. Filled circles over the curve of the low-redshift galaxies indicate colors from redshift 0.0 to 3.0. The dashed line is indicating the region used \citet{Ouchi04a} to select LBGs in their work.\\}
\label{fig:evoltrack_all_ouchi} 
\end{figure}

However Fig.~\ref{fig:evoltrack_all} shows that in our NB filter set,
some of the low-redshift
galaxies have similar colors as $z=3.78$ LBGs, which suggests that our new
filter configuration could make it challenging to
select a sample of LBGs at $z=3.78$ with high completeness and at the same time high
purity. 
When we use NB filters the low-redshift galaxy colors are located in a wider region in the
color-color plot in comparison with the location of the color locus of contaminants
when broad band filters are used. We attribute this to sensitivity of the NB filters
to features in the galaxy spectra such as emission or absorption lines. In the case of
broad bands these features are diluted by averaging over large regions of spectra, but for
NB
the features result in large excursions in color with changing
redshift, making the low-redshift galaxy locus extend over a larger
region of color space that overlaps with the colors of $z=3.78$ LBGs.

Given that LBG colors at $z=3.78$ span the range $\rm
NB_{571}-NB_{596} \gtrsim 0.5$ and $-0.6 \lesssim \rm
NB_{596}-r_{GUNN} \lesssim 0.8$ (see Fig.~\ref{fig:colors_yerr}),
there are several types of contaminants that could be affecting our
LBG selection. Their colors are indicated by points on the
respective low-redshift galaxy evolutionary tracks
are labeled by with letters in
Fig.~\ref{fig:evoltrack_all}, and some examples are shown in
Fig.~\ref{fig:contaminants}. The first type are red galaxies at
$z\sim0.45$ having a large $\lambda_{RF}\sim4000$\AA\, Balmer break
and strong Calcium H \& K absorption. This break is located just
between our two NBs, so they present red colors (point C in
brown curve in Fig.~\ref{fig:evoltrack_all}).
The second type of interlopers are star-forming
galaxies at $z\sim0.60$ with strong [OII] 3727\AA\, emission
lines. If the $\rm NB_{596}$ is located just over this line, and $\rm
NB_{571}$ over the continuum,
we again detect red colors (point A on green curve in 
Fig.~\ref{fig:evoltrack_all}). The third type of interlopers are
galaxies at $z\sim1.04$ with strong MgI and MgII absorption lines at
$\lambda_{RF}=2852$\AA, and $\lambda_{RF}=2799$\AA\, respectively in
combination with the $\lambda_{RF}\sim2900$\AA\, break. When the $\rm
NB_{571}$ is located over this absorption the filter $\rm NB_{596}$
falls on the continuum, then, red colors are detected (points E and G
on the magenta and brown curves, respectively, in Fig.~\ref{fig:evoltrack_all}). Other
interlopers are galaxies with
strong flux breaks redshifted just between our NB
filters. One example are galaxies at $z\sim1.23$ with a large
$\lambda_{RF}\sim2640$\AA\, break (points D and F on the
magenta and brown curves, respectively, in Fig.~\ref{fig:evoltrack_all}) and galaxies at
$z\sim1.83$ with a strong break at $\lambda_{RF}\sim2085$\AA\, (point B on the
brown curve in Fig.~\ref{fig:evoltrack_all}).

\subsection{Selection Region and LBG Sample}
\label{sec:sample}
 
As we are interested in measuring the clustering properties of LBGs at
$z=3.78$, we need to select a sample with high completeness and purity.
In order to avoid
low-redshift contaminants, we were forced to choose a smaller
selection region in the color-color diagram, which results in
relatively low completeness,
 but it ensures that the sample is not highly contaminated. 

First, we defined two vertical color cuts in Fig.~\ref{fig:colors_yerr}, one to the left of the median
LBG colors at $z=3.78$ and one to the right. The first cut is
meant to exclude LBGs located in the upper left region of the diagram, which mostly
corresponds to LBGs at $z\sim3.9$ with strong Ly$\alpha$ line
emission.
The second cut avoids LBGs at $z>3.9$.  A third color
cut defines a lower limit for $\rm NB_{571}-NB_{596}$, which
ensures we are detecting the Ly$\alpha$ break, while
at the same time avoiding LBGs at $z\lesssim3.7$. We used a diagonal color cut, to 
most effectively avoid the contamination of low-redshift galaxies (see Fig.~\ref{fig:evoltrack_all}),
while at the same time including most of the LBGs at $z=3.78$, thus maintaining
the highest completeness possible.

We also tested several different color criteria to select LBGs. In section
\S~\ref{sec:robustness} we will further discuss our color selection,
contamination by low-redshift galaxies, and the impact that contamination
can have on our clustering measurements. There we argue that the choice
of color selection that we present here selects a reasonably complete LBG sample with
high purity.  Our final set of color cuts are shown in \ref{fig:colors_yerr}, and
defined by the following relations:
\begin{eqnarray}
& & \rm NB_{571}-NB_{596}> 1.2 \nonumber \\
& & -0.6 <   \rm NB_{596}-r_{GUNN} <0.8 \nonumber \\
& & \rm NB_{571}-NB_{596} >  0.7 (NB_{596}-r_{GUNN}) + 0.9 
\label{eq:colorcut}
\end{eqnarray}

We selected LBGs based on our galaxy photometry, but required sources
to have a ${\rm S\slash N}\ge 4.0$ in both the $\rm NB_{596}$ and $\rm
r_{GUNN}$ filters, to ensure a solid detection of the LBG
continuum. 
In order to reduce contamination by false detections, we only
considered objects that have FLAGS$=0$ in SExtractor, which excluded
objects that were blended, saturated, truncated (too close to an image
boundary), or affected by very bright neighboring objects.  Bright
stars in our images were masked in order to avoid spurious object
detection due to contamination from their stellar flux. This procedure
results in a set of masks indicating where we were able to detect
galaxies, which we use later in our clustering analysis to compute the
effective area of our survey.

We also imposed a lower limit on the magnitude in order to exclude
bright low-redshift interloper galaxies from our selection.  Thus we
only considered objects with magnitudes fainter than $\rm r_{GUNN} =
23.97$, corresponding to LBGs with $L\sim3.5\,\rm L_{*}$. We chose
this value by computing the LBG luminosity function at $z\sim4$, and
finding the bright end cut at which we would lose no more than 1\% of
the galaxies. In other words, 99\% of the total number of LBGs have
magnitudes between our bright end cut of $\rm r_{GUNN} = 24.0$ and
the limiting magnitude $\rm r_{GUNN} = 25.82$ (mean limit magnitude at
$4\sigma$ for a $2\arcsec$ diameter aperture) of our images, which
corresponds to $L=0.76\,\rm L_{*}$. In this way we can safely assume
we are excluding only extremely rare bright LBGs.  For the LBG
luminosity function we used the Schechter parameters from
\citet{Ouchi04a} who studied the photometric properties based on a
large sample of $\sim 2200$ LBGs at $z\sim4$.
The values used are
$\phi^{*}=2.8\times10^{-3}\,h^{3}_{70}$\,Mpc$^{-3}$,
M$^{*}_{1700}=-20.6$ mag, and $\alpha = -1.6$.

\begin{figure}
\centering{\epsfig{file=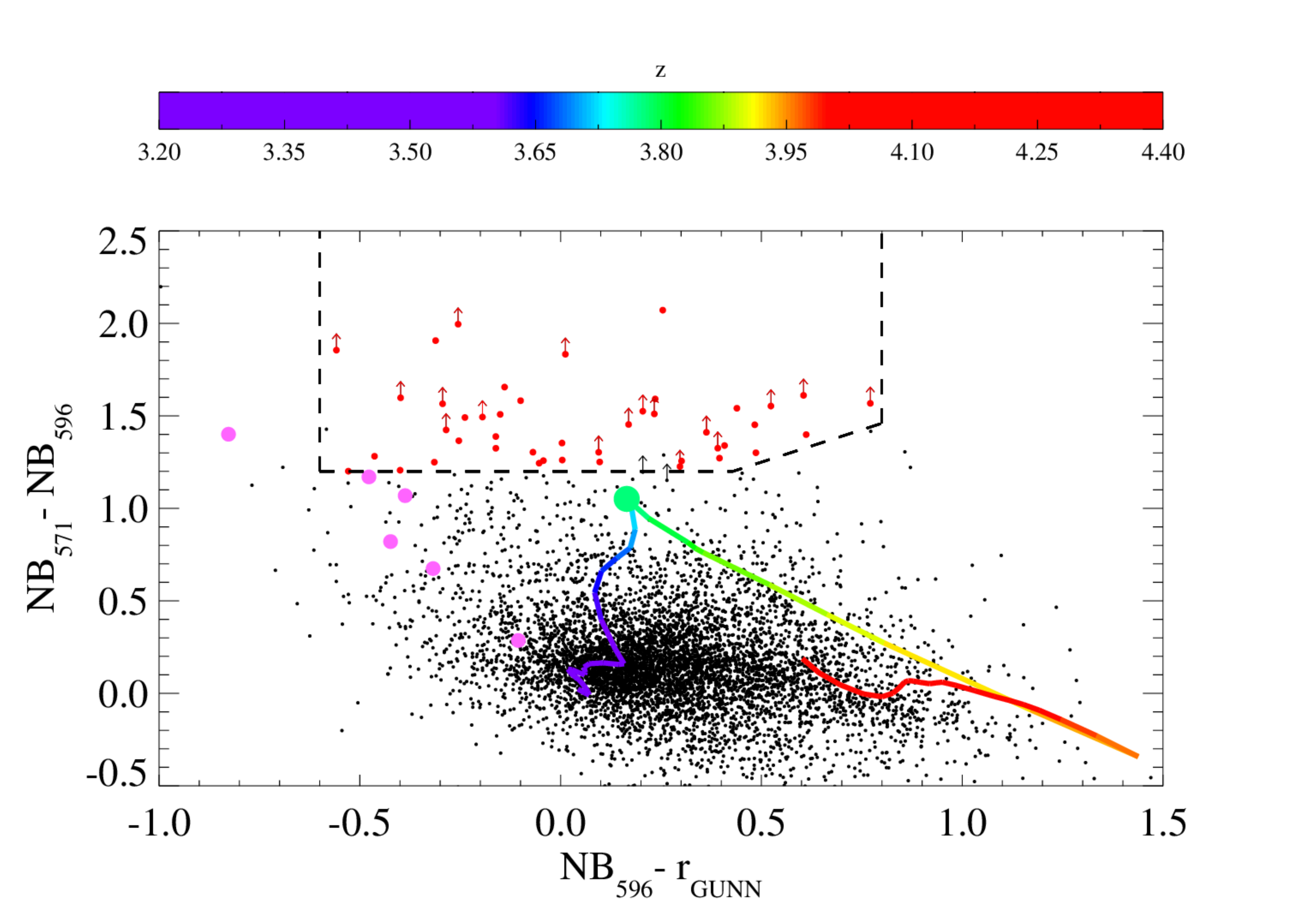, width=\columnwidth}}
\caption{Color-color diagram for the six stacked QSO fields. Here the evolutionary track showed in Fig.~\ref{fig:evoltrack_all} is plotted as redshift color-coded track according to the color bar. We have highlighted the selected LBGs as red points. The magenta points indicate the color of each QSO in the filters. Arrows indicate lower limits for $\rm NB_{571}-NB_{596}$ color. These are cases in which the object was not detected in $\rm NB_{571}$ filter at 2$\sigma$ level and magnitude was replaced by the corresponding limit magnitude.\\} 
\label{fig:colorcolor_together}
\end{figure}

\begin{figure}
\centering{\epsfig{file=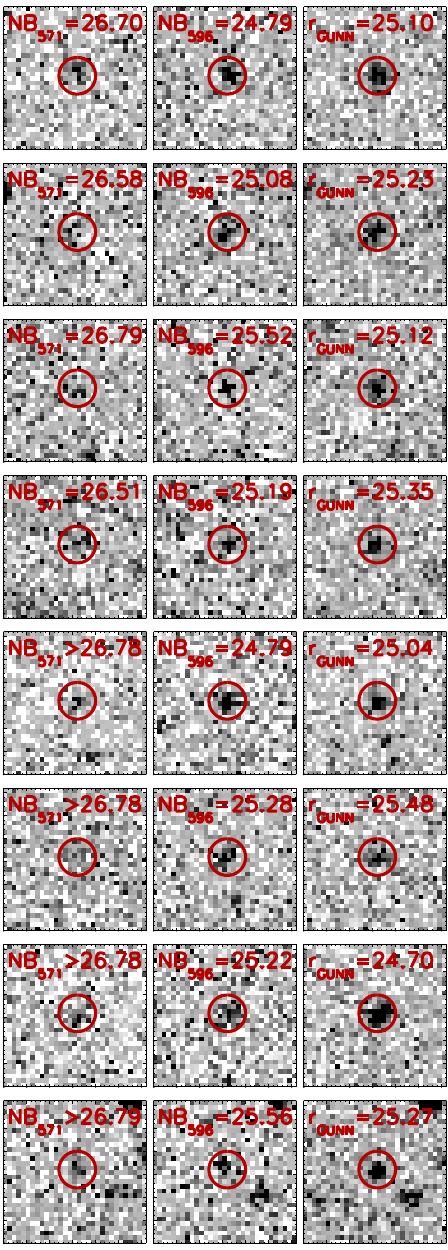, width=6.9cm}}
\caption{Images of some selected LBGs. From left to right we show the $\rm NB_{571}$, $\rm NB_{596}$ and $\rm r_{GUNN}$ images. Each panel is $7.5\arcsec$ on a side. The red circle show the position of the detected object, and its size correspond to the region in which the photometry was done ($2\arcsec$ in diameter). The magnitudes are indicated in each panel.\\} 
\label{fig:dropouts}
\end{figure}

\begin{figure}
\centering{\epsfig{file=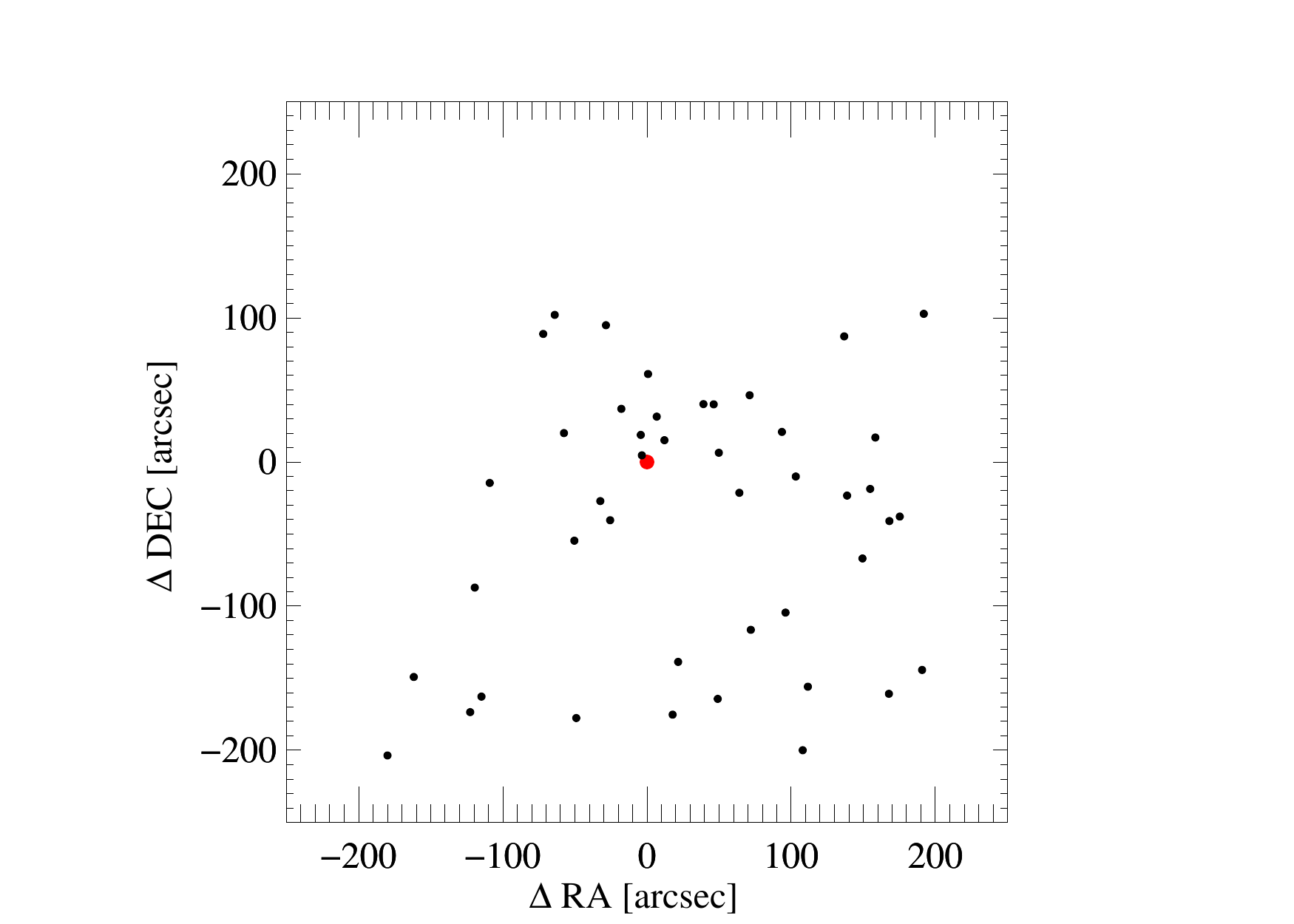, width=\columnwidth}}
\caption{The distribution of LBGs around the QSO in the plane of the sky for the six stacked fields. The central QSOs is located in 0.0 and is plotted by a large red circle. \\} 
\label{fig:distribution} 
\end{figure}

\begin{figure*}
\centering{\epsfig{file=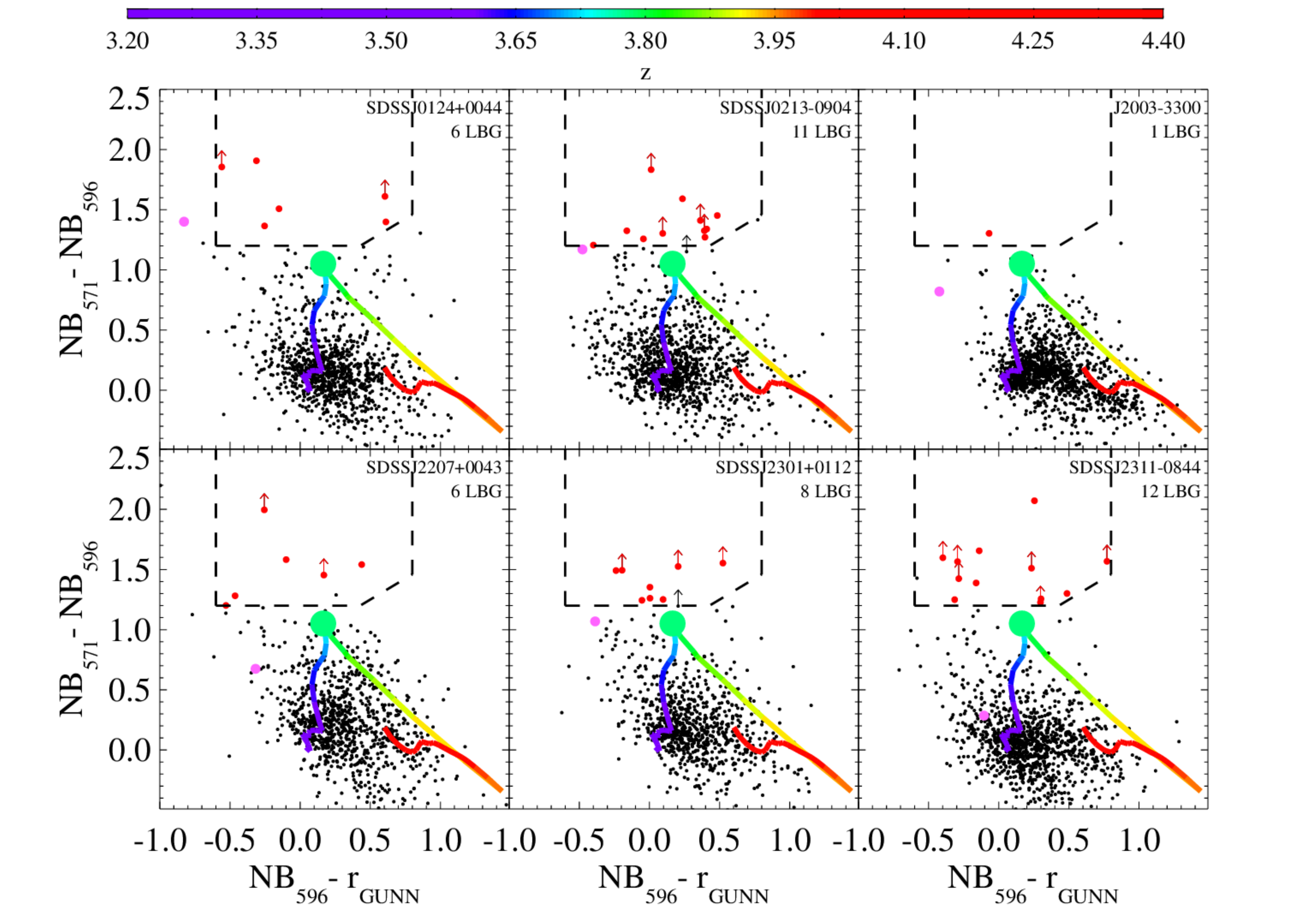, width=\textwidth}}
\caption{The same as in Fig.~\ref{fig:colorcolor_together} but for the six individual QSO fields. At the top right of each plot the number of LBGs found is shown.\\} 
\label{fig:colorcolor}
\end{figure*}

\begin{figure}
\centering{\epsfig{file=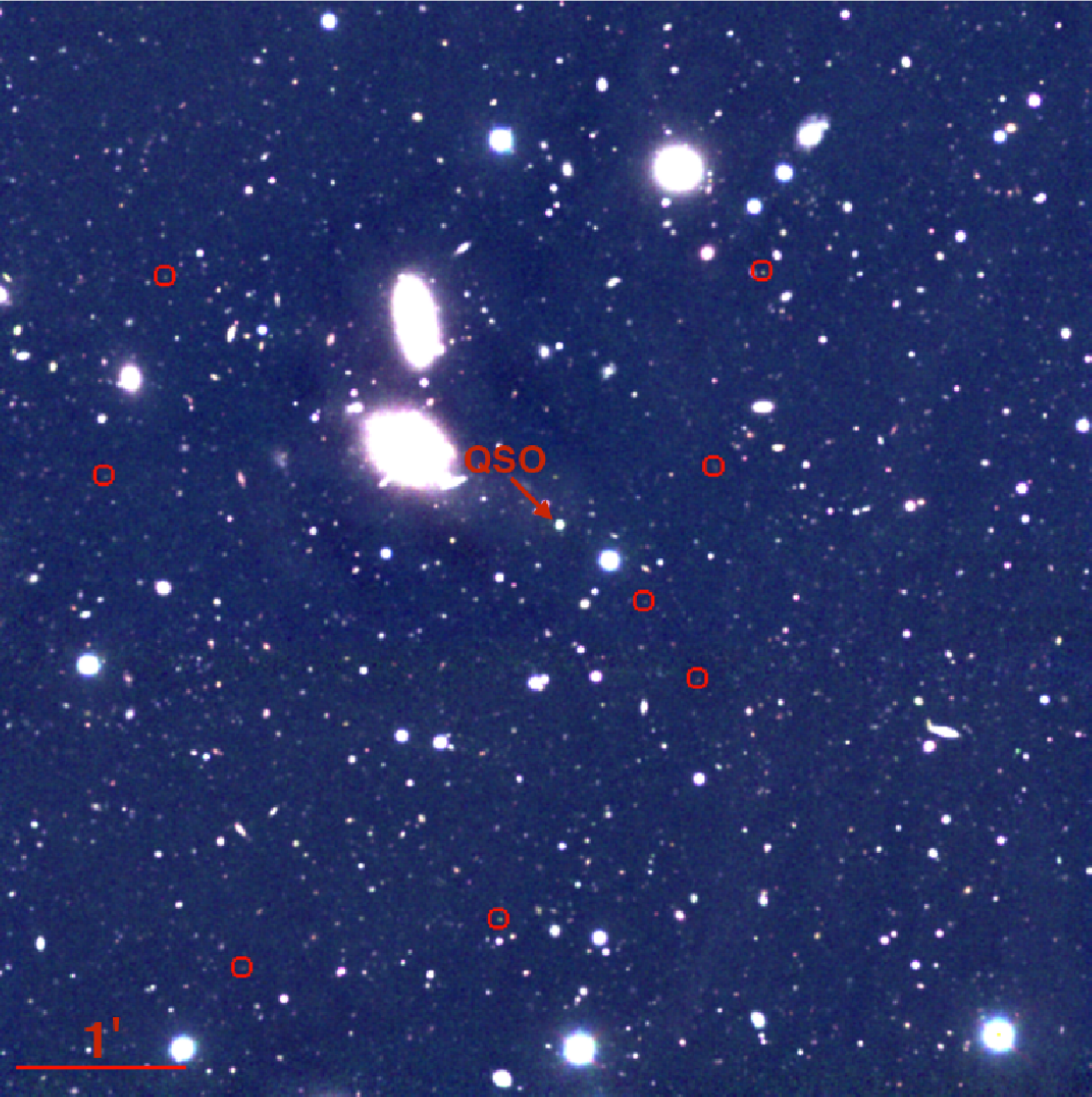, width=\columnwidth}}
\caption{False color image of 42 arcmin$^{2}$ of the field SDSSJ2301$+$0112. Red circles indicate the LBGs candidates positions. \\} 
\label{fig:image_color} 
\end{figure}

Given all of these selection criteria and the color cuts defined in
eqn.~(\ref{eq:colorcut}), we selected LBGs in each of our
fields. We compute the total area of our survey by adding the
effective area of each individual field, which is defined by
subtracting the masked area from the total area of the image. The
the total area of our survey is 232.7 arcmin$^{2}$ corresponding to
an average area per image of 38.79 arcmin$^{2}$ (recall the FOV of
FORS1 is  $6.8\times6.8$ arcmin$^{2}$ or 46.24 arcmin$^{2}$). 
We show color-color diagrams of
objects detected in all six of our fields in Fig.~\ref{fig:colorcolor_together}.
We found a total of 44 LBGs (see
Table \ref{table:LBGsample}) corresponding to a mean number density of
0.19 LBGs arcmin$^{-2}$.
Image cutouts in our three filters for several of our selected LBGs 
are shown in Fig.~\ref{fig:dropouts}. In Fig.~\ref{fig:distribution} we show the spatial distribution of the LBGs relative to the QSO
(red dot at zero) for our six fields.
We also show the individual color-color diagrams and indicate the number of LBGs found in each
individual field in Fig.~\ref{fig:colorcolor}. Note that the number of LBGs in the fields
cannot be directly compared because each image has different limiting magnitude and different 
effective area (different reduced image size, masked region, etc).
In Fig.~\ref{fig:image_color} we show a false color image of the field around QSO
SDSS~J2301$+$0112 with the LBG candidate positions indicated.

\begin{deluxetable*}{lrrrrr}
\tabletypesize{\scriptsize}
\tabletypesize{\small}
\tablecaption{LBGs sample. The magnitudes correspond to AB magnitudes measured in a $2\arcsec$ diameter aperture for each band.\label{table:LBGsample}\\}
\tablewidth{0pt}
\tablewidth{0.8\textwidth}
\tablehead{
\colhead{ID}&
\colhead{RA}&
\colhead{DEC}&
\colhead{$\rm r_{GUNN}$}&
\colhead{$\rm NB_{571}$}&
\colhead{$\rm NB_{596}$}\\
\colhead{}&
\colhead{(J2000)}&
\colhead{(J2000)}&
\colhead{}&
\colhead{}&
\colhead{}
}
\startdata
SDSSJ0124$+$0044$\_$1& 20.9854&  0.7384&24.58&$>$26.80&25.19\\  
SDSSJ0124$+$0044$\_$2& 21.0644&  0.7319&25.50&$>$26.80&24.94\\  
SDSSJ0124$+$0044$\_$3& 21.0086&  0.7312&25.24& 26.35&24.98\\      
SDSSJ0124$+$0044$\_$4& 21.0624&  0.7310&25.10& 26.70&24.79\\      
SDSSJ0124$+$0044$\_$5& 21.0159&  0.7594&25.23& 26.58&25.08\\      
SDSSJ0124$+$0044$\_$6& 21.0145&  0.7476&24.69& 26.70&25.30\\      
SDSSJ0213$-$0904$\_$1& 33.3153& -9.1322&25.09&$>$26.94&25.10\\   
SDSSJ0213$-$0904$\_$2& 33.3341& -9.1315&24.49& 26.32&24.73\\      
SDSSJ0213$-$0904$\_$3& 33.2946& -9.1311&24.50& 26.43&24.98\\      
SDSSJ0213$-$0904$\_$4& 33.3429& -9.1285&25.16&$>$26.94&25.53\\   
SDSSJ0213$-$0904$\_$5& 33.3763& -9.1275&24.60& 25.41&24.20\\      
SDSSJ0213$-$0904$\_$6& 33.2968& -9.1281&25.54&$>$26.94&25.63\\   
SDSSJ0213$-$0904$\_$7& 33.2836& -9.1243&25.22&$>$26.94&25.61\\   
SDSSJ0213$-$0904$\_$8& 33.3310& -9.0741&25.12& 26.79&25.52\\      
SDSSJ0213$-$0904$\_$9& 33.2955& -9.1070&25.39& 26.61&25.35\\      
SDSSJ0213$-$0904$\_$10&33.3111& -9.0545&25.35& 26.51&25.19\\      
SDSSJ0213$-$0904$\_$11&33.3831& -9.0543&24.88& 26.62&25.28\\      
J2003$-$3300$\_$1&300.8540&-32.8583&25.06& 26.29&24.99\\           
SDSSJ2207$+$0043$\_$1&331.8270&  0.6683&25.93& 26.75&25.47\\     
SDSSJ2207$+$0043$\_$2&331.9070&  0.6693&25.16&$>$26.78&25.33\\  
SDSSJ2207$+$0043$\_$3&331.9030&  0.7306&25.04&$>$26.78&24.79\\  
SDSSJ2207$+$0043$\_$4&331.9200&  0.7197&24.71& 26.69&25.14\\     
SDSSJ2207$+$0043$\_$5&331.8970&  0.6925&24.03& 24.70&23.50\\
SDSSJ2207$+$0043$\_$6&331.9300&  0.6848&25.01& 26.49&24.91\\
SDSSJ2301$+$0112$\_$1&345.3280&  1.1687&25.23& 26.49&25.23\\
SDSSJ2301$+$0112$\_$2&345.2810&  1.2176&25.05&$>$26.78&25.25\\
SDSSJ2301$+$0112$\_$3&345.3410&  1.2168&25.26& 26.45&25.21\\
SDSSJ2301$+$0112$\_$4&345.2880&  1.2045&25.48&$>$26.78&25.28\\
SDSSJ2301$+$0112$\_$5&345.2830&  1.1969&24.70&$>$26.78&25.22\\
SDSSJ2301$+$0112$\_$6&345.3030&  1.1735&24.49& 25.74&24.25\\
SDSSJ2301$+$0112$\_$7&345.3350&  1.2363&24.91& 26.26&25.01\\
 SDSSJ2301$+$0112$\_$8&345.2770&  1.2367&24.16& 25.51&24.16\\
 SDSSJ2311$-$0844$\_$1 &347.8960& -8.7096&25.69& 26.63&25.38\\
 SDSSJ2311$-$0844$\_$2 &347.9240& -8.7231&24.45&$>$26.79&25.22\\
 SDSSJ2311$-$0844$\_$3 &347.9170& -8.7249&25.65&$>$26.79&25.36\\
SDSSJ2311$-$0844$\_$4 &347.9030& -8.7347&25.05&$>$26.79&25.28\\
SDSSJ2311$-$0844$\_$5 &347.9330& -8.7388&24.72& 26.24&24.59\\
SDSSJ2311$-$0844$\_$6 &347.9220& -8.7419&25.52&$>$26.79&25.22\\
SDSSJ2311$-$0844$\_$7 &347.9430& -8.7424&25.27&$>$26.79&25.56\\
 SDSSJ2311$-$0844$\_$8 &347.9460& -8.7546&24.71& 26.50&25.20\\
 SDSSJ2311$-$0844$\_$9 &347.9310& -8.7650&25.59&$>$26.79&25.19\\
  SDSSJ2311$-$0844$\_$10&347.9150& -8.7248&24.32& 26.65&24.58\\
  SDSSJ2311$-$0844$\_$11&347.8990& -8.7257&24.30& 25.86&24.60\\
 SDSSJ2311$-$0844$\_$12&347.9180& -8.7342&25.37& 26.60&25.21
\enddata
\tablenotetext{}{\\}
\end{deluxetable*}

\subsection{Redshift Selection Function}
\label{sec:comple}

\begin{figure}
\centering{\epsfig{file=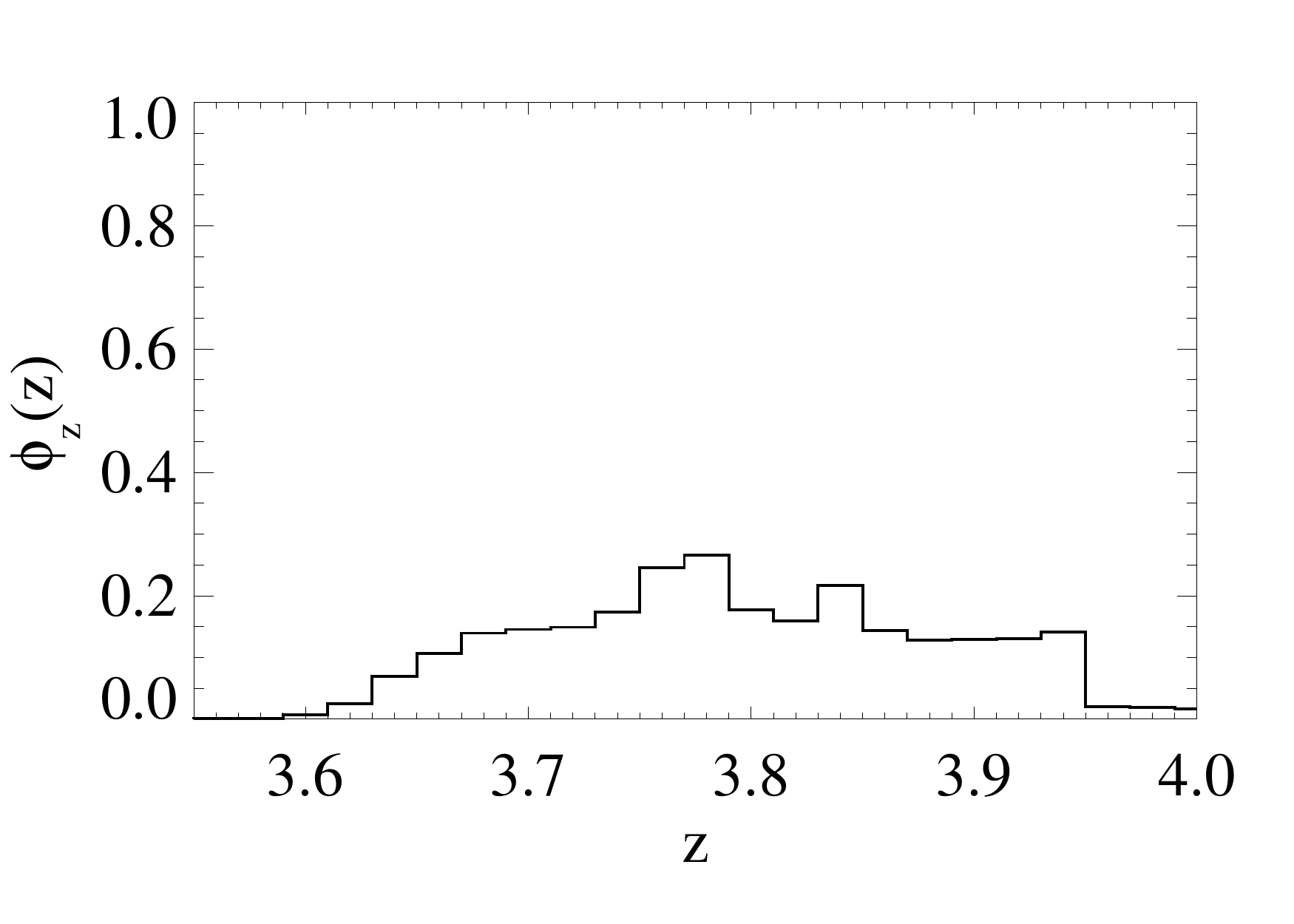, width=\columnwidth}}
\caption{Redshift selection function of the LBGs selection. The completeness per redshift bin was determined from 1000 simulated LBGs spectra with different EW$_{\rm Ly\alpha}$, UV continuum slope, and $\rm r_{GUNN}$ magnitudes. This is calculated by computing the fraction of the LBG simulated colors, per redshift bin, which was selected by the selection region. This defines the redshift selection function which is used for the clustering measurements in section \S~\ref{sec:clustering}.\\}
\label{fig:completeness}
\end{figure}

We used the LBG color modeling machinery described
in \S~\ref{sec:modeling} to compute the redshift
selection function $\phi_{z}(z)$ of our LBG color-selection criteria. At each redshift
step, we redshift the 1000 rest-frame simulated LBG spectra into the observed frame,
draw luminosities from the luminosity function, compute  magnitudes and colors,
and add photometric errors. 
We then compute the completeness at each redshift by calculating the
fraction of simulated LBGs that satisfy the selection criteria defined
in \S~\ref{sec:sample}, namely: fulfill the color criteria in
eqn.~(\ref{eq:colorcut}), and fulfill the magnitude constraints (given
by the $4\sigma$ limiting magnitudes for $\rm NB_{596}$ and $\rm
r_{GUNN}$ and by the bright end cut imposed for our selection, $\rm
r_{GUNN}>24.0$).
Note that as the limiting magnitude of our fields
are slightly different, we computed $\phi_{z}(z)$ for each individual field, using their corresponding 
$\rm NB_{596}$ and $\rm r_{GUNN}$ limiting magnitudes. 
The final $\phi_{z}(z)$ varied from field to field by a small amount,
then we computed the median of $\phi_{z}(z)$ over the six fields, 
which is shown in Fig.~\ref{fig:completeness}.

From Fig.~\ref{fig:completeness} we see that our selection criteria
recover a peak value of 
$\simeq 26\%$ of LBGs at $z\sim3.78$ over a small $\Delta z\sim0.04$
interval. Our criteria also select $\gtrsim10\%$ of LBGs
over  a wider redshift interval, ranging from $z\sim3.65$ to $z\sim3.95$
($\Delta z\sim0.3$) which corresponds to $\sim18,800\,{\rm
  km\,s}^{-1}$ ($\sim 167\,h^{-1}$\,cMpc) at $z=3.78$. The NB technique selects 
  LBGs over a much narrower redshift range compared to broad band
LBG  selection, which typically selects galaxies over a range $\Delta z\sim 1.0$ \citep[e.g.][]{Ouchi04a, Bouwens07, Bouwens10}, or $\sim3.3$ times larger than our selection. 
However, as low-redshift galaxies have similar colors as the LBGs in our filters, we
have to adopt relatively conservative color cuts making our completeness
relatively low. 

 \section{Clustering Analysis}
 \label{sec:clustering}
 
In this section we analyze the clustering of LBGs around
our QSOs at $z=3.78$. First, we present the measurement of the QSO-LBG
cross-correlation function in \S~\ref{sec:CCF}.
We then estimate the correlation function parameters ($r_{0},\gamma$)
assuming a power law form $\xi(r)=(r/r_{0})^{-\gamma}$. 
Our
results are compared with theoretical expectations based
on the  auto-correlation of both LBGs and QSOs at $z\sim4$ assuming linear bias.  In
\S~\ref{sec:ACF} we present the LBG auto-correlation computed from our QSO fields,
and finally in \S~\ref{sec:disc} we compare our results with previous measurements. 

\subsection{The QSO-LBG Cross-Correlation Measurement}
\label{sec:CCF}

Following convention, we study clustering using the two-point correlation function $\xi(r)$,
which measures the excess probability over a random distribution, of finding an object at separation $r$ from another random object, in a volume element $dV$ \citep{Peebles80}. For the case of galaxies around QSOs
this correlation function is defined by 
\begin{equation}
dP = n_{\rm G} [1+\xi_{QG} (r)]dV
\label{eq:prob}
\end{equation} 
where $\xi_{QG}(r)$ is the QSO-galaxy cross-correlation function and
$n_{\rm G}$ is the mean number density of galaxies in the universe. Here
$r$ is real space comoving distance, which is however not the
observable even when redshift information is available, as peculiar velocities
induce redshift space distortions along the line-of-sight
\citep{Sargent77}. Typically LBG clustering studies that
lack redshifts measure the angular correlation function \citep[e.g.][]{Giavalisco98, Ouchi04b,
  Lee06}. Although we do not have redshifts of our LBGs, our
NB selection technique selects LBGs over a narrow
redshift interval $\Delta z \simeq 0.3$ (see Fig.~\ref{fig:completeness} in \S~\ref{sec:comple})
which allows us to measure clustering as a function of transverse comoving
distance instead of angular distance
(at $z=3.78$, the angular diameter distance changes by just
$\simeq 3\%$ over this redshift interval). 
Thus we write the real space separation $r$ as $r^{2}=R^{2}+Z^{2}$, where
$R$ is the transverse comoving distance between the QSO and the galaxy,
and $Z$ is the radial comoving distance between them approximately given by:
\begin{equation}
\Delta Z=\frac{c}{H(z)} \Delta z
\label{eq:comoving}
\end{equation}
where $H(z)$ the Hubble constant evaluated at redshift $z=3.78$, which we take to be a
constant over the redshift interval considered (an approximation valid to $\simeq 5\%$).  

We measure the volume-averaged projected cross-correlation
function between QSOs and LBGs $\chi (R_{\rm min}, R_{\rm max})$,
which is a dimensionless quantity defined as the real space
QSO-LBG cross-correlation function $\xi_{QG} (R,Z)$ integrated over a volume
and then normalized by it \citep[e.g.][]{Hennawi06, Hennawi15}:
\begin{equation}
\chi (R_{\rm min}, R_{\rm max}) = \frac{\int \xi_{QG} (R,Z)  dV_{\rm eff}}{V_{\rm eff}}, 
\label{eq:estimator2_simpler}
\end{equation}
where $V_{\rm eff}$ is a cylindrical volume defined by the radial bin $[R_{\rm min}, R_{\rm max}]$,
the height $\Delta Z$ probed by our filter configuration, and modulated by the selection function
of our survey. We measure  $\chi (R_{\rm min}, R_{\rm max})$ in logarithmically spaced radial bins centered on
the QSO for all fields using the estimator:
\begin{equation}
\chi (R_{\rm min}, R_{\rm max})=  \frac{\langle QG\rangle}{\langle QR\rangle} -1 
\label{eq:estimator}
\end{equation}
where $\langle QG\rangle$ and $\langle QR\rangle$ are the number of QSO-LBG and QSO-random pairs in this
cylindrical volume.
The quantity $\langle QG\rangle$ is directly measured by counting the QSO-LBG
pairs found in our images.

The quantity $\langle QR\rangle$ is the expected random number of QSO-LBG pairs, which is given by:
\begin{equation}
\langle QR\rangle =  n_{\rm G}(z,{\rm r_{GUNN}^{lower}}<{\rm r_{GUNN}}<{\rm r_{GUNN}^{limit}}) V_{\rm eff}, 
\label{eq:NLBG2}
\end{equation}
where $V_{\rm eff}$ is the effective volume of the radial bin in
question and $n_{\rm G}(z,{\rm r_{GUNN}^{lower}}<{\rm r_{GUNN}}<{\rm
  r_{GUNN}^{limit}})$ is the mean number density of LBGs at redshift
$z$ in the magnitude range of our survey, which will be henceforth
referred to as $n_{\rm G}$ to simplify notation. Given that galaxy
clustering measurements are normally performed in
random locations of the universe, 
the mean number density
measured from the survey is typically a good proxy for the
mean number density of the universe, provided the survey
volume is large enough. In such cases, the galaxy number
counts for the random sample can be computed
from the data itself, and one typically constructs
random catalogs with a number density determined from the survey
to estimate $\langle QR\rangle$. However, in our case we are pointed
towards a QSO situated in what is likely to be an overdensity. 
Therefore the mean number density
of galaxies in our survey is not representative of the mean in random locations and
 we cannot follow the standard procedure for computing $\langle QR\rangle$.

If we had observations of control fields (i.e not centered on QSOs)
with our same filter configuration, then it would be possible to
measure the background number density of LBGs directly and determine
$\langle QR\rangle$. Another alternative would be to measure this
quantity directly from the outer parts of images, where the clustering
becomes negligible, given a sufficiently large FOV instrument.
Unfortunately, we do not have images of control fields, and the
FOV of FORS1 is too small to provide a reliable measurement of the
background. Thus our only alternative is to estimate $\langle
QR\rangle$ from eqn.~(\ref{eq:NLBG2}), where $n_G$ is calculated from
the $z\sim 4$ LBG luminosity function, and the effective volume
$V_{\rm eff}$ is determined from our Monte Carlo simulations of our
selection function (see \S~\ref{sec:comple}) and the effective area
covered by our survey. We provide further details of these
computations in what follows.

To calculate $n_{\rm G}$ we used the Schechter parameters from
\citet{Ouchi04a} $z\sim 4$ LBG luminosity function. We integrated
this luminosity function over the magnitude limits given by our LBG selection, and this magnitude integral was
weighted by the photometric completeness fraction of our source detection following the same
procedure described in \S~\ref{sec:modeling}.
Given that our fields all have slightly different limiting magnitudes
and different source completeness, we compute the expected $n_{\rm G}$ for
each field. We assumed that $n_{\rm G}$ is constant over the redshift
ranged considered, which is a good approximation given the narrow
redshift range $\Delta z \simeq 0.3$ that we probe. The expected mean
number density of LBGs in random fields, $n_{\rm G}$, computed for the
magnitude range of each field is given in Table
\ref{table:field_param}.

We define the effective volume of a radial bin as:
\begin{equation} \label{eq:veff}
V_{\rm eff} =  \int_{Z_{\rm min}}^{Z_{\rm max}} \int_{R_{\rm min}}^{R_{\rm max}} \phi(R,Z) 2\pi R dR dZ 
\end{equation}
where $\phi(R,Z)$ encodes the geometry of the survey, which can be separated into the radial $R$
and the redshift (line-of-sight) $Z$ selection function as $\phi(R,Z)=\phi_{Z}(Z)\phi_{R}(R)$.
The redshift selection function of our survey $\phi_{z}(z)$ was modeled in \S~\ref{sec:comple}
and we convert it to a redshift selection function in comoving units $\phi_{Z}(Z)$ using eqn.~(\ref{eq:comoving}). Then, we integrate it over the redshift range covered by our Monte Carlo modeling (corresponding to $3.2 \leq z \leq 4.4$). 

The radial selection
function $\phi_{R}(R)$ is easily calculated using the detection masks for our images. We created catalogs
with randomly distributed galaxies with number density $n_{\rm ran}$ such that
we had $\sim 100,000$ sources in the entire image.
Then we computed $\phi_{R}(R)$ in radial bins as the ratio between the number of randomly distributed
galaxies and the expected number without masking $n_{\rm ran}\pi (R^{2}_{\rm max}-R^{2}_{\rm min})$.
The resulting $\phi_{R}(R)$ then quantifies the fraction of the bin area where we could have detected
LBGs. We computed the value of $V_{\rm eff}$ for each radial bin in each field using eqn.~(\ref{eq:veff}).
Summing the $V_{\rm eff}$ over the radial bins, we obtain the total volume covered by each of our six
fields $V_{\rm field}$, given in Table~\ref{table:field_param}. We obtained that the total volume of 
our survey is 14,782 $h^{-3}$\,cMpc$^{3}$.

To obtain a rough estimate of the LBG overdensity in our QSO fields, we calculated the expected
number of random QSO-LBGs pairs $\langle QR\rangle_{\rm field}$ for each of our fields and compare
to the number we find $\langle QG\rangle_{\rm field}$. These results are tabulated in
Table~\ref{table:field_param}, where we also show the overdensity per field $\langle QG\rangle_{\rm field}/\langle QR\rangle_{\rm field}$. We see that five of our six fields exhibit an LBG overdensity of LBGs, while
one appears to be underdense. Adding up the results for all six fields, we find that the
random expectation is $\langle QR\rangle=28.6$ LBGs, whereas we detected a total of $\langle QG\rangle = 44$
LBGs, giving an overall overdensity of 1.5, and indicating that our fields are on average overdense.

\begin{deluxetable*}{l c c c c c}
\tabletypesize{\small}
\tabletypesize{\scriptsize}
\tablecaption{LBG Overdensity in each individual field.\label{table:field_param}}
\tablewidth{0pt}
\tablewidth{0.8\textwidth}
\tablehead{
\colhead{Field}&
\colhead{$n_{\rm G}$}&
\colhead{$V_{\rm field}$}&
\colhead{$\langle QR\rangle_{\rm field}$}&
\colhead{$\langle QG\rangle_{\rm field}$}&
\colhead{Overdensity}\\
\colhead{(1)}&
\colhead{(2)}&
\colhead{(3)}&
\colhead{(4)}&
\colhead{(5)}&
\colhead{(6)}
}
\startdata												
SDSSJ0124+0044 & 2.15&  2600.20&5.60& 6  & 1.07\\
SDSSJ0213--0904 & 1.79&  2860.50&5.12& 11& 2.15\\
J2003--3300 & 1.71&  2303.21&3.94& 1 & 0.25\\
SDSSJ2207+0043 & 1.93&  2032.63&3.92& 6 & 1.53\\
SDSSJ2311--0844 & 2.13&  2504.83&5.34& 8 & 1.50\\
SDSSJ2301+0112 & 1.88& 2480.15&4.66 & 12 & 2.58
\enddata
\tablecomments{
(1) Field ID, 
(2) The mean number density of $z\sim4$ LBGs in units of $(10^{-3}\,h^{3}$\,cMpc$^{-3})$, in the magnitude range of the survey $r_{GUNN}^{lower}<{\rm r_{GUNN}}<{\rm r_{GUNN}^{limit}}$. Given that $\rm r_{GUNN}^{limit}$ and completeness in the source detection are different for each field, we obtain a number density slightly different for each one, 
(3) Total volume of the field in units of $(h^{-3}$\,cMpc$^{3})$, computed as $V_{\rm field} = \sum_{i=1}^{N_{\rm bins}} V_{\rm eff, i}$,  
(4) Total number of expected LBGs on the whole field computed as $\langle QR\rangle_{\rm field} = n_{\rm G}V_{\rm field}$, 
(5) Total number of observed QSO-LBG pairs on the whole field,  
(6) Total overdensity per field, computed as $\langle QG\rangle_{\rm field}/\langle QR\rangle_{\rm field}$.\\
}
\end{deluxetable*}

\begin{figure}[t!]
 \centering{\epsfig{file=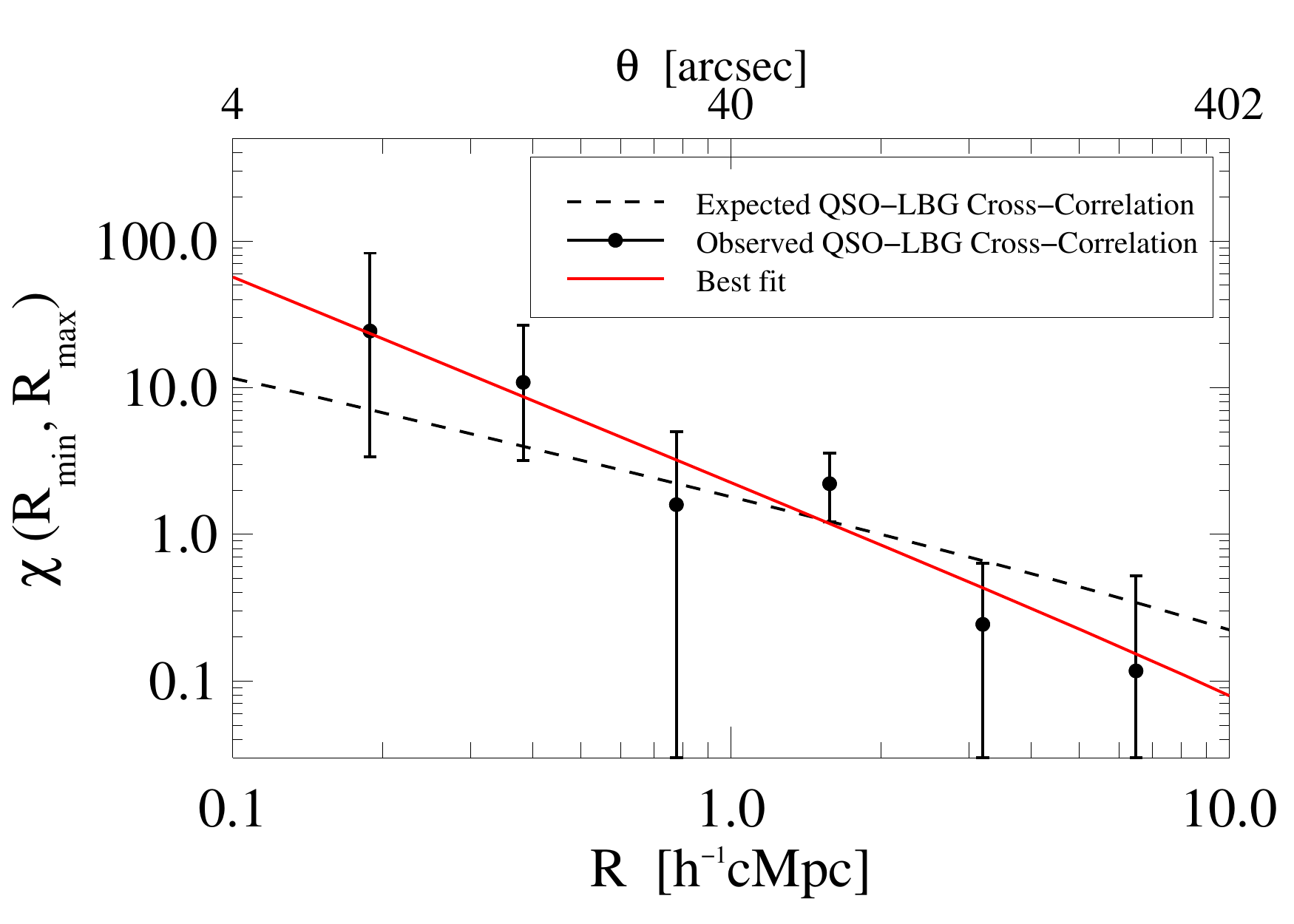, width=\columnwidth}}
\caption{\label{fig:MLE} QSO-LBG cross-correlation function and its maximum likelihood model. The filled circles are showing our measurement described in \S~\ref{sec:CCF} with $1\sigma$ Poisson error bars. The solid red curve shows the best MLE for both $r^{QG}_{0}$ and $\gamma$ as free parameters. We obtain $r^{QG}_{0}=6.93 \,h^{-1}\,{\rm cMpc}$ and $\gamma=2.4$. The dashed black line shows the theoretical expectation of $\chi (R_{\rm min}, R_{\rm max})$ for the six stacked fields calculated from the independently determined QSO and LBGs auto-correlation functions, assuming a linear bias model.\\}
\end{figure}

To explore the profile of this overdensity around QSOs, we computed
$\langle QG\rangle$ and $\langle QR\rangle$ in bins of transverse distance for each of our six fields,
and then summed them to determine the binned volume averaged cross-correlation function $\chi (R_{\rm min}, R_{\rm max})$ according to eqn.~(\ref{eq:estimator}). These results are tabulated in Table \ref{table:chi}
and plotted in Fig.~\ref{fig:MLE}. We estimate errors on  $\chi (R_{\rm min}, R_{\rm max})$ assuming
that shot-noise dominates the error budget, and use the one-sided Poisson confidence intervals for small number statistics from \citet{Gehrels86}. 

\begin{deluxetable*}{c c r r r r}
\tabletypesize{\small}
\tabletypesize{\scriptsize}
\tablecaption{QSO-LBG Cross-Correlation Function.\label{table:chi}}
\tablewidth{0pt}
\tablewidth{0.8\textwidth}
\tablehead{
\colhead{$R_{\rm min}$}&
\colhead{$R_{\rm max}$}&
\colhead{$\langle QG\rangle$}&
\colhead{$\langle QR\rangle$}&
\colhead{$\chi (R_{\rm min}, R_{\rm max})$}&
\colhead{$V_{\rm eff, total}$}\\ 
\colhead{$(h^{-1}$\,cMpc)}&
\colhead{$(h^{-1}$\,cMpc)}&
\colhead{}&
\colhead{}&
\colhead{}&
\colhead{$(h^{-3}$\,cMpc$^{3})$}
}
\startdata
 0.124& 0.252&           1& 0.039&24.362$^{+58.332}_{-20.974}$ &20.84  \\
 0.252& 0.513&           2& 0.168&10.883$^{+15.674}_{- 7.676}$  &88.48\\
 0.513& 1.041&           2& 0.771& 1.594$^{+ 3.422}_{- 1.676}$  &400.18\\
 1.041& 2.112&          10& 3.110& 2.216$^{+ 1.373}_{- 1.000}$  &1609.04\\
 2.112& 4.288&          16&12.868& 0.243$^{+ 0.395}_{- 0.308}$  & 6644.21\\
 4.288& 8.706&          13&11.637& 0.117$^{+ 0.404}_{- 0.306}$  &6018.75
\enddata
\tablecomments{We present the data for the volume-averaged projected cross-correlation function between QSOs and LBGs $\chi (R_{\rm min}, R_{\rm max})$ shown in Fig.~\ref{fig:MLE}. This is measured in radial bins defined by $R_{\rm min}$ and $R_{\rm max}$. $\langle QG\rangle$ is the observed number of QSO-LBG pairs per bin, and $\langle QR\rangle$ is the expected number of QSO-random pairs per bin, computed from eqn.~(\ref{eq:NLBG2}). We also show the total volume of the bin added over the fields, computed as $V_{\rm eff, total}=\sum_{i=0}^{N_{\rm fields}} V_{\rm eff, i}$.\\}
\end{deluxetable*}
 
Given that the auto-correlation functions of both LBGs and QSOs at $z\sim 4$
have been previously measured, we can compute the expected volume
averaged QSO-LBG cross-correlation function $\chi(R_{\rm min}, R_{\rm max})$
assuming
linear bias and compare it to our measurements. Since we are probing
non-linear scales in our measurement where the linear bias assumption
surely breaks down, the expected cross-correlation obtained in this manner
is approximate, but nevertheless a useful reference.
If we assume that both LBGs and QSOs trace the same underlying dark matter, and assume linear bias such that $\delta_{G}=b_{G}\delta_{\rm DM}$,  and $\delta_{Q}=b_{Q}\delta_{\rm DM}$, then we can write
$\xi_{QG}=\sqrt{\xi_{QQ} \xi_{GG}}$. Assuming a power law form $\xi = (r/r_0)^\gamma$ for the respective
auto-correlations of QSOs and LBGs, and that they have identical slopes $\gamma$, then the cross-correlation length can be written as $r^{QG}_{0} = \sqrt{ r^{QQ}_{0} r^{GG}_{0}}$.  
To compute $\xi_{QG}$
we use respective measurements of the auto-correlation lengths of LBGs
and QSOs at $z\sim4$ from the literature.  For LBGs \citet{Ouchi04b}
measured $r^{GG}_{0} = 4.1\,h^{-1}$\,cMpc and $\gamma=1.8$, whereas
for QSOs we adopt $r^{QQ}_{0} = 22.3\,h^{-1}$\,cMpc, which was
measured by \citet{Shen07} for $z > 3.5$ QSOs assuming a fixed
$\gamma=1.8$. Combining these implies $r^{QG}_{0}=9.6\,h^{-1}$\,cMpc
for $\gamma=1.8$.  Plugging this power law LBG-QSO cross-correlation
function into eqn.~(\ref{eq:estimator2_simpler}) and integrating over
the effective survey volume gives us the expected value of
$\chi(R_{\rm min}, R_{\rm max})$, which is shown as a dashed line in
Fig.~\ref{fig:MLE}. One sees that our QSO-LBG cross-correlation
measurement is in reasonable agreement with the expected value of
$\chi (R_{\rm min}, R_{\rm max})$ combining auto-correlation measurements and assuming linear
bias. In \S~\ref{sec:fitting} we quantify this agreement by fitting
our cross-correlation function.

\subsubsection{Fitting the Cross-Correlation Function}
\label{sec:fitting}

\begin{figure}[t!]
\centering{\epsfig{file=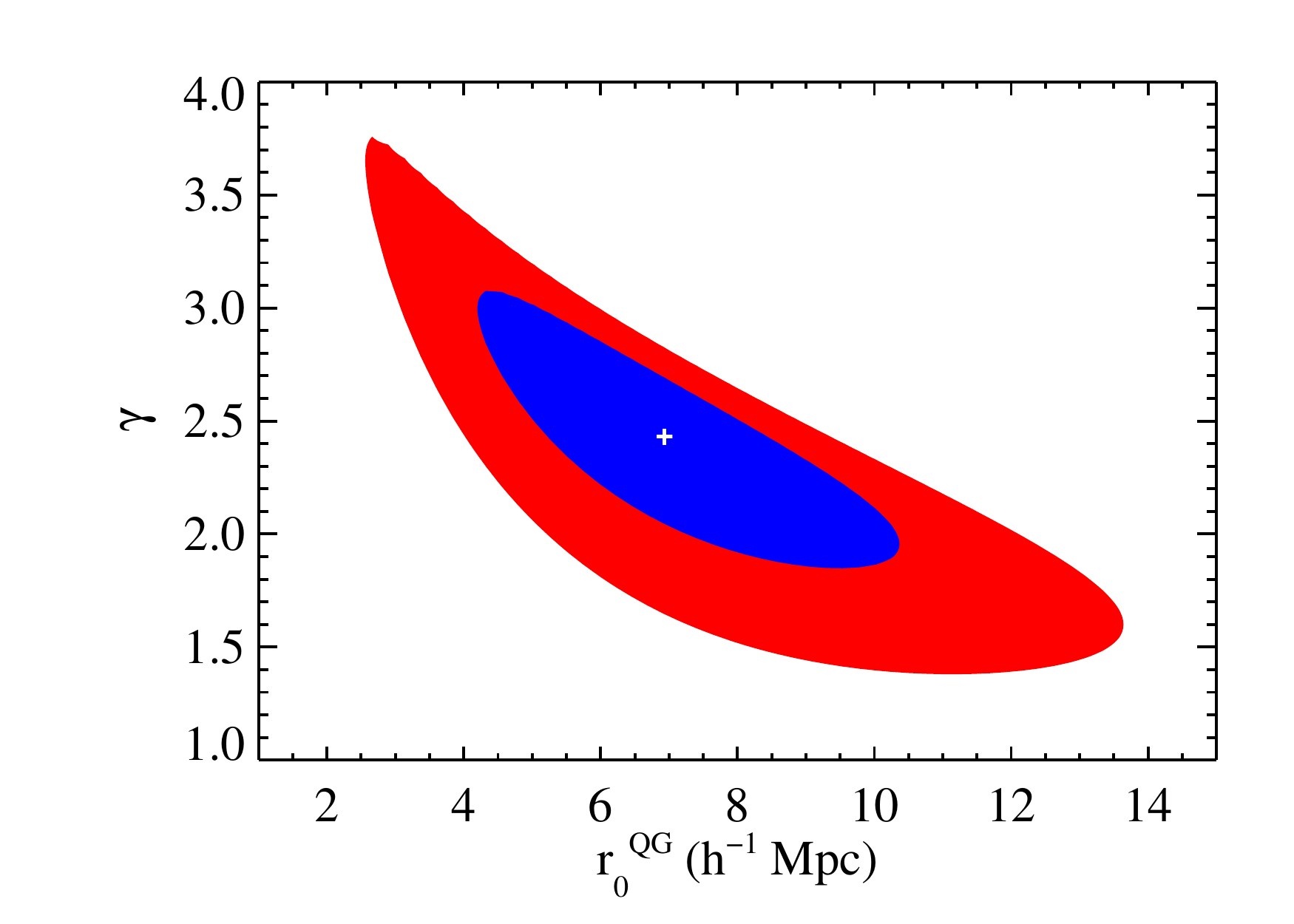, width=\columnwidth}}
\caption{\label{fig:confidence} $1\sigma$ and  $2\sigma$ confidence regions of $r^{QG}_{0}$ and $\gamma$ parameters (in blue and red respectively), determined using a maximum likelihood estimator. The best estimation is shown as a white cross.\\}
\end{figure}

Given the projected cross-correlation
function measurement, we now determine the real-space cross-correlation parameters
$r_{0}^{QG}$ and $\gamma$ that best fit our data. To this end we use
maximum likelihood estimator (MLE), and fit for the parameters which maximize the
probability of the data we observe.  Since we are dealing with a counting
process with small number counts in each  bin (see Table \ref{table:chi}),
we can assume that Poisson error dominates the error budget. 
Adopting the Poisson distribution for the counts in our cross-correlation function bins,
we can write the likelihood of our data as
\begin{equation}
\mathcal{L} = \prod_{i=1}^{N_{\rm bins}}\frac{e^{-\lambda_{i}}\lambda_{i}^{x_{i}}}{x_{i}!}
\label{eq:like}
\end{equation}
where the product is over the $N_{\rm bins}$ radial cross-correlation function
bins, $x_{i}$ is the number counts measured in the $i$th bin and
$\lambda_{i}$ is the expected number counts in the $i$th bin for a
given set of model parameters. In our case we have defined $x =\langle
QG\rangle$ and $\lambda=\langle QG\rangle^{\rm exp}$, where
\begin{multline} 
	\langle QG\rangle^{\rm exp} =  \\  
	 n_{\rm G} \int_{Z_{\rm min}}^{Z_{\rm max}} \int_{R_{\rm min}}^{R_{\rm max}} \phi(R,Z)[1+\xi_{QG} (R,Z)] 2\pi R dR dZ  
	 \label{eq:NLBG1b} 
\end{multline}
Here $\xi_{QG} (R,Z)= \left( \frac{\sqrt{R^{2} + Z^{2}}}{r^{QG}_{0}} \right)^{-\gamma}$
and is determined by the
 model parameters $r_{0}^{QG}$
and $\gamma$. 
Taking the
natural logarithm of both sides of eqn.~(\ref{eq:like}), we
obtain:
\begin{equation}
{\rm ln} \mathcal{L} \propto  \sum_{i=1}^{N_{\rm bins}} \left[ \langle QG\rangle_{i}\, {\rm ln} \left(\langle QG\rangle_{i}^{\rm exp}\right)-\langle QG\rangle_{i}^{\rm exp} \right], 
\label{eq:lnlike_my} 
\end{equation}
where model independent terms have been dropped. We calculated the
log-likelihood for a grid of ($r^{\rm QG}_{0}, \gamma$) values which defines an uniform prior, ranging from
$1.0 \leq \gamma \leq 5.0$ and $1.0 \leq r^{QG}_{0}
\leq 15.0$ and maximized the likelihood to obtain
$r^{QG}_{0}=6.93 \,h^{-1}\,{\rm cMpc}$ and
$\gamma=2.4$. These values were used in eqn.~(\ref{eq:estimator2_simpler}) to
calculate the expected $\chi (R_{\rm min}, R_{\rm max})$ value shown as the
red line in Fig.~\ref{fig:MLE}.
We also computed the 1$\sigma$ and 2$\sigma$ 2D confidence regions for these parameters, 
shown in the $r^{QG}_{0}-\gamma$ plane in Fig.~\ref{fig:confidence}. 
We determined errors on the parameters by marginalization.  Given that our grid of values is uniform, the normalized likelihood is the joint
posterior distribution of the parameters $P(r^{QG}_{0}, \gamma)$.
Therefore, we marginalized out $r^{QG}_{0}$ and $\gamma$ to obtain the probability distributions
$P(\gamma)$ and $P(r^{QG}_{0})$, respectively.
From those probability distributions we computed 68\% confidence regions about our MLE to define the error on
the parameters. 
We find $r^{QG}_{0}=6.93^{+2.13}_{-1.89} \,h^{-1}\,{\rm cMpc}$ and $\gamma=2.4^{+0.3}_{-0.5}$.

As shown in Fig.~\ref{fig:confidence}, our measurements are relatively
noisy when we fit $r^{QG}_{0}$ and $\gamma$ simultaneously, and there
is a clear degeneracy between these parameters. For that reason,
following common practice, we also fit the correlation function with
$\gamma$ fixed.  Independent measurements of QSO auto-correlation suggest a slope
of $\gamma=2.0$ \citep{Shen07}, which lies
within the  $1\sigma$ confidence region of our measurement (see
Fig.~\ref{fig:confidence}). Thus if we choose to fix
the slope to this value, the maximum likelihood and the $1\sigma$
confidence interval for the cross-correlation length is
$r^{QG}_{0}=8.83^{+1.39}_{-1.51}\,h^{-1}$\,cMpc.

Note that in the analysis described above we have assumed that the error
bars on our cross-correlation function are dominated by Poisson counting errors. This
ignores cosmic variance fluctuations, and also assumes that the positions of the LBGs
around the QSO are uncorrelated. However, as galaxies are not randomly distributed in the
universe, but rather have significant auto-correlations, our binned measurements of
$\langle QG\rangle$ are not truly independent. 
Given these correlations, our results could be somewhat sensitive to
our choice of binning, and our error bars could also be somewhat underestimated.
In principle, we should include the 
correlations and cosmic variance in our likelihood, analogous to computing the
non-diagonal elements of the covariance matrix for a multivariate
Gaussian likelihood. However, there is no simple analytical expression
for the likelihood of a correlated Poisson process, and furthermore
correctly modeling the cosmic variance would require the use of N-body
simulations of massive QSO halos at $z\sim4$. Note however that while
the positions of LBGs in the same field will be correlated, our QSO fields
are separated by Gpc distances, and hence the positions of LBGs in
different fields are completely independent. Given that our
correlation function comes from six distinct fields, and the relatively large Poisson
error bars, we believe that ignoring correlations and cosmic variance is a reasonable
approximation.

Our measurement indicates a strong cross-correlation between QSOs and
LBGs at $z\sim4$, implying that QSOs
trace massive dark matter halos in the early universe, with detectable
enhancements of LBGs.  We expect that those halos evolve to the most 
massive cluster of galaxies at $z=0$.
Our results are in agreement with the expected cross-correlation
function ($r^{QG}_{0}=9.6\,h^{-1}$\,cMpc for $\gamma=1.8$) computed from the individual QSO
and LBG auto-correlation functions assuming linear bias, as shown by the
dashed line in Fig.~\ref{fig:MLE}. 

\subsection{Auto-Correlation of LBGs in QSO Fields}
\label{sec:ACF}

Another measure of the clustering of LBGs in QSO environments is the
LBG auto-correlation function in our fields. If QSOs trace highly
biased locations of the universe, then we expect the LBGs around them to be more
highly clustered than LBGs in random fields, resulting in an enhancement of the LBG auto-correlation function. 
The auto-correlation function of $z\sim4$ LBGs in random
fields was measured by \citet{Ouchi04b}, which we compare to our
results.

To measure the LBG auto-correlation function we adopt the estimator:
\begin{equation}
\chi(R_{\rm min}, R_{\rm max}) = \frac{\langle GG\rangle}{\langle RR\rangle} -1
\label{eq:estimator_auto}
\end{equation}
where $\langle GG\rangle$ is the number of observed LBG-LBG pairs, and
$\langle RR\rangle$ is expected random number of LBG-LBG pairs,
in a cylindrical volume defined
by the radial bin $[R_{\rm min}, R_{\rm max}]$ and the height $\Delta
Z$. We measured $\langle GG\rangle$ directly from the images by
counting the LBG pairs in each radial bin. Following the same argument in
\S~\ref{sec:CCF}, we used the LBG luminosity function to compute the background
number density $n_{\rm G}$, rather than estimating it from our survey images. 

We computed the expected random number of LBG pairs as \citep[see e.g.][]{Padmanabhan07}
\begin{equation}
\langle RR\rangle = N_{\rm G} n_{\rm G} V_{\rm eff}
\label{eq:RR}
\end{equation}
where $n_{\rm G}$ is the same quantity defined in \S~\ref{sec:CCF} and $V_{\rm eff}$ is given by eqn.~(\ref{eq:veff}),
 but in this case using a different radial selection function $\phi_{R}(R)$, because of the different
binning used.  Here $N_{\rm G}$ is the expected number of LBGs for the entire
volume in question in a random region of the universe, which is computed for each of our six fields as
$N_{\rm G} = n_{\rm G} V_{\rm field}$.
The radial selection function $\phi_{R}(R)$ in this case is computed in an analogous way as for the cross-correlation: we created catalogs with $N_{\rm ran}\sim 100,000$ randomly distributed galaxies on our masked images,
and then we computed $\phi_{R}(R)$ as the ratio between the observed number of random galaxy pairs over the expected number of random galaxy pairs per radial bin. Here, the expected number of galaxy pairs per bin is computed by $N_{\rm ran}n_{\rm ran}\pi (R^{2}_{\rm max}-R^{2}_{\rm min})$. Note according to
eqn.~(\ref{eq:RR}) $\langle RR\rangle$ is proportional to the square of the LBG number density $n_{\rm G}$ and to the square of the redshift selection function $\phi_{Z}(Z)$ such that:
\begin{equation}
\langle RR\rangle \propto n^{2}_{G} \left(\int_{Z_{\rm min}}^{Z_{\rm max}} \phi_{Z}(Z) dZ\right)^{2}
\label{eq:RR_final}
\end{equation}

\begin{figure}[t!]
\begin{center}
\centering{\epsfig{file=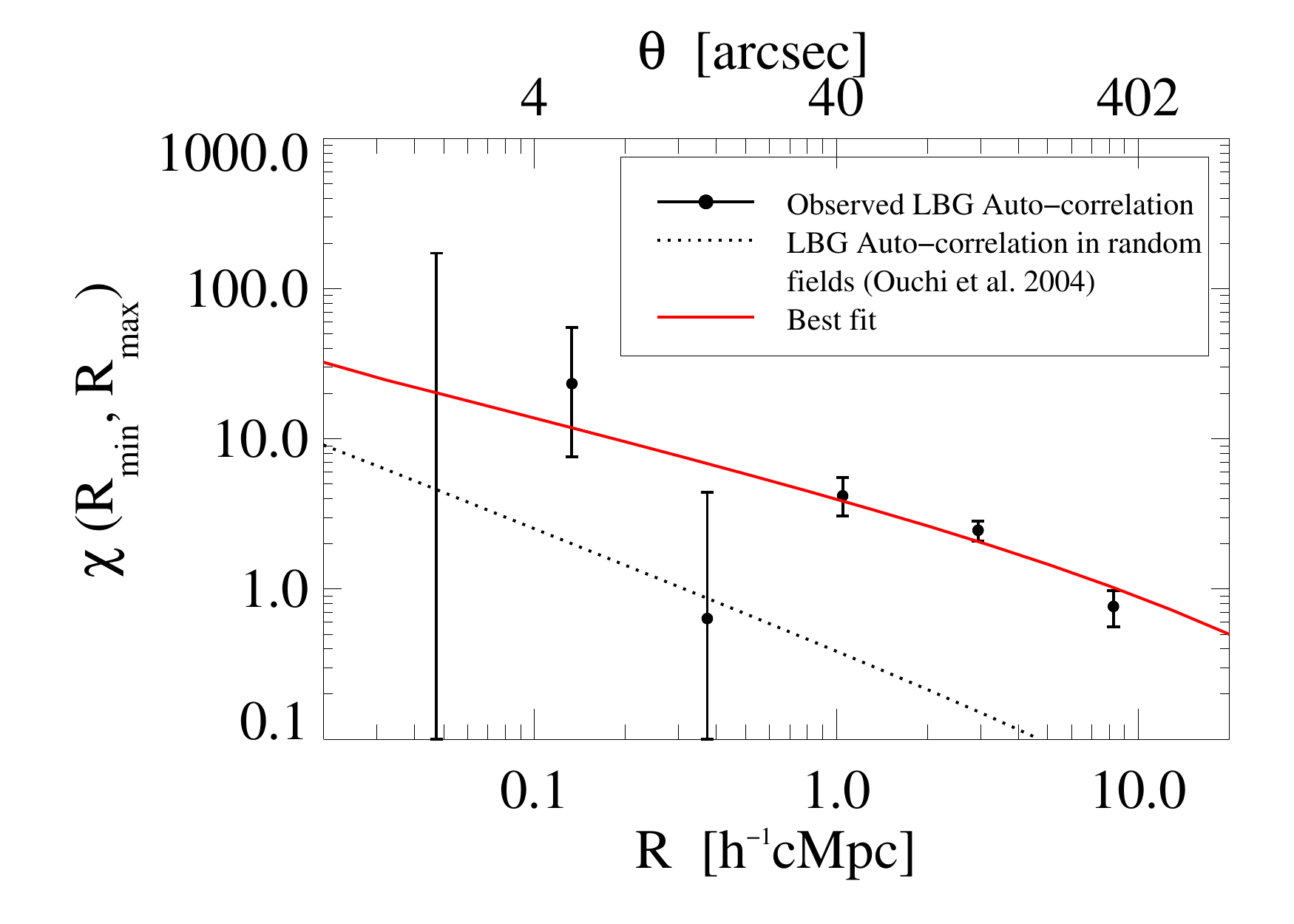, width=\columnwidth}}
\caption{\label{fig:autocorr} The data points are showing the LBGs auto-correlation measurement in QSO fields as we describe in section \S~\ref{sec:ACF}. The solid red curve shows the best fit for our measurements given by $r^{GG}_{0}=21.59 \,h^{-1}\,{\rm cMpc}$ and $\gamma=1.5$. The dotted black curve shows the LBGs auto-correlation in blank fields at $z\sim4$ measured by \citet{Ouchi04b}. We find a stronger clustering in our fields in comparison with random fields, which suggests that QSOs are located in overdense regions.\\}
\end{center}
\end{figure}

We computed $\langle GG\rangle$ and $\langle RR\rangle$ for each individual field and then we stacked the counts to measure the binned $\chi (R_{\rm min}, R_{\rm max})$ value as in eqn.~(\ref{eq:estimator_auto}). We show the results in Fig.~\ref{fig:autocorr} and the numerical values are given in Table \ref{table:auto}. We estimate errors on  $\chi (R_{\rm min}, R_{\rm max})$ using 
the one-sided Poisson confidence intervals for small number statistics in the same way as in \S~\ref{sec:CCF}.

\begin{deluxetable*}{c c r r r r}
\tabletypesize{\small}
\tabletypesize{\scriptsize}
\tablecaption{LBGs Auto-Correlation Function.\label{table:auto}}
\tablewidth{0pt}
\tablewidth{0.8\textwidth}
\tablehead{
\colhead{$R_{\rm min}$}&
\colhead{$R_{\rm max}$}&
\colhead{$\langle GG\rangle$}&
\colhead{$\langle RR\rangle$}&
\colhead{$\chi (R_{\rm min}, R_{\rm max})$}&
\colhead{$V_{\rm eff, total}^{a}$}\\ 
\colhead{$(h^{-1}$\,cMpc)}&
\colhead{$(h^{-1}$\,cMpc)}&
\colhead{}&
\colhead{}&
\colhead{}&
\colhead{$(h^{-3}$\,cMpc$^{3})$}
}
\startdata
  0.025&  0.070&           0&  0.011& -1.000$^{+173.441}_{-  0.000}$   &   1.13689\\
  0.070&  0.196&           2&  0.082& 23.336$^{+ 32.099}_{- 15.721}$  &  8.78984\\
  0.196&  0.551&           1&  0.611&  0.636$^{+  3.763}_{-  1.353}$  &  65.3105\\
  0.551&  1.546&          22&  4.251&  4.175$^{+  1.355}_{-  1.094}$  &  454.160\\
  1.546&  4.341&          82& 23.688&  2.462$^{+  0.382}_{-  0.382}$  &  2528.37\\
  4.341& 12.188&          72& 40.779&  0.766$^{+  0.208}_{-  0.208}$   &  4335.68 
\enddata    
\tablecomments{We present the data for the LBG auto-correlation function in QSOs fields $\chi (R_{\rm min}, R_{\rm max})$ shown in Fig.~\ref{fig:autocorr}. This is measured in radial bins defined by $R_{\rm min}$ and $R_{\rm max}$. $\langle GG\rangle$ is the observed number of LBG-LBG pairs per bin, and $\langle RR\rangle$ is the expected number of random-random pairs per bin, computed from eqn.~(\ref{eq:RR}). We also show the total volume of the bin added over the fields, computed as $V_{\rm eff, total}=\sum_{i=0}^{N_{\rm fields}} V_{\rm eff, i}$.\\}
\end{deluxetable*}

Analogous to our approach for the cross-correlation, we used a MLE to fit our auto-correlation function.
In this case 
the expected number of LBG-LBG pairs $\langle GG\rangle^{\rm exp}$ is modeled as:
\begin{multline} 
	\langle GG\rangle^{\rm exp} = n^{2}_{G} V_{\rm field} \\  
	 \times    \int_{Z_{\rm min}}^{Z_{\rm max}} \int_{R_{\rm min}}^{R_{\rm max}} \phi(R,Z)[1+\xi_{GG} (R,Z)] 2\pi R dR dZ\label{eq:GG} 
\end{multline}
where $\xi_{GG}$ is the LBG auto-correlation function assumed to have a power law form with correlation length $r^{GG}_{0}$.

\begin{figure}[t!]
\begin{center}
\centering{\epsfig{file=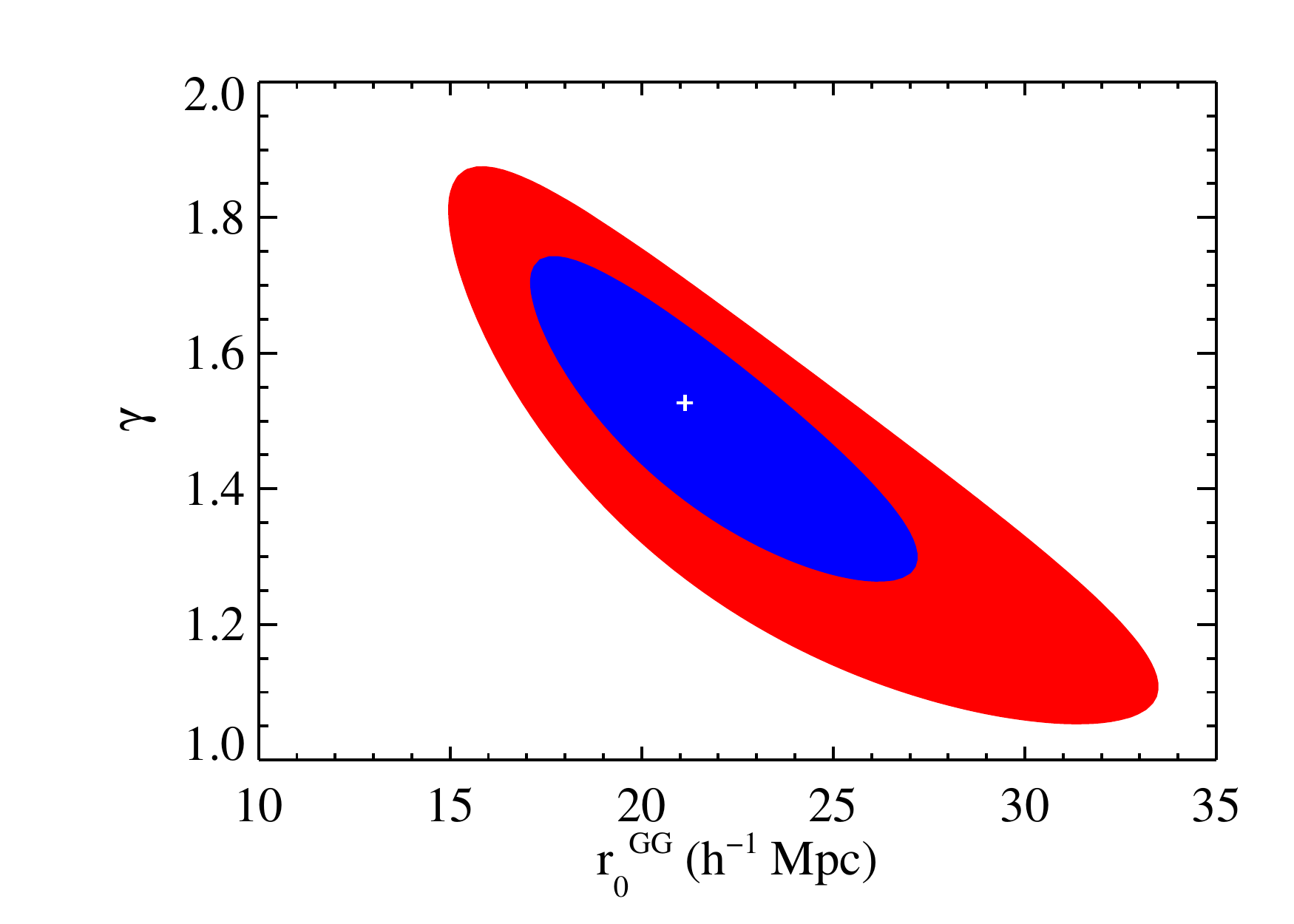, width=\columnwidth}}
\caption{\label{fig:confidence_auto} The same as Fig.~\ref{fig:confidence} but for the $r^{GG}_{0}$ and $\gamma$ parameters corresponding to the LBG auto-correlation function.\\}
\end{center}
\end{figure}

For the fitting we used an uniform prior
defined
by $1.0\leq \gamma \leq 2.5$ and $5.0\leq r^{GG}_{0} \leq 60.0$. We show 
the 1$\sigma$ and 2$\sigma$ 
2D confidence regions for the parameters in Fig.~\ref{fig:confidence_auto}. 
We obtained that the maximum
likelihood and the 1$\sigma$ confidence intervals are 
$r^{GG}_{0}=21.59^{+3.73}_{-2.96} \,h^{-1}\,{\rm cMpc}$ and
$\gamma=1.5^{+0.1}_{-0.2}$, which is
plotted as the red line in Fig.~\ref{fig:autocorr}.
Following the same arguments as in \S~\ref{sec:CCF}, 
we also fit the auto-correlation function with
$\gamma$ fixed. LBG auto-correlation function measured in random locations at $z\sim4$ 
suggest a slope of $\gamma=1.8$ \citep{Ouchi04b}, which lies
outside the  $1\sigma$ confidence region of our measurement (see
Fig.~\ref{fig:confidence_auto}). We then prefer to fix 
$\gamma$ to its maximum
likelihood value $\gamma=1.5$, which agrees with
the LBG auto-correlation function slope measured in QSO fields at $z=2.7$ \citep{Trainor12}. 
After fixing $\gamma$, we obtain
$r^{GG}_{0}=21.59^{+1.72}_{-1.69} \,h^{-1}\,{\rm cMpc}$.

In order to
compare this clustering signal with the one computed in random fields,
we use the LBG auto-correlation at
$z\sim4$ measured by \citet{Ouchi04b}.
Plugging their best fit values ($r^{GG}_{0}=4.1 \,h^{-1}\,{\rm cMpc}$ and
$\gamma=1.8$) into eqn.~(\ref{eq:estimator2_simpler}) using a power law form for $\xi_{GG} (R,Z)$, gives the
dotted line plotted in  Fig.~\ref{fig:autocorr}. 
To better compare our auto-correlation measurement with the \citet{Ouchi04b} random field
values we performed a fit with fixed $\gamma=1.8$, obtaining a maximum likelihood value and 1$\sigma$ confidence interval given by $r^{GG}_{0}=16.86^{+1.17}_{-1.14} \,h^{-1}\,{\rm cMpc}$, which is $\sim4$ times higher than the correlation length in random fields.
The fact that our LBG auto-correlation measurement is higher, suggests that the LBGs in our fields are more clustered than LBGs in random fields, which provides another
indication that QSOs fields trace regions of the universe that are denser than the cosmic average,
confirming our findings from the cross-correlation measurement in section \S~\ref{sec:CCF}.

While it may at first seem counter-intuitive that the LBG
auto-correlation is enhanced by such a large factor ($\sim 4$) in QSO
environs, this is actually exactly the expected behavior as we clarify
here.  The LBG auto-correlation function measures the radially binned
profile of galaxy pairs, and it is not trivial to relate the
auto-correlation to the cross-correlation.  In order to build
intuition, we will think in terms of the total number of galaxies
detected in our survey (see Table \ref{table:field_param}). On
average, we found 1.5 times more galaxies in QSO fields compared with
the number of galaxies expected in random locations (which is
estimated from our selection function and the number density of LBGs
$n_{\rm G}$), and that means that we should detect at least $1.5^{2}$
more galaxy pairs in our fields compared with the expectation in
random fields, simply because we are overdense by that factor. This
implies that the auto-correlation function will never be less than
$1.5^{2} - 1= 1.25$.

To better illustrate what happens to the auto-correlation in an
overdensity, we will consider a hypothetical scenario where galaxies
are randomly distributed in the universe with number density $n_{\rm
  G}$, and are clustered \emph{only} around QSOs, which are however
rare objects in the universe. Now imagine that the number
density of galaxies around QSOs is enhanced within a sphere of radius $R_{\rm QSO}$,
but that galaxies are otherwise randomly distributed within the sphere. In other
words, we imagine that in QSO fields the number density of galaxies is simply
increased by a factor $X$, but that within the sphere they are unclustered.
If QSOs are rare structures, then when averaging over large volumes of the universe,
we expect that the observed number of galaxy pairs
$\langle GG\rangle$ will be very close to the random expectation
$\langle RR\rangle$ (computed using $n_{\rm G}$), and
then the galaxy auto-correlation function measured from random locations
will be flat and very close to zero on small scales, $r< R_{\rm QSO}$. On larger scales, 
$r\sim R_{\rm QSO}$, it would deviate more from zero,
reflecting the clustering due to the top-hat overdensities around QSOs 
(however, if QSOs are very rare objects this positive correlation function
would be diluted and could still be quite small). 
Note however that if we measure the galaxy auto-correlation around QSOs
at radii $r < R_{\rm QSO}$, then $\langle GG\rangle$
will be $X^{2}$ times larger than $\langle RR\rangle$ (which is again
computed using $n_{\rm G}$), and then we would measure an auto-correlation
function of roughly $X^{2} -1$. This could be much larger
than the value measured in random locations which would be close to zero.
This simple example illustrates that because
$\langle RR\rangle$ is computed from the number density of galaxies in random regions $n_{\rm G}$,
overdense fields will always result in an enhanced auto-correlation
relative to that in random locations, and that these enhancements can be quite large.
The situation clearly becomes more
complicated if galaxies are intrinsically clustered with a power law
profile, and there is no simple analytical relationship between the cross and auto
correlation functions. To fully quantitatively understand the relationship between the cross and auto
correlation functions in QSO environments, one would need to analyze cosmological simulations
\citep[see e.g.][]{White12}. But the generic expectation is an enhancement of
the auto-correlation function in QSO environs compared with blank field pointings, which 
exactly what we see in Fig.~\ref{fig:autocorr}.

\subsection{Comparison with Previous Measurements}
\label{sec:disc}

The highest redshift for which the QSO-LBG cross-correlation has been measured
before is at $z\sim3$ by \citet{Trainor12}, who reported an overdensity of galaxies in QSO fields, and found
a cross-correlation length 
of $r^{QG}_{0, z\sim3}= 7.3\pm1.3 \,h^{-1}\,{\rm cMpc}$ for
a fixed $\gamma=1.5$. 
At $z\sim 4$ we find a steeper slope (we fixed $\gamma=2.0$)
than \citet{Trainor12}, but in order to facilitate a comparison with their results, we fit our cross-correlation measurement
for their same fixed $\gamma=1.5$. This $\gamma$ value, is near the border of our 2$\sigma$
confidence region (see Fig.~\ref{fig:confidence}) and thus disfavored by our measurements, but
we nevertheless proceed with this for comparison purposes. We obtain a cross-correlation length of
$r^{QG}_{0}=10.73^{+2.20}_{-2.41}\,h^{-1}$\,cMpc, which is $\sim1.5$
times higher than their cross-correlation length at $z\sim3$, 
indicating that halos hosting QSOs 
are considerably more biased and highly clustered at $z\sim4$.
 This agrees with the result reported by \citet{Shen07} who find that the
 QSO auto-correlation increase significantly from $z = 3$ to $z = 4$,
  and as such we expect to have found a larger cross-correlation. 

At $z>4$ only individual QSO fields have been studied so far. Some studies of QSO environments at
$z\sim6-7$
find no enhancements of galaxies compared with the background \citep[e.g ][]{Willott05, Banados13, Simpson14, Mazzucchelli16},
which could be suggesting that the strong QSO-galaxy cross-correlation breaks down at those redshifts. The lack of QSO auto-correlation measurements at these high redshifts makes impossible to know masses
of dark matter halos hosting $z\sim 6$ QSOs, but if their masses
are comparable to those hosting QSOs at $z\sim4$ (i.e $M_{\rm halo} \gtrsim 10^{12}\,\rm M_{\odot}$ as suggested
by the \citealt{Shen07} auto-correlation), then one would generically expect a strong QSO-galaxy clustering
signal  as we have detected here at $z\sim 4$.

\section{Testing the Robustness of our Results}
\label{sec:testing}

Two requirements must be fulfilled to ensure a robust clustering measurement: 
we need a low contamination level in the LBG sample and an accurate knowledge 
of the background number density of LBGs. Given that we used a novel NB 
technique to select LBGs, we need to carefully consider those requirements. 
In this section, we first discuss caveats related to the 
the use of this selection technique. Then we consider the effects of
using a contaminated sample for clustering measurements, and finally we
explore the impact of using different LBGs selection criteria on our
results.

\subsection{The Use of a NB Technique For LBG Selection: Caveats} 
\label{sec:technique_caveats}

A first complication of using our novel method for
color-selecting LBGs is that the level of contamination of our sample is
unknown.  
In principle, the
purity of the sample can only be determined with follow-up spectroscopy, or detailed
modeling of the population of contaminant galaxies.
Both alternatives would be challenging to implement and beyond the scope of
this paper, but as a compromise we qualitatively discuss the impact of
contamination on the correlation function (see
\S~\ref{sec:robustness}) and we demonstrate the robustness of our
results against contamination by exploring their sensitivity
to the color-selection criteria  (see \S~\ref{sec:different_selections}).
Note however that we could excise contamination in our LBG sample if 
we had additional imaging on our fields using traditional broad band filters. This would allow us to confirm
 the presence of the Ly$\alpha$ break in our LBG candidates.

Another complication of using our novel color-selection is that we did not have
an independent measurement of the background number density of LBG required to compute the clustering. This implied
that we had to rely on Monte Carlo simulations to determine the LBG selection function $\phi_{Z}(Z)$,
and then our clustering results are sensitive to errors in this modeling.
 If the completeness of the sample were close to 100\%, then 10\% errors on $\phi_{Z}(Z)$ would impact our measurement at the 10\%
level, whereas if the completeness were $\sim20$\% (as is the case), then there could be 100\% error on $\phi_{Z}(Z)$, which could
 strongly impact the amplitude of the measured clustering. 
Note that the auto-correlation is even more sensitive to
this quantity compared to the cross-correlation, because while
$\langle QR\rangle$ is proportional to $\phi_{Z}(Z)$, $\langle RR\rangle$
is proportional to the square of this quantity. In \S~\ref{sec:different_selections}
we test our redshift selection function to demonstrate that it is accurate and correctly modeled.

\subsection{Impact of Contamination on the Clustering Measurements}
\label{sec:robustness}

One method to qualitatively check the contamination level in the
sample is by studying the shape of the measured correlation
function. For example, if we measure the cross-correlation function via
eqn.~(\ref{eq:estimator}) using a highly contaminated sample, the
numerator in that equation would be overestimated, because of
the inclusion of low-redshift contaminants which are taken to be
real LBGs. However, since the denominator $\langle QR\rangle$ is simply
computed from the LBG luminosity function and on our redshift selection function, this
value does not include the extra number counts due to contamination. This implies
that the measured cross-correlation will not behave like a power-law, 
but rather it will flatten toward larger scales. Quantitatively, for
a contaminated sample, what we would actually measure is
\begin{equation}
\chi (R_{\rm min}, R_{\rm max})=  \frac{\langle QG\rangle + N_{\rm cont}}{\langle QR\rangle} -1 
\label{eq:estimator_contam}
\end{equation}
where $\langle QG\rangle $ and $\langle QR\rangle$ are given by
eqns.~(\ref{eq:NLBG1b}) and ~(\ref{eq:NLBG2}) respectively, and
$N_{\rm cont}$ is the number of contaminants in the bin. Given that
the contaminants are galaxies at different redshifts, the
cross-correlation between them and the $z\sim4$ QSO is zero, then the
number of contaminants will be given by $N_{\rm cont} = n_{\rm
  cont}V_{\rm eff, cont}$, where $n_{\rm cont}$ is the number density
of contaminants and $V_{\rm eff, cont}$ is the effective volume of the
bin,
which is given by eqn.~(\ref{eq:veff}), but with the redshift
selection function of the contaminants $\phi_{Z, \rm cont}(Z)$.
Then the
eqn.~(\ref{eq:estimator_contam}) reduces to:
\begin{eqnarray}
  \chi && (R_{\rm min},  R_{\rm max})=  \chi^{\rm true}(R_{\rm min},  R_{\rm max}) \nonumber \\
&&+ \frac{n_{\rm cont}\int_{Z_{\rm min, cont}}^{Z_{\rm max, cont}} \phi_{Z, \rm cont}(Z)dZ }{n_{\rm G} \int_{Z_{\rm min}}^{Z_{\rm max}} \phi_{Z}(Z)dZ} \frac{D_{C}^{2}(z_{\rm cont})}{D_{C}^{2}(z_{\rm LBG})} 
\label{eq:estimator_contam2}
\end{eqnarray}
where $\chi^{\rm true} (R_{\rm min},  R_{\rm max})$ is the correlation function that we would measure from a non-contaminated sample (i. e, here $\chi^{\rm true}(R_{\rm min},  R_{\rm max})=\langle QG\rangle/\langle QR\rangle -1$), and $D_{C}(z)$ is the transverse comoving distance at redshift $z$. In the absence of contaminants, the second term in this equation would be zero, 
and we recover the correlation function defined in
eqn.~(\ref{eq:estimator2_simpler}), which has a power law shape. However, if a
large number of contaminants are included which span a large range in
redshift, the second term becomes important, and given that it does not
depend on radius, this same constant is added everywhere to the
cross-correlation function flattening its shape, with the degree of flattening
dependent on the level of contamination.

This flattening effect will be even stronger for the auto-correlation
function  since it is proportional to the square of both the
number density of contaminants and redshift range they cover. Then for a contaminated
sample one obtains
\begin{eqnarray}
\chi &&(R_{\rm min}, R_{\rm max})= \chi^{\rm true}(R_{\rm min},  R_{\rm max}) \nonumber \\
&&+ \frac{n^{2}_{\rm cont}\left(\int_{Z_{\rm min, cont}}^{Z_{\rm max, cont}} \phi_{Z, \rm cont}(Z)dZ\right)^{2} }{n^{2}_{G} \left(\int_{Z_{\rm min}}^{Z_{\rm max}} \phi_{Z}(Z)dZ\right)^{2}} \frac{D_{C}^{4}(z_{cont})}{D_{C}^{4}(z_{LBG})} 
\label{eq:estimator_contam_auto}
\end{eqnarray}
where ignored the clustering of the contaminants, which should be
greatly diluted in projection if the contaminants span a large range
of redshifts.  To take clustering of contaminants into account, an
additional term should be added to this equation to account for their
auto-correlation. Therefore, the smoking gun of high contamination in
the LBG sample would be a flat cross-correlation and auto-correlation
function in Figs.~\ref{fig:MLE} and \ref{fig:autocorr}, respectively.
Given that we measured a power law shape for both correlations, we believe 
that our LBG sample is not strongly affected by contamination. 

We have explored a third way to check contamination
which is also independent of our estimate of $n_{\rm G}$ and $\phi_{Z}(Z)$.
For a highly contaminated sample that includes galaxies over a wide range
of redshifts, it would be  more appropriate to measure angular distances instead of
transverse comoving distances.  We thus compute the angular correlation function $\omega(\theta)$ using the standard
procedure, where $\langle RR\rangle$ is determined from the angular number density of the
data itself, and we do not assume anything about the number density or selection function.
 In this case we only measure how clustered is our sample in comparison to a random distribution
  with the same number density as our sample. This angular correlation function calculation thus differs from our LBG
auto-correlation function in \S~\ref{sec:ACF}, where $\langle RR\rangle$ was
computed from $n_{\rm G}$ and our selection function $\phi_{Z}(Z)$.
For a highly contaminated sample we
expect the angular correlation function to be close to
zero on all angular scales, because the inclusion of uncorrelated
galaxies over a broad redshift range, would dilute
any real clustering signal.
On the other hand, for a relatively pure sample composed primarily of LBGs at $z=3.78\pm0.3$,  we
expect to measure a power law angular auto-correlation because we would be
selected only highly biased  galaxies in a small volume. Note however that
even for a pure LBG sample, the $\omega(\theta)$ computed in this way is not the true angular
correlation function of LBGs, because we are pointing towards overdense regions around QSOs.

We estimated the angular auto-correlation function of the LBGs as:
\begin{equation}
\omega (\theta) = \frac{\langle GG (\theta)\rangle}{\langle RR (\theta)\rangle} - 1
\end{equation}
where $\langle GG (\theta)\rangle$ is the number of LBG-LBG pairs per
angular bin, which is directly measured from our images, and $\langle
RR(\theta)\rangle$ is the number of random-random pairs per angular
bin.
The $\langle RR (\theta)\rangle$ quantity was estimated using a
random catalog of sources created as follows. First, we computed the
total number of LBG candidates in all the fields, then we divided
that by the total unmasked area to get the average angular number density of LBGs. 
Second, we multiplied the unmasked area per image by this
average number density to determine the number of galaxies expected in
each field. Finally, we increased the number of galaxies by a large
factor $F$ in order to decrease the noise in the measurement, and we
randomly distributed those sources on the image and then we measured
$\langle RR (\theta)\rangle$ by counting the pairs of simulated
galaxies per angular bin. We then re-scaled $\langle RR
(\theta)\rangle$ down by $F^{2}$.

\begin{figure}[t!]
\begin{center}
\centering{\epsfig{file=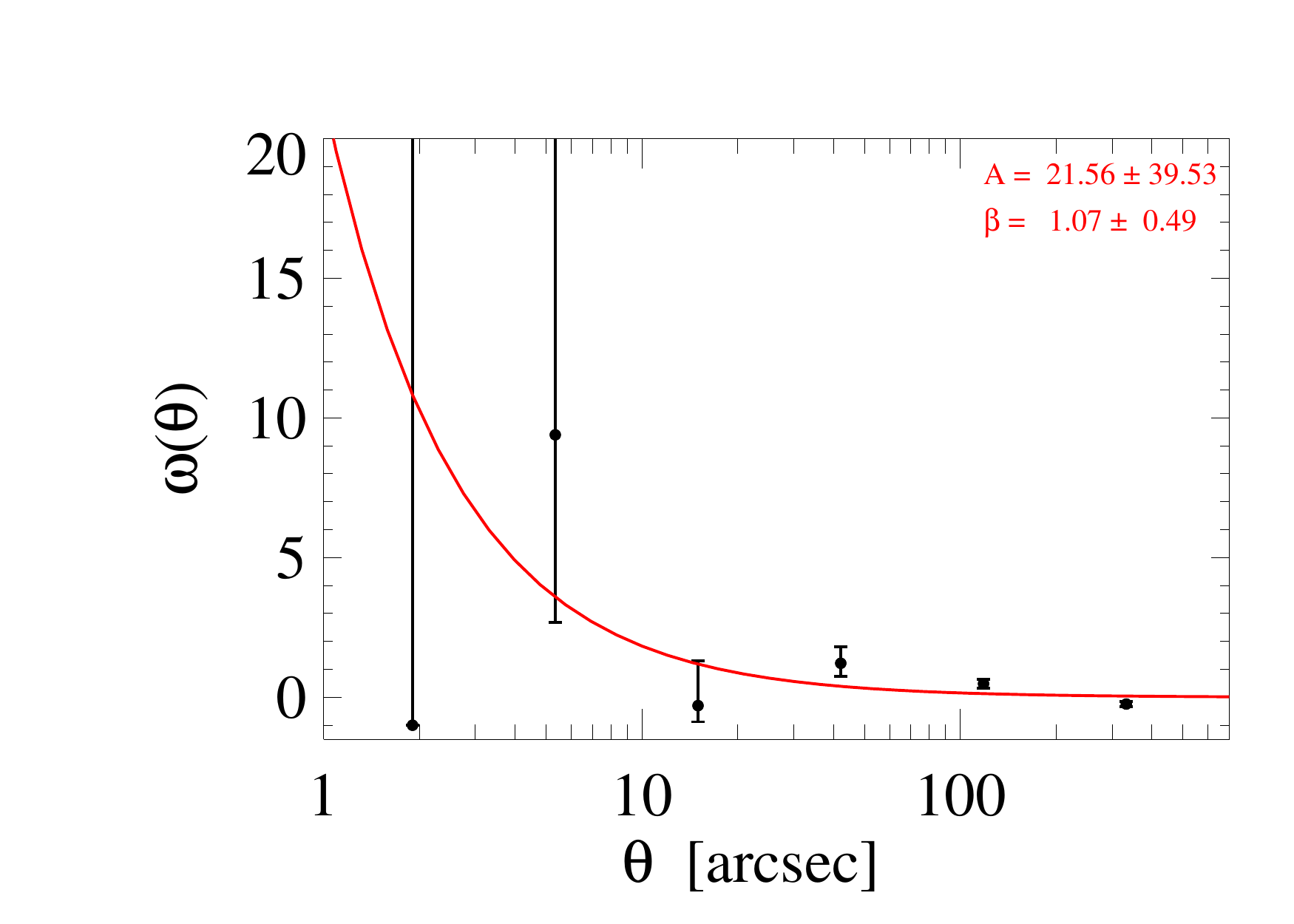, width=\columnwidth}}
\caption{\label{fig:omega} Angular auto-correlation function measurement for the LBGs sample. This measurement is used to test if the sample is contaminated. If the sample were highly contaminated $\omega(\theta)=0$ at every scale and a power law fit with slope $\beta$ would be consistent with zero. We find that the LBGs sample is not highly contaminated, given their power law shape in this plot, which is well fitted by $\omega(\theta)=21.56\,\theta^{-1.07}$ (red line).\\}  
\end{center}
\end{figure}

Our measurement of the angular correlation function is shown in Fig.~\ref{fig:omega}. We see a non-flat correlation
 function which suggest that our LBG sample is not highly contaminated.
 Assuming a power law form given by $\omega(\theta)=A\theta^{-\beta}$ we performed a Levenberg-Marquardt
least-squares fit to these data to quantify how consistent the measurement is with a flat shape
(where $\beta=0$). We obtain best fit parameters of $A=21.56\pm39.54$ and $\beta=1.07\pm0.49$. 
Given the large error bars in the measurement we are not able to discard a correlation function consistent 
with zero, however, as we show in the next subsection, if the LBG sample were highly contaminated then the angular correlation function would be much flatter. The fact that we measure a signal in Fig.~\ref{fig:omega} suggest that we are measuring real LBG clustering.

\subsection{Robustness of Clustering Measurements Against Changes in Color-Selection}

\label{sec:different_selections}

\begin{turnpage}
\begin{figure*}[t!]
\begin{center}
%\vspace{13cm}
%\hspace{18cm}
\includegraphics[angle=0, width=1.15\textwidth]{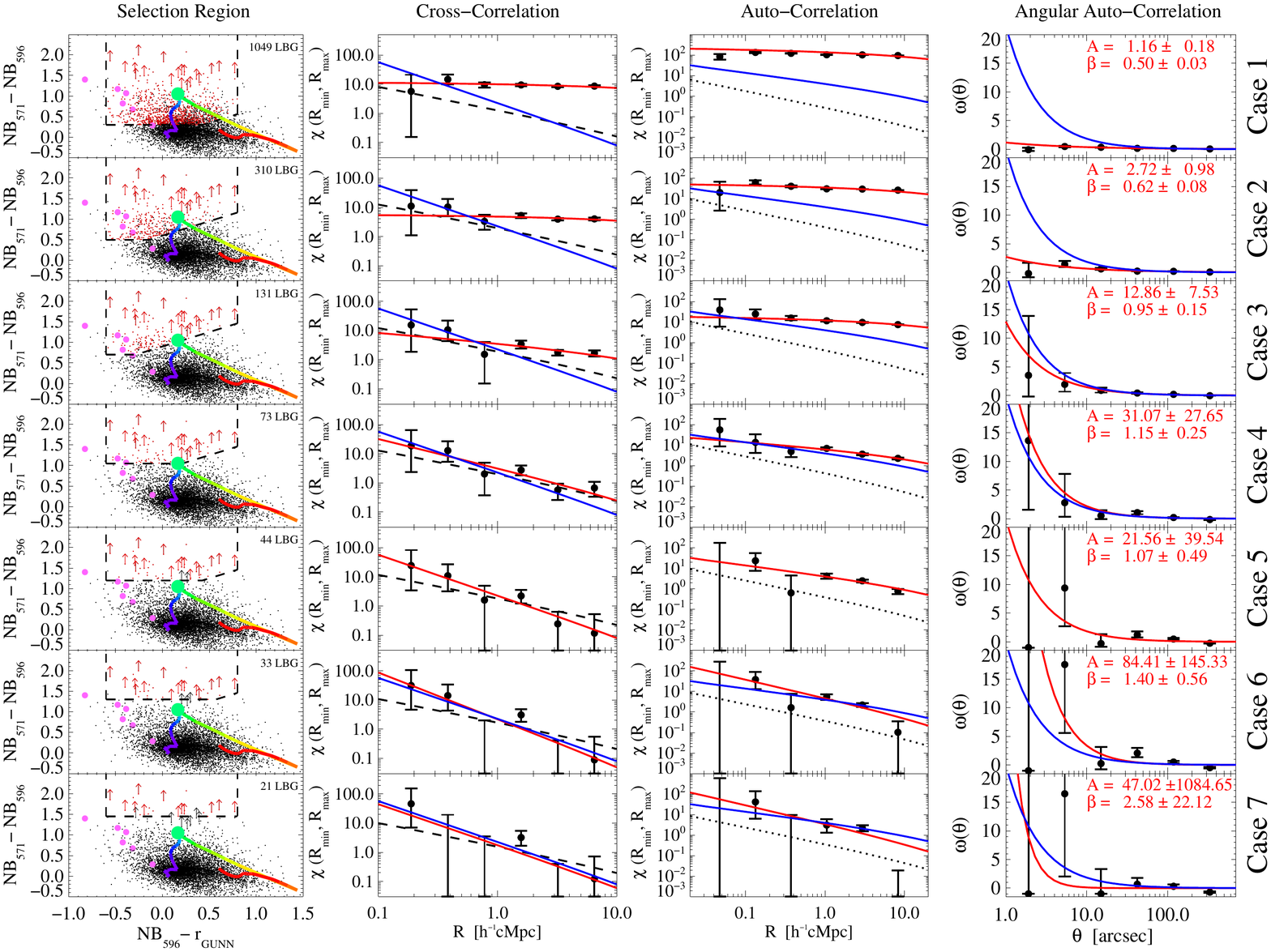} 
\caption{\label{fig:compar}
  Impact of contaminants on the clustering
  measurements. We show seven different selection criteria and their
  respective clustering measurements. From left to right we show the
  color-color plot showing the color cuts used and the photometry. The
  cross-correlation including the best fit according to our MLE
  estimator and the theoretical expectation of $\chi (R_{\rm min},  R_{\rm max})$ calculated from 
  the QSO and LBGs auto-correlation functions and assuming a linear 
  bias model (dashed line). The auto-correlation measurement including the best fit according to our
  MLE estimator, and the LBGs auto-correlation in blank fields at $z\sim4$ measured by 
  \citet{Ouchi04b} (dotted line). Finally, the angular auto-correlation function, with a
  power law fit $\omega(\theta)=A\theta^{-\beta}$, with the $A, \beta$
  values indicated in the top right corner. In the three correlation
  function plots we show the best fit in each case as a red curve and
  the best fit for our fiducial selection (i. e Case 5) as a blue
  curve. From top to bottom we show selections progressively more
  conservative, and then less contaminated. The selection used in this
  paper to measure the clustering properties of $z\sim4$ LBGs in QSO
  environments correspond to the Case 5. We detected a convergence in
  the clustering measurements for the three last cases.\\}
\end{center}
\end{figure*} 
\end{turnpage}

Here, we study the impact of using different color-selections on our
clustering measurements to demonstrate 
that our results are not significantly impacted by contamination, and to
show that our Monte Carlo simulation of the completeness is robust if we change the color-selection
criteria. 
To this end we have defined different LBGs selection criteria,
and for each one we compute the cross-correlation, auto-correlation,
and angular auto-correlation function. The cross-correlation and
auto-correlation functions were fitted in each case using a MLE
following the same procedure described in \S~\ref{sec:clustering}.
We consider a progression of seven different selections, from the most permissive Case 1,
which selects the majority of $z\sim4$ LBGs, but also likely
incurs a large fraction of low-redshift contaminants,
to the most conservative Case 7, which results in a low completeness
for  $z\sim4$ LBGs, but ensures low contamination.
These results are shown in Fig.~\ref{fig:compar}, and we tabulate the best
fit values for each case in Table \ref{table:clustering}. There
we also tabulate the best fit correlation lengths for a fixed $\gamma=2.0$ for the
QSO-LBG cross-correlation, and $\gamma=1.5$ for the LBG auto-correlation, in
order to study how $r_{0}$ varies for the different cases. 
Note that for the three most permissive Cases 1-3, we measure a flat correlation
function and hence do not quote fits for fixed $\gamma$.  Additionally for those cases we had to 
use a different prior for the MLE fit, since flat correlation functions result in
small values for the slope and large values for the correlation length,
which are not covered by the prior used for the other cases. We only quote the best fit parameters for 
Cases 1-3 because the 1$\sigma$ confidence region extends beyond the prior, precluding reliable
error estimates.

\begin{deluxetable*}{ll    |    lll    |    lll }
\tabletypesize{\small}
\tabletypesize{\scriptsize}
\tablecaption{Best fitted parameters for the cross-correlation and auto-correlation functions for the seven cases showed in Fig.~\ref{fig:compar}.\label{table:clustering}}
\tablewidth{0pt}
\tablewidth{1.0\textwidth}
\tablehead{
\colhead{}&
\colhead{}&
\multicolumn{3}{|  c |}{Cross-Correlation}&
\multicolumn{3}{c }{Auto-Correlation}\\
\colhead{}&
\colhead{Selection criteria}&
\colhead{$r^{QG}_{0}$}&
\colhead{$\gamma$}&
\colhead{$r^{QG}_{0, \gamma=2.0}$}&
\colhead{$r^{GG}_{0}$}&
\colhead{$\gamma$}&
\colhead{$r^{GG}_{0, \gamma=1.5}$}
}
\startdata
 & $\rm (NB_{571}-NB_{596})> 0.30 $&  &  &  &  & &     \\
1& $ -0.6  < \rm (NB_{596}-r_{GUNN}) <0.8$&$ 873.03$ &$0.7$ & &$5456.67$ &$0.9$   \\
& $\rm (NB_{571}-NB_{596}) > $&  &  &  &  & &  \\ 
& $0.7 \rm (NB_{596}-r_{GUNN}) + 0.0$&  &  &  &  &  \\ \hline
 & $\rm (NB_{571}-NB_{596})> 0.50 $&  &  &  &  & &     \\
2& $ -0.6  < \rm (NB_{596}-r_{GUNN}) <0.8$&$ 198.96$ &$0.6$ & &$1100.00$ &$0.8$   \\
& $\rm (NB_{571}-NB_{596}) > $&  &  &  &  & &  \\ 
& $0.7 \rm (NB_{596}-r_{GUNN}) + 0.6$&  &  &  &  &  \\ \hline
 & $\rm (NB_{571}-NB_{596})> 0.70 $&  &  &  &  & &     \\
3& $ -0.6  < \rm (NB_{596}-r_{GUNN}) <0.8$&$  25.67$ &$1.3$ & &$ 261.08$ &$0.8$   \\
& $\rm (NB_{571}-NB_{596}) > $&  &  &  &  & &  \\ 
& $0.7 \rm (NB_{596}-r_{GUNN}) + 0.9$&  &  &  &  &  \\ \hline
 & $\rm (NB_{571}-NB_{596})> 1.05 $&  &  &  &  & &     \\
4& $ -0.6  < \rm (NB_{596}-r_{GUNN}) <0.8$&$  10.25^{+   2.18}_{-   2.08}$ &$2.0^{+0.3}_{-0.3}$ &$  10.25^{+   1.13}_{-   1.19}$ &$  41.17^{+   8.52}_{-   4.19}$ &$1.3^{+0.1}_{-0.1}$  &$  31.23^{+   1.27}_{-   1.32}$  \\
& $\rm (NB_{571}-NB_{596}) > $&  &  &  &  & &   \\ 
& $0.7 \rm (NB_{596}-r_{GUNN}) + 0.9$&  &  &  &  &    \\ \hline
 & $\rm (NB_{571}-NB_{596})> 1.20 $&  &  &  &  & &     \\
5& $ -0.6  < \rm (NB_{596}-r_{GUNN}) <0.8$&$   6.93^{+   2.13}_{-   1.89}$ &$2.4^{+0.3}_{-0.5}$ &$   8.83^{+   1.39}_{-   1.51}$ &$  21.59^{+   3.73}_{-   2.96}$ &$1.5^{+0.1}_{-0.2}$  &$  21.59^{+   1.72}_{-   1.69}$  \\
& $\rm (NB_{571}-NB_{596}) > $&  &  &  &  & &   \\ 
& $0.7 \rm (NB_{596}-r_{GUNN}) + 0.9$&  &  &  &  &    \\ \hline
 & $\rm (NB_{571}-NB_{596})> 1.30 $&  &  &  &  & &     \\
6& $ -0.6  < \rm (NB_{596}-r_{GUNN}) <0.8$&$   6.22^{+   2.53}_{-   1.92}$ &$2.6^{+0.3}_{-0.6}$ &$   8.83^{+   1.61}_{-   1.77}$ &$  14.96^{+   2.83}_{-   1.93}$ &$1.9^{+0.1}_{-0.2}$  &$  19.94^{+   2.20}_{-   2.17}$  \\
& $\rm (NB_{571}-NB_{596}) > $&  &  &  &  & &   \\ 
& $0.7 \rm (NB_{596}-r_{GUNN}) + 0.9$&  &  &  &  &    \\ \hline
 & $\rm (NB_{571}-NB_{596})> 1.45 $&  &  &  &  & &     \\
7& $ -0.6  < \rm (NB_{596}-r_{GUNN}) <0.8$&$   6.46^{+   2.95}_{-   3.61}$ &$2.4^{+0.6}_{-0.8}$ &$   7.88^{+   2.15}_{-   2.46}$ &$  13.06^{+   4.21}_{-   2.93}$ &$1.9^{+0.2}_{-0.4}$  &$  16.62^{+   3.42}_{-   3.25}$  \\
& $\rm (NB_{571}-NB_{596}) > $&  &  &  &  & &   \\ 
& $0.7 \rm (NB_{596}-r_{GUNN}) + 0.9$&  &  &  &  &   
\enddata
\tablecomments{$r_{0}$ is shown in $(h^{-1}$\,cMpc) units.\\}
\end{deluxetable*}

We find that the cross-correlation function flattens and its amplitude
increases for more permissive selections that increase the level
of contamination, and the
auto-correlation function shows a similar but even stronger tendency.
 This is the behavior that we expected as we describe in \S~\ref{sec:robustness} and according to 
 eqns. ~(\ref{eq:estimator_contam2}) and ~(\ref{eq:estimator_contam_auto}).  
As for the angular correlation function, we find
that the more conservative the selection, the steeper the slope of
$\omega(\theta)$ and the more significant its departure from
zero. These are again the trends we expect because reduced
contamination results in a more strongly clustered sample of $z\sim 4$
galaxies, selected from a narrow redshift slice reducing the amount
that the clustering is diluted by projection.
Note however, that for the less conservative cases (i.e, Case 1 and Case 2),
where the sample
is dominated by contaminants, the angular correlation function is
close to zero, but not perfectly consistent with $\beta=0$. We believe
that the measurement of a weak clustering signal in these cases
results from the actual clustering of foreground contaminants,
which is diluted by the line-of-sight projection, but nevertheless
remains strong enough to not be perfectly consistent with zero.

\begin{figure}[t!]
\begin{center}
\centering{\epsfig{file=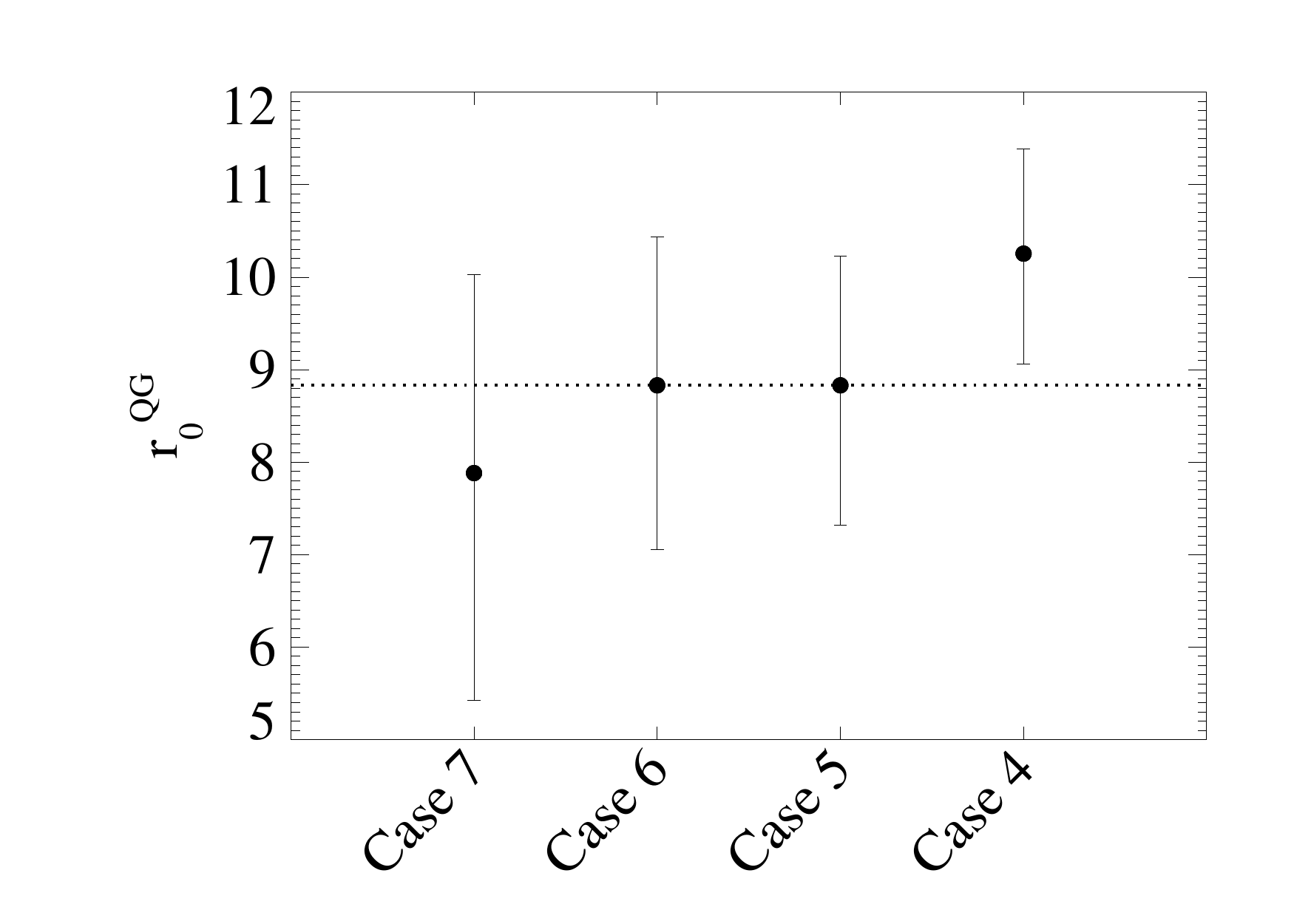, width=\columnwidth}}
\caption{\label{fig:r0_values} Best fitted $r^{QG}_{0}$ values for a fixed $\gamma=2.0$ for the four more conservative selections shown in Fig.~\ref{fig:compar}. We detect a convergence of the correlation length. The horizontal dashed line indicate the best fitted $r^{QG}_{0}$ value for the Case 5. \\}  
\end{center}
\end{figure}

\begin{figure}[t!]
\begin{center}
\centering{\epsfig{file=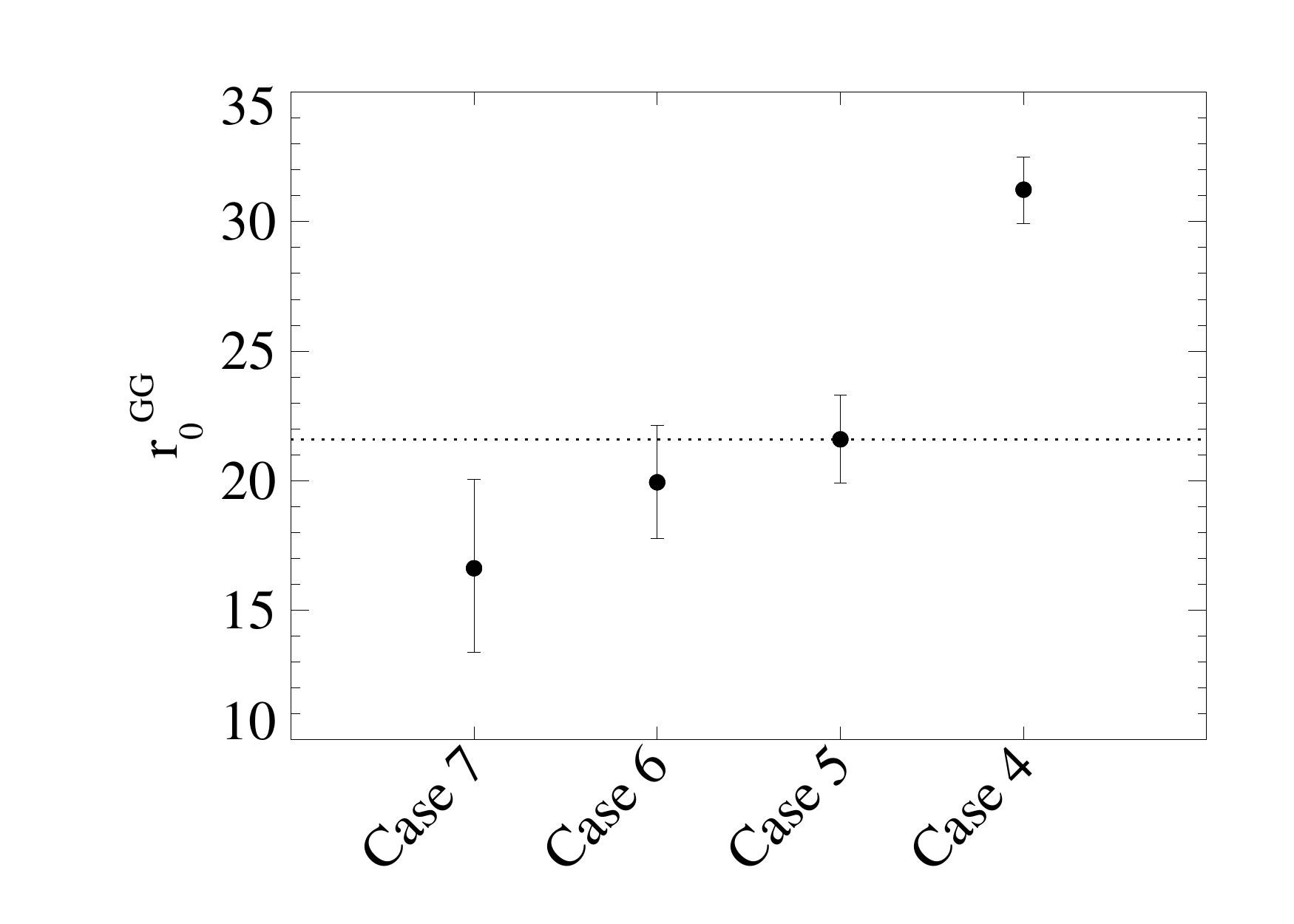, width=\columnwidth}}
\caption{\label{fig:r0_values_auto} Best fitted $r^{GG}_{0}$ values for a fixed $\gamma=1.5$ for the four more conservative selections shown in Fig.~\ref{fig:compar}. We detected a convergence of the correlation length. The horizontal dashed line indicate the best fitted $r^{GG}_{0}$ value for the Case 5.\\}  
\end{center}
\end{figure}

The takeaway message from Fig.~\ref{fig:compar} is that we observe
convergence of both the cross-correlation and auto-correlation
functions for the more conservative selections.  Specifically, we find
stable results for Cases 5-7, with the only significant difference
being the signal-to-noise ratio of the clustering measurements,
resulting from the smaller sample of LBGs selected in the more
conservative cases.  In Fig.~\ref{fig:r0_values} and
\ref{fig:r0_values_auto} we plot the values of the
cross-correlation length $r^{QG}_{0}$ (for fixed $\gamma=2.0$) and
auto-correlation length $r^{GG}_{0}$ (for fixed $\gamma=1.5$),
respectively, for the four most conservative selections. The convergence of the
correlation lengths demonstrates that: 1) we do not suffer
large contamination and hence our results are robust against contamination, 2) that our Monte Carlo
simulation of the selection function is reliable, since it results in consistent
measurements as the color-selection and selection function are varied, 3) our results are largely
independent of the exact color-selection region adopted. For these reasons
we simply adopt Case 5 to present the final results in this paper.

Finally, we performed one last test to establish that the
  redshift selection function modeled from our Monte Carlo is essentially correct. 
We compared the total observed QSO-LBGs pairs in all the fields
$\langle QG\rangle^{\rm obs}$ for each selection with the
expected value $\langle QG\rangle^{\rm exp}$ based on our
clustering measurements, and our Monte Carlo determination of the
redshift selection function.  Specifically, for each selection the
$\langle QG\rangle^{\rm obs}$ was measured by
summing the observed QSO-LBGs pairs over the fields up to scales of
$R\sim 9\,h^{-1}\,{\rm cMpc}$, and Poisson errors were computed for
this measurement.  The expected value $\langle QG\rangle^{\rm exp}$
for each field
was computed
using eqn.~(\ref{eq:NLBG1b}), where we computed the corresponding
$\phi_{Z}(Z)$ using our Monte Carlo simulation method described in
\S~\ref{sec:modeling} for each selection criteria. For all the cases,
we used $r^{QG}_{0}=6.93\,h^{-1}\,{\rm cMpc}$ and $\gamma=2.4$,
which are the best fit parameters for our fiducial
color-selection (Case 5; see Fig.~\ref{fig:compar}) in the computation of
$\langle QG\rangle^{\rm exp}$.
The total expected number of QSO-LBG pairs in the whole survey $\langle QG\rangle^{\rm exp}$, was
computed by summing $\langle QG\rangle^{\rm exp}$ over the bins and over the fields.

\begin{figure}[t!]
\begin{center}
\centering{\epsfig{file=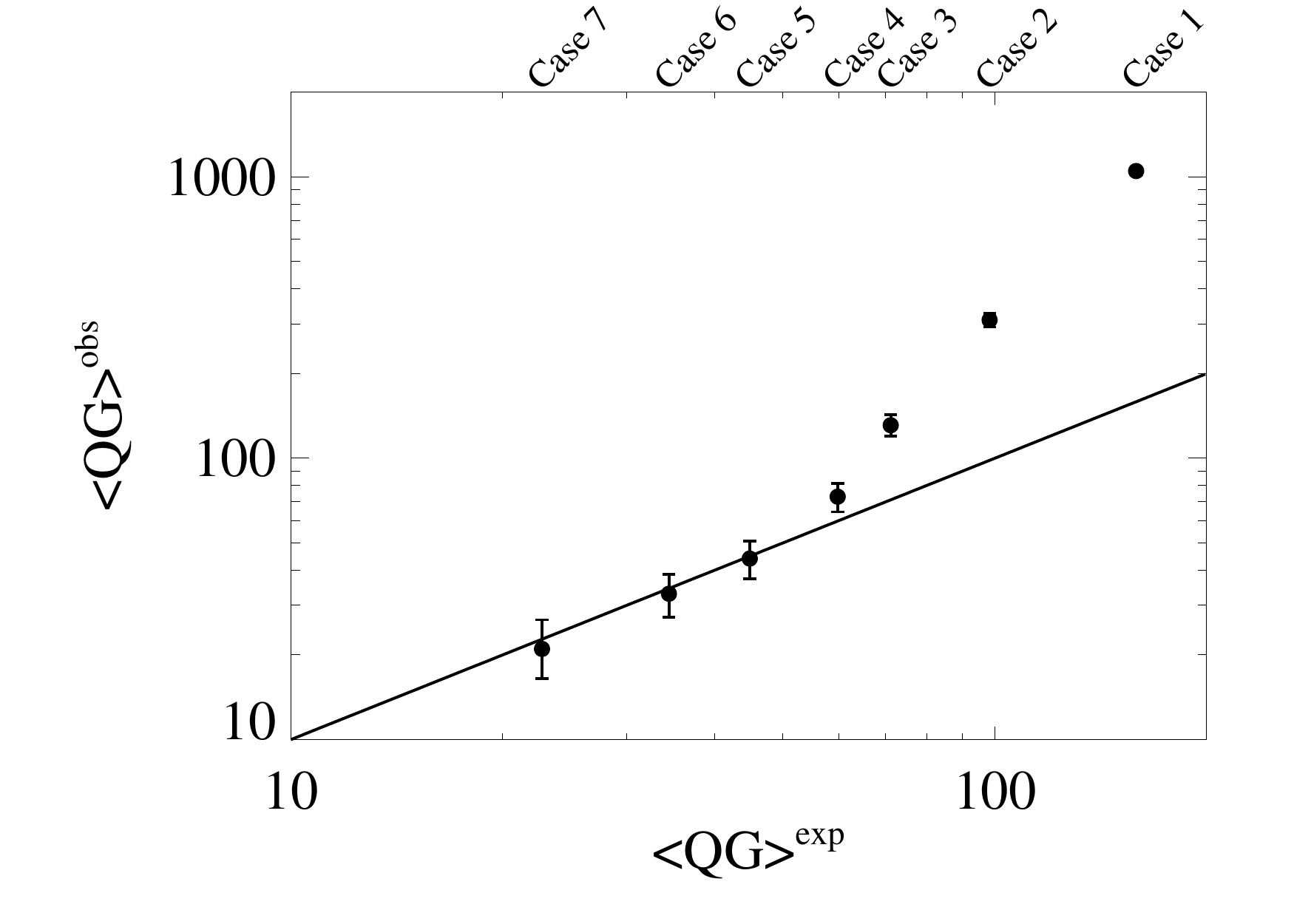, width=\columnwidth}}
\caption{The total observed QSO-LBGs pairs in the whole survey for our selections with Poisson error bars compared with the expected number of QSO-LBGs pairs, assuming the cross-correlation parameters for the Case 5 and using the redshift selection function $\phi_{Z}(Z)$ computed for each case as we described in \S~\ref{sec:LBG_sel}. The solid line indicates the case where $\langle
QG\rangle^{\rm obs}=\langle QG\rangle^{\rm exp}$. Our results show that the three more conservative cases (non-contaminated samples) are consistent with the expectation, which means that the modeled redshift selection function $\phi_{Z}(Z)$  is sensitive and that the level of contamination in those selections are negligible. \\}   
\label{fig:compar_QG}
\end{center}
\end{figure}

If the contamination is low,  and the redshift
selection function $\phi_{Z}(Z)$ is correctly computed for each case, we
expect that $\langle QG\rangle^{\rm exp}$ should equal to
$\langle QG\rangle^{\rm obs}$. As the sample becomes more
contaminated we expect that $\langle QG\rangle^{\rm obs}$
will exceed $\langle QG\rangle^{\rm exp}$ and 
increasingly deviate from it for more permissive selections.
The results of this test are shown in Fig.~\ref{fig:compar_QG}, where we plot
$\langle QG\rangle^{\rm obs}$ versus $\langle QG\rangle^{\rm exp}$
for the seven color-selections we considered, and compare to the line 
$\langle QG\rangle^{\rm exp}=\langle QG\rangle^{\rm obs}$ (solid line). 
We find that the
total number of observed QSO-LBGs pairs is consistent with our
expectations for the three
more conservative selections Cases 5-7, but that $\langle
QG\rangle^{\rm obs}$ exceeds $\langle QG\rangle^{\rm exp}$ for more permissive selections, with the
deviations progressively increasing as more contaminants are
included. Note that by construction we will have $\langle
QG\rangle^{\rm obs}=\langle QG\rangle^{\rm
  exp}$ for Case 5, since the clustering measurements ($\langle
QG\rangle^{\rm obs}$) were fit to determine
the correlation function parameters, which go into the
computation of $\langle QG\rangle^{\rm exp}$. But the
fact that expected $\langle QG\rangle^{\rm exp}$ matches
the observed $\langle QG\rangle^{\rm obs}$ for the more conservative Cases 6 and 7
demonstrates that 1) the modeling of the redshift selection function
$\phi_{Z}(Z)$ is correct, 
 2) the contamination is insignificant, 
 and 3) our clustering measurements are robust.

\section{Summary and Conclusions}
\label{sec:summ}

The strong observed auto-correlation of QSOs at $z>3.5$ indicates that they inhabit massive dark matter 
halos with $M_{\rm halo} > 10^{12}\,\rm M_{\odot}$, which implies
QSO environments should exhibit an enhancement of galaxies manifest
as a strong QSO-galaxy cross-correlation function.
We characterized the environments of six
QSO fields at $z=3.78$ that were chosen to host massive BHs
($\gtrsim 10^{9}\,\rm M_{\odot}$). The fields were imaged
using
VLT/FORS1 with two custom NB filters, and the broad band $\rm r_{GUNN}$,
to identify LBGs using a novel technique which selects them in a redshift
range $\sim3.3$ times smaller than the range typically
probed when selecting LBGs with broad band filters. This significantly
reduces the line-of-sight projection effects that have hampered previous
searches for overdensities around $z\gtrsim 5$ QSOs.

Since we used a non-standard filter set to select LBGs, we performed
detailed Monte Carlo simulations to model LBG colors, define
our selection criteria, and compute the
redshift selection function and volume probed by our survey.
This new method effectively selects LBGs in a narrow redshift range, but the color loci of
$z \simeq 3.78$ LBGs and low-redshift galaxies overlap more than with
traditional LBG selection using broader filters.  Defining a pure sample
free low-redshift contaminants required adopting stricter color cuts, which decreased the
completeness of the resulting LBG sample. We devised selection criteria
which resulted in $\sim26\%$ completeness at $z=3.78$, and detected 44 LBGs in our six
fields, corresponding to a number density of 0.19 LBGs arcmin$^{-2}$. 
Our survey probes $\Delta z\simeq 0.3$, and covers a volume equal to 14,782 $h^{-3}$\,cMpc$^{3}$
within $R < 9\,h^{-1}\,{\rm cMpc}$
from the QSO, and we find  on average 1.5 times more galaxies than expected in
random locations of the universe. 

Our work resulted in the first volume-averaged projected QSO-LBG cross-correlation function at $z\sim 4$. We fit our measurements with a
(real-space) power-law cross-correlation function, and found
$r^{QG}_{0}=6.93^{+2.13}_{-1.89} \,h^{-1}\,{\rm cMpc}$ and $\gamma=2.4^{+0.3}_{-0.5}$.
When we fix the slope at $\gamma =2.0$ we find $r^{QG}_{0}=8.83^{+1.39}_{-1.51}\,h^{-1}\,{\rm cMpc}$.
This strong cross-correlation function is in agreement with the
theoretical expectation for the cross-correlation assuming a linear bias
model, which can be estimated
using the auto-correlation of both LBGs and QSOs at $z\sim4$.

We also measured the auto-correlation function of LBGs near these QSOs and
found an auto-correlation length 
of $r^{GG}_{0}=21.59^{+1.72}_{-1.69} \,h^{-1}\,{\rm cMpc}$ for
a fixed slope of $\gamma=1.5$ which is $\sim 4$ times higher than the
measured auto-correlation length of LBGs
in random fields at the same
redshift. Our measurement of an enhanced LBG auto-correlation in QSO environments,
and the strong QSO-LBG cross-correlation both indicate that QSOs at $z\sim4$ trace massive dark
matter halos in the early universe, which are the likely progenitors of massive cluster of galaxies at $z=0$.

We demonstrated that our results are robust against
contamination and that our selection function modeling is reliable,
by varying our color-selection criteria and showing that 
the cross-correlation and auto-correlation
functions are converged.
Spectroscopic follow-up of our LBGs candidates would provide an additional
and definitive test of the reliability of our novel color-selection technique.
However, because the colors of LBGs in our filters do not separate as cleanly
from contaminants as for broad-band LBG selection, we had to choose relatively
conservative color cuts that recovered only 26\% of LBGs. As such, we believe
that the preferred approach to search for overdensities around $z\sim 4-6$ QSOs
using NB filters is to perform traditional LAE selection. Although LAE selection
also only selects a fraction of the total population of high-redshift galaxies \citep{Stark10, Stark11, Curtis12},
the primary advantages are: 1) contamination (from low-redshift line-emitters) is very low,
2) the background number density is known from wide-field observations of blank fields \citep{Hu04, Shimasaku06, Murayama07, Ouchi08}. 

The challenge for the future is to perform similar QSO-galaxy
clustering analyses at higher redshifts. Indeed, if QSOs at $z\sim 5-6$ trace
halos of similar masses as those at $z\sim4$ (i.e. $M_{\rm
  halo}\gtrsim 10^{12}\,\rm M_{\odot}$), then we expect a
strong QSO-galaxy cross-correlation function. Clustering 
studies based on  both broad-band and narrow-band imaging, as well as follow-up
spectroscopy, are now needed to search for these overdensities
around QSOs, and clarify the relationship between early supermassive BHs
and the formation of structure in the early universe.

\acknowledgments
We acknowledge Gabor Worseck, Yue Shen, Arjen van der Wel and Bram Venemans for kindly providing useful data and material used in this paper. We thank the members of the ENIGMA group\footnote{http://www.mpia.de/ENIGMA/} at the Max Planck Institute for Astronomy (MPIA) for useful discussions and comments, in particular we thank Fabrizio Arrigoni for his help in the process of data reduction. We thank Martin White for helpful discussion about clustering. Cristina Garc\'ia-Vergara acknowledges the support from CONICYT Doctoral Fellowship Programme (CONICYT-PCHA/Doctorado Nacional 2012-21120442), DAAD in the context of the PUC-HD Graduate Exchange Fellowship, proyecto financiamiento BASAL PFB06, and proyecto FONDECYT 1120676. 

\bibliography{my_bib}

\begin{thebibliography}{}
\expandafter\ifx\csname natexlab\endcsname\relax\def\natexlab#1{#1}\fi

\bibitem[{{Adelberger} \& {Steidel}(2005)}]{Adelberger05}
{Adelberger}, K.~L., \& {Steidel}, C.~C. 2005, \apj, 630, 50

\bibitem[{{Adelberger} {et~al.}(2003){Adelberger}, {Steidel}, {Shapley}, \&
  {Pettini}}]{Adelberger03}
{Adelberger}, K.~L., {Steidel}, C.~C., {Shapley}, A.~E., \& {Pettini}, M. 2003,
  \apj, 584, 45

\bibitem[{{Appenzeller} \& {Rupprecht}(1992)}]{Appenzeller92}
{Appenzeller}, I., \& {Rupprecht}, G. 1992, The Messenger, 67, 18

\bibitem[{{Ba{\~n}ados} {et~al.}(2013){Ba{\~n}ados}, {Venemans}, {Walter},
  {Kurk}, {Overzier}, \& {Ouchi}}]{Banados13}
{Ba{\~n}ados}, E., {Venemans}, B., {Walter}, F., {et~al.} 2013, \apj, 773, 178

\bibitem[{{Ba{\~n}ados} {et~al.}(2016){Ba{\~n}ados}, {Venemans}, {Decarli},
  {Farina}, {Mazzucchelli}, {Walter}, {Fan}, {Stern}, {Schlafly}, {Chambers},
  {Rix}, {Jiang}, {McGreer}, {Simcoe}, {Wang}, {Yang}, {Morganson}, {De Rosa},
  {Greiner}, {Balokovi{\'c}}, {Burgett}, {Cooper}, {Draper}, {Flewelling},
  {Hodapp}, {Jun}, {Kaiser}, {Kudritzki}, {Magnier}, {Metcalfe}, {Miller},
  {Schindler}, {Tonry}, {Wainscoat}, {Waters}, \& {Yang}}]{Banados16}
{Ba{\~n}ados}, E., {Venemans}, B.~P., {Decarli}, R., {et~al.} 2016, \apjs, 227,
  11

\bibitem[{{Bahcall} {et~al.}(1997){Bahcall}, {Kirhakos}, {Saxe}, \&
  {Schneider}}]{Bahcall97}
{Bahcall}, J.~N., {Kirhakos}, S., {Saxe}, D.~H., \& {Schneider}, D.~P. 1997,
  \apj, 479, 642

\bibitem[{{Becker} {et~al.}(1995){Becker}, {White}, \& {Helfand}}]{Becker95}
{Becker}, R.~H., {White}, R.~L., \& {Helfand}, D.~J. 1995, \apj, 450, 559

\bibitem[{{Bertin}(2006)}]{Bertin06}
{Bertin}, E. 2006, in Astronomical Society of the Pacific Conference Series,
  Vol. 351, Astronomical Data Analysis Software and Systems XV, ed.
  C.~{Gabriel}, C.~{Arviset}, D.~{Ponz}, \& S.~{Enrique}, 112

\bibitem[{{Bertin} \& {Arnouts}(1996)}]{Bertin96}
{Bertin}, E., \& {Arnouts}, S. 1996, \aaps, 117, 393

\bibitem[{{Bertin} {et~al.}(2002){Bertin}, {Mellier}, {Radovich}, {Missonnier},
  {Didelon}, \& {Morin}}]{Bertin02}
{Bertin}, E., {Mellier}, Y., {Radovich}, M., {et~al.} 2002, in Astronomical
  Society of the Pacific Conference Series, Vol. 281, Astronomical Data
  Analysis Software and Systems XI, ed. D.~A. {Bohlender}, D.~{Durand}, \&
  T.~H. {Handley}, 228

\bibitem[{{Bouwens} {et~al.}(2007){Bouwens}, {Illingworth}, {Franx}, \&
  {Ford}}]{Bouwens07}
{Bouwens}, R.~J., {Illingworth}, G.~D., {Franx}, M., \& {Ford}, H. 2007, \apj,
  670, 928

\bibitem[{{Bouwens} {et~al.}(2009){Bouwens}, {Illingworth}, {Franx}, {Chary},
  {Meurer}, {Conselice}, {Ford}, {Giavalisco}, \& {van Dokkum}}]{Bouwens09}
{Bouwens}, R.~J., {Illingworth}, G.~D., {Franx}, M., {et~al.} 2009, \apj, 705,
  936

\bibitem[{{Bouwens} {et~al.}(2010){Bouwens}, {Illingworth}, {Oesch},
  {Stiavelli}, {van Dokkum}, {Trenti}, {Magee}, {Labb{\'e}}, {Franx},
  {Carollo}, \& {Gonzalez}}]{Bouwens10}
{Bouwens}, R.~J., {Illingworth}, G.~D., {Oesch}, P.~A., {et~al.} 2010, \apjl,
  709, L133

\bibitem[{{Brammer} {et~al.}(2008){Brammer}, {van Dokkum}, \&
  {Coppi}}]{Brammer08}
{Brammer}, G.~B., {van Dokkum}, P.~G., \& {Coppi}, P. 2008, \apj, 686, 1503

\bibitem[{{Bruzual} \& {Charlot}(2003)}]{Bruzual03}
{Bruzual}, G., \& {Charlot}, S. 2003, \mnras, 344, 1000

\bibitem[{{Calzetti} {et~al.}(2000){Calzetti}, {Armus}, {Bohlin}, {Kinney},
  {Koornneef}, \& {Storchi-Bergmann}}]{Calzetti00}
{Calzetti}, D., {Armus}, L., {Bohlin}, R.~C., {et~al.} 2000, \apj, 533, 682

\bibitem[{{Cardelli} {et~al.}(1989){Cardelli}, {Clayton}, \&
  {Mathis}}]{Cardelli89}
{Cardelli}, J.~A., {Clayton}, G.~C., \& {Mathis}, J.~S. 1989, \apj, 345, 245

\bibitem[{{Chabrier}(2003)}]{Chabrier03}
{Chabrier}, G. 2003, \pasp, 115, 763

\bibitem[{{Ciardullo} {et~al.}(2012){Ciardullo}, {Gronwall}, {Wolf},
  {McCathran}, {Bond}, {Gawiser}, {Guaita}, {Feldmeier}, {Treister}, {Padilla},
  {Francke}, {Matkovi{\'c}}, {Altmann}, \& {Herrera}}]{Ciardullo12}
{Ciardullo}, R., {Gronwall}, C., {Wolf}, C., {et~al.} 2012, \apj, 744, 110

\bibitem[{{Coil} {et~al.}(2007){Coil}, {Hennawi}, {Newman}, {Cooper}, \&
  {Davis}}]{Coil07}
{Coil}, A.~L., {Hennawi}, J.~F., {Newman}, J.~A., {Cooper}, M.~C., \& {Davis},
  M. 2007, \apj, 654, 115

\bibitem[{{Curtis-Lake} {et~al.}(2012){Curtis-Lake}, {McLure}, {Pearce},
  {Dunlop}, {Cirasuolo}, {Stark}, {Almaini}, {Bradshaw}, {Chuter}, {Foucaud},
  \& {Hartley}}]{Curtis12}
{Curtis-Lake}, E., {McLure}, R.~J., {Pearce}, H.~J., {et~al.} 2012, \mnras,
  422, 1425

\bibitem[{{Dodelson}(2003)}]{Dodelson03}
{Dodelson}, S. 2003, {Modern cosmology}

\bibitem[{{Fan} {et~al.}(2000){Fan}, {White}, {Davis}, {Becker}, {Strauss},
  {Haiman}, {Schneider}, {Gregg}, {Gunn}, {Knapp}, {Lupton}, {Anderson},
  {Anderson}, {Annis}, {Bahcall}, {Boroski}, {Brunner}, {Chen}, {Connolly},
  {Csabai}, {Doi}, {Fukugita}, {Hennessy}, {Hindsley}, {Ichikawa},
  {Ivezi{\'c}}, {Loveday}, {Meiksin}, {McKay}, {Munn}, {Newberg}, {Nichol},
  {Okamura}, {Pier}, {Sekiguchi}, {Shimasaku}, {Stoughton}, {Szalay},
  {Szokoly}, {Thakar}, {Vogeley}, \& {York}}]{Fan00}
{Fan}, X., {White}, R.~L., {Davis}, M., {et~al.} 2000, \aj, 120, 1167

\bibitem[{{Fanidakis} {et~al.}(2013){Fanidakis}, {Macci{\`o}}, {Baugh},
  {Lacey}, \& {Frenk}}]{Fanidakis13}
{Fanidakis}, N., {Macci{\`o}}, A.~V., {Baugh}, C.~M., {Lacey}, C.~G., \&
  {Frenk}, C.~S. 2013, \mnras, 436, 315

\bibitem[{{Fern{\'a}ndez-Soto} {et~al.}(2003){Fern{\'a}ndez-Soto}, {Lanzetta},
  \& {Chen}}]{Fernandez03}
{Fern{\'a}ndez-Soto}, A., {Lanzetta}, K.~M., \& {Chen}, H.-W. 2003, \mnras,
  342, 1215

\bibitem[{{Ferrarese}(2002)}]{Ferrarese02}
{Ferrarese}, L. 2002, \apj, 578, 90

\bibitem[{{Ferrarese} \& {Merritt}(2000)}]{Ferrarese00}
{Ferrarese}, L., \& {Merritt}, D. 2000, \apjl, 539, L9

\bibitem[{{Fukugita} {et~al.}(1995){Fukugita}, {Shimasaku}, \&
  {Ichikawa}}]{Fukugita95}
{Fukugita}, M., {Shimasaku}, K., \& {Ichikawa}, T. 1995, \pasp, 107, 945

\bibitem[{{Gebhardt} {et~al.}(2000){Gebhardt}, {Bender}, {Bower}, {Dressler},
  {Faber}, {Filippenko}, {Green}, {Grillmair}, {Ho}, {Kormendy}, {Lauer},
  {Magorrian}, {Pinkney}, {Richstone}, \& {Tremaine}}]{Gebhardt00}
{Gebhardt}, K., {Bender}, R., {Bower}, G., {et~al.} 2000, \apjl, 539, L13

\bibitem[{{Gehrels}(1986)}]{Gehrels86}
{Gehrels}, N. 1986, \apj, 303, 336

\bibitem[{{Giavalisco} {et~al.}(1998){Giavalisco}, {Steidel}, {Adelberger},
  {Dickinson}, {Pettini}, \& {Kellogg}}]{Giavalisco98}
{Giavalisco}, M., {Steidel}, C.~C., {Adelberger}, K.~L., {et~al.} 1998, \apj,
  503, 543

\bibitem[{{Gioia} {et~al.}(2001){Gioia}, {Henry}, {Mullis}, {Voges}, {Briel},
  {B{\"o}hringer}, \& {Huchra}}]{Gioia01}
{Gioia}, I.~M., {Henry}, J.~P., {Mullis}, C.~R., {et~al.} 2001, \apjl, 553,
  L105

\bibitem[{{Hennawi} {et~al.}(2015){Hennawi}, {Prochaska}, {Cantalupo}, \&
  {Arrigoni-Battaia}}]{Hennawi15}
{Hennawi}, J.~F., {Prochaska}, J.~X., {Cantalupo}, S., \& {Arrigoni-Battaia},
  F. 2015, Science, 348, 779

\bibitem[{{Hennawi} {et~al.}(2006){Hennawi}, {Strauss}, {Oguri}, {Inada},
  {Richards}, {Pindor}, {Schneider}, {Becker}, {Gregg}, {Hall}, {Johnston},
  {Fan}, {Burles}, {Schlegel}, {Gunn}, {Lupton}, {Bahcall}, {Brunner}, \&
  {Brinkmann}}]{Hennawi06}
{Hennawi}, J.~F., {Strauss}, M.~A., {Oguri}, M., {et~al.} 2006, \aj, 131, 1

\bibitem[{{Hennawi} {et~al.}(2010){Hennawi}, {Myers}, {Shen}, {Strauss},
  {Djorgovski}, {Fan}, {Glikman}, {Mahabal}, {Martin}, {Richards}, {Schneider},
  \& {Shankar}}]{Hennawi10}
{Hennawi}, J.~F., {Myers}, A.~D., {Shen}, Y., {et~al.} 2010, \apj, 719, 1672

\bibitem[{{Hinshaw} {et~al.}(2013){Hinshaw}, {Larson}, {Komatsu}, {Spergel},
  {Bennett}, {Dunkley}, {Nolta}, {Halpern}, {Hill}, {Odegard}, {Page}, {Smith},
  {Weiland}, {Gold}, {Jarosik}, {Kogut}, {Limon}, {Meyer}, {Tucker}, {Wollack},
  \& {Wright}}]{Hinshaw13}
{Hinshaw}, G., {Larson}, D., {Komatsu}, E., {et~al.} 2013, \apjs, 208, 19

\bibitem[{{Hu} {et~al.}(2004){Hu}, {Cowie}, {Capak}, {McMahon}, {Hayashino}, \&
  {Komiyama}}]{Hu04}
{Hu}, E.~M., {Cowie}, L.~L., {Capak}, P., {et~al.} 2004, \aj, 127, 563

\bibitem[{{Husband} {et~al.}(2013){Husband}, {Bremer}, {Stanway}, {Davies},
  {Lehnert}, \& {Douglas}}]{Husband13}
{Husband}, K., {Bremer}, M.~N., {Stanway}, E.~R., {et~al.} 2013, \mnras, 432,
  2869

\bibitem[{{Intema} {et~al.}(2006){Intema}, {Venemans}, {Kurk}, {Ouchi},
  {Kodama}, {R{\"o}ttgering}, {Miley}, \& {Overzier}}]{Intema06}
{Intema}, H.~T., {Venemans}, B.~P., {Kurk}, J.~D., {et~al.} 2006, \aap, 456,
  433

\bibitem[{{Jones} {et~al.}(2012){Jones}, {Stark}, \& {Ellis}}]{Jones12}
{Jones}, T., {Stark}, D.~P., \& {Ellis}, R.~S. 2012, \apj, 751, 51

\bibitem[{{Kashikawa} {et~al.}(2007){Kashikawa}, {Kitayama}, {Doi}, {Misawa},
  {Komiyama}, \& {Ota}}]{Kashikawa07}
{Kashikawa}, N., {Kitayama}, T., {Doi}, M., {et~al.} 2007, \apj, 663, 765

\bibitem[{{Kim} {et~al.}(2009){Kim}, {Stiavelli}, {Trenti}, {Pavlovsky},
  {Djorgovski}, {Scarlata}, {Stern}, {Mahabal}, {Thompson}, {Dickinson},
  {Panagia}, \& {Meylan}}]{Kim09}
{Kim}, S., {Stiavelli}, M., {Trenti}, M., {et~al.} 2009, \apj, 695, 809

\bibitem[{{Kormendy} \& {Bender}(2011)}]{Kormendy11}
{Kormendy}, J., \& {Bender}, R. 2011, \nat, 469, 377

\bibitem[{{Lacey} \& {Cole}(1993)}]{Lacey93}
{Lacey}, C., \& {Cole}, S. 1993, \mnras, 262, 627

\bibitem[{{Lee} {et~al.}(2006){Lee}, {Giavalisco}, {Gnedin}, {Somerville},
  {Ferguson}, {Dickinson}, \& {Ouchi}}]{Lee06}
{Lee}, K.-S., {Giavalisco}, M., {Gnedin}, O.~Y., {et~al.} 2006, \apj, 642, 63

\bibitem[{{Magorrian} {et~al.}(1998){Magorrian}, {Tremaine}, {Richstone},
  {Bender}, {Bower}, {Dressler}, {Faber}, {Gebhardt}, {Green}, {Grillmair},
  {Kormendy}, \& {Lauer}}]{Magorrian98}
{Magorrian}, J., {Tremaine}, S., {Richstone}, D., {et~al.} 1998, \aj, 115, 2285

\bibitem[{{Mazzucchelli} {et~al.}(2016){Mazzucchelli}, {Ba\~nados}, {Decarli},
  {Farina}, {Venemans}, {Walter}, \& {Overzier}}]{Mazzucchelli16}
{Mazzucchelli}, C., {Ba\~nados}, E., {Decarli}, R., {et~al.} 2016, submitted to
  \apj

\bibitem[{{McLeod} \& {Bechtold}(2009)}]{McLeod09}
{McLeod}, K.~K., \& {Bechtold}, J. 2009, \apj, 704, 415

\bibitem[{{Morselli} {et~al.}(2014){Morselli}, {Mignoli}, {Gilli}, {Vignali},
  {Comastri}, {Sani}, {Cappelluti}, {Zamorani}, {Brusa}, {Gallozzi}, \&
  {Vanzella}}]{Morselli14}
{Morselli}, L., {Mignoli}, M., {Gilli}, R., {et~al.} 2014, \aap, 568, A1

\bibitem[{{Murayama} {et~al.}(2007){Murayama}, {Taniguchi}, {Scoville},
  {Ajiki}, {Sanders}, {Mobasher}, {Aussel}, {Capak}, {Koekemoer}, {Shioya},
  {Nagao}, {Carilli}, {Ellis}, {Garilli}, {Giavalisco}, {Kitzbichler}, {Le
  F{\`e}vre}, {Maccagni}, {Schinnerer}, {Smol{\v c}i{\'c}}, {Tribiano},
  {Cimatti}, {Komiyama}, {Miyazaki}, {Sasaki}, {Koda}, \&
  {Karoji}}]{Murayama07}
{Murayama}, T., {Taniguchi}, Y., {Scoville}, N.~Z., {et~al.} 2007, \apjs, 172,
  523

\bibitem[{{Oesch} {et~al.}(2010){Oesch}, {Bouwens}, {Illingworth}, {Carollo},
  {Franx}, {Labb{\'e}}, {Magee}, {Stiavelli}, {Trenti}, \& {van
  Dokkum}}]{Oesch10}
{Oesch}, P.~A., {Bouwens}, R.~J., {Illingworth}, G.~D., {et~al.} 2010, \apjl,
  709, L16

\bibitem[{{Oke}(1974)}]{Oke74}
{Oke}, J.~B. 1974, \apjs, 27, 21

\bibitem[{{Ota} {et~al.}(2008){Ota}, {Kashikawa}, {Malkan}, {Iye}, {Nakajima},
  {Nagao}, {Shimasaku}, \& {Gandhi}}]{Ota08}
{Ota}, K., {Kashikawa}, N., {Malkan}, M.~A., {et~al.} 2008, ArXiv e-prints,
  arXiv:0804.3448

\bibitem[{{Ouchi} {et~al.}(2004{\natexlab{a}}){Ouchi}, {Shimasaku}, {Okamura},
  {Furusawa}, {Kashikawa}, {Ota}, {Doi}, {Hamabe}, {Kimura}, {Komiyama},
  {Miyazaki}, {Miyazaki}, {Nakata}, {Sekiguchi}, {Yagi}, \&
  {Yasuda}}]{Ouchi04a}
{Ouchi}, M., {Shimasaku}, K., {Okamura}, S., {et~al.} 2004{\natexlab{a}}, \apj,
  611, 660

\bibitem[{{Ouchi} {et~al.}(2004{\natexlab{b}}){Ouchi}, {Shimasaku}, {Okamura},
  {Furusawa}, {Kashikawa}, {Ota}, {Doi}, {Hamabe}, {Kimura}, {Komiyama},
  {Miyazaki}, {Miyazaki}, {Nakata}, {Sekiguchi}, {Yagi}, \&
  {Yasuda}}]{Ouchi04b}
---. 2004{\natexlab{b}}, \apj, 611, 685

\bibitem[{{Ouchi} {et~al.}(2005){Ouchi}, {Shimasaku}, {Akiyama}, {Sekiguchi},
  {Furusawa}, {Okamura}, {Kashikawa}, {Iye}, {Kodama}, {Saito}, {Sasaki},
  {Simpson}, {Takata}, {Yamada}, {Yamanoi}, {Yoshida}, \& {Yoshida}}]{Ouchi05}
{Ouchi}, M., {Shimasaku}, K., {Akiyama}, M., {et~al.} 2005, \apjl, 620, L1

\bibitem[{{Ouchi} {et~al.}(2008){Ouchi}, {Shimasaku}, {Akiyama}, {Simpson},
  {Saito}, {Ueda}, {Furusawa}, {Sekiguchi}, {Yamada}, {Kodama}, {Kashikawa},
  {Okamura}, {Iye}, {Takata}, {Yoshida}, \& {Yoshida}}]{Ouchi08}
---. 2008, \apjs, 176, 301

\bibitem[{{Overzier} {et~al.}(2008){Overzier}, {Bouwens}, {Cross}, {Venemans},
  {Miley}, {Zirm}, {Ben{\'{\i}}tez}, {Blakeslee}, {Coe}, {Demarco}, {Ford},
  {Homeier}, {Illingworth}, {Kurk}, {Martel}, {Mei}, {Oliveira},
  {R{\"o}ttgering}, {Tsvetanov}, \& {Zheng}}]{Overzier08}
{Overzier}, R.~A., {Bouwens}, R.~J., {Cross}, N.~J.~G., {et~al.} 2008, \apj,
  673, 143

\bibitem[{{Padmanabhan} {et~al.}(2007){Padmanabhan}, {White}, \&
  {Eisenstein}}]{Padmanabhan07}
{Padmanabhan}, N., {White}, M., \& {Eisenstein}, D.~J. 2007, \mnras, 376, 1702

\bibitem[{{Padmanabhan} {et~al.}(2009){Padmanabhan}, {White}, {Norberg}, \&
  {Porciani}}]{Padmanabhan09}
{Padmanabhan}, N., {White}, M., {Norberg}, P., \& {Porciani}, C. 2009, \mnras,
  397, 1862

\bibitem[{{Padmanabhan}(2006)}]{Padmanabhan06}
{Padmanabhan}, T. 2006, in American Institute of Physics Conference Series,
  Vol. 843, Graduate School in Astronomy: X, ed. S.~{Daflon}, J.~{Alcaniz},
  E.~{Telles}, \& R.~{de la Reza}, 111--166

\bibitem[{{Patat} {et~al.}(2011){Patat}, {Moehler}, {O'Brien}, {Pompei},
  {Bensby}, {Carraro}, {de Ugarte Postigo}, {Fox}, {Gavignaud}, {James},
  {Korhonen}, {Ledoux}, {Randall}, {Sana}, {Smoker}, {Stefl}, \&
  {Szeifert}}]{Patat11}
{Patat}, F., {Moehler}, S., {O'Brien}, K., {et~al.} 2011, \aap, 527, A91

\bibitem[{{Peebles}(1980)}]{Peebles80}
{Peebles}, P.~J.~E. 1980, {The large-scale structure of the universe}

\bibitem[{{Sargent} \& {Turner}(1977)}]{Sargent77}
{Sargent}, W.~L.~W., \& {Turner}, E.~L. 1977, \apjl, 212, L3

\bibitem[{{Schlegel} {et~al.}(1998){Schlegel}, {Finkbeiner}, \&
  {Davis}}]{Schlegel98}
{Schlegel}, D.~J., {Finkbeiner}, D.~P., \& {Davis}, M. 1998, \apj, 500, 525

\bibitem[{{Schneider}(2015)}]{Schneider15}
{Schneider}, P. 2015, {Extragalactic Astronomy and Cosmology: An Introduction},
  doi:10.1007/978-3-642-54083-7

\bibitem[{{Shapley} {et~al.}(2003){Shapley}, {Steidel}, {Pettini}, \&
  {Adelberger}}]{Shapley03}
{Shapley}, A.~E., {Steidel}, C.~C., {Pettini}, M., \& {Adelberger}, K.~L. 2003,
  \apj, 588, 65

\bibitem[{{Shapley} {et~al.}(2006){Shapley}, {Steidel}, {Pettini},
  {Adelberger}, \& {Erb}}]{Shapley06}
{Shapley}, A.~E., {Steidel}, C.~C., {Pettini}, M., {Adelberger}, K.~L., \&
  {Erb}, D.~K. 2006, \apj, 651, 688

\bibitem[{{Shen} {et~al.}(2007){Shen}, {Strauss}, {Oguri}, {Hennawi}, {Fan},
  {Richards}, {Hall}, {Gunn}, {Schneider}, {Szalay}, {Thakar}, {Vanden Berk},
  {Anderson}, {Bahcall}, {Connolly}, \& {Knapp}}]{Shen07}
{Shen}, Y., {Strauss}, M.~A., {Oguri}, M., {et~al.} 2007, \aj, 133, 2222

\bibitem[{{Shen} {et~al.}(2009){Shen}, {Strauss}, {Ross}, {Hall}, {Lin},
  {Richards}, {Schneider}, {Weinberg}, {Connolly}, {Fan}, {Hennawi}, {Shankar},
  {Vanden Berk}, {Bahcall}, \& {Brunner}}]{Shen09}
{Shen}, Y., {Strauss}, M.~A., {Ross}, N.~P., {et~al.} 2009, \apj, 697, 1656

\bibitem[{{Shen} {et~al.}(2010){Shen}, {Hennawi}, {Shankar}, {Myers},
  {Strauss}, {Djorgovski}, {Fan}, {Giocoli}, {Mahabal}, {Schneider}, \&
  {Weinberg}}]{Shen10}
{Shen}, Y., {Hennawi}, J.~F., {Shankar}, F., {et~al.} 2010, \apj, 719, 1693

\bibitem[{{Shen} {et~al.}(2011){Shen}, {Richards}, {Strauss}, {Hall},
  {Schneider}, {Snedden}, {Bizyaev}, {Brewington}, {Malanushenko},
  {Malanushenko}, {Oravetz}, {Pan}, \& {Simmons}}]{Shen11}
{Shen}, Y., {Richards}, G.~T., {Strauss}, M.~A., {et~al.} 2011, \apjs, 194, 45

\bibitem[{{Shimasaku} {et~al.}(2006){Shimasaku}, {Kashikawa}, {Doi}, {Ly},
  {Malkan}, {Matsuda}, {Ouchi}, {Hayashino}, {Iye}, {Motohara}, {Murayama},
  {Nagao}, {Ohta}, {Okamura}, {Sasaki}, {Shioya}, \& {Taniguchi}}]{Shimasaku06}
{Shimasaku}, K., {Kashikawa}, N., {Doi}, M., {et~al.} 2006, \pasj, 58, 313

\bibitem[{{Simpson} {et~al.}(2014){Simpson}, {Mortlock}, {Warren}, {Cantalupo},
  {Hewett}, {McLure}, {McMahon}, \& {Venemans}}]{Simpson14}
{Simpson}, C., {Mortlock}, D., {Warren}, S., {et~al.} 2014, \mnras, 442, 3454

\bibitem[{{Stark} {et~al.}(2010){Stark}, {Ellis}, {Chiu}, {Ouchi}, \&
  {Bunker}}]{Stark10}
{Stark}, D.~P., {Ellis}, R.~S., {Chiu}, K., {Ouchi}, M., \& {Bunker}, A. 2010,
  \mnras, 408, 1628

\bibitem[{{Stark} {et~al.}(2011){Stark}, {Ellis}, \& {Ouchi}}]{Stark11}
{Stark}, D.~P., {Ellis}, R.~S., \& {Ouchi}, M. 2011, \apjl, 728, L2

\bibitem[{{Steidel} {et~al.}(1999){Steidel}, {Adelberger}, {Giavalisco},
  {Dickinson}, \& {Pettini}}]{Steidel99}
{Steidel}, C.~C., {Adelberger}, K.~L., {Giavalisco}, M., {Dickinson}, M., \&
  {Pettini}, M. 1999, \apj, 519, 1

\bibitem[{{Steidel} {et~al.}(2003){Steidel}, {Adelberger}, {Shapley},
  {Pettini}, {Dickinson}, \& {Giavalisco}}]{Steidel03}
{Steidel}, C.~C., {Adelberger}, K.~L., {Shapley}, A.~E., {et~al.} 2003, \apj,
  592, 728

\bibitem[{{Steidel} {et~al.}(1996){Steidel}, {Giavalisco}, {Pettini},
  {Dickinson}, \& {Adelberger}}]{Steidel96}
{Steidel}, C.~C., {Giavalisco}, M., {Pettini}, M., {Dickinson}, M., \&
  {Adelberger}, K.~L. 1996, \apjl, 462, L17

\bibitem[{{Steidel} {et~al.}(1995){Steidel}, {Pettini}, \&
  {Hamilton}}]{Steidel95}
{Steidel}, C.~C., {Pettini}, M., \& {Hamilton}, D. 1995, \aj, 110, 2519

\bibitem[{{Stiavelli} {et~al.}(2005){Stiavelli}, {Djorgovski}, {Pavlovsky},
  {Scarlata}, {Stern}, {Mahabal}, {Thompson}, {Dickinson}, {Panagia}, \&
  {Meylan}}]{Stiavelli05}
{Stiavelli}, M., {Djorgovski}, S.~G., {Pavlovsky}, C., {et~al.} 2005, \apjl,
  622, L1

\bibitem[{{Stone}(1996)}]{Stone96}
{Stone}, R.~P.~S. 1996, \apjs, 107, 423

\bibitem[{{Toshikawa} {et~al.}(2012){Toshikawa}, {Kashikawa}, {Ota},
  {Morokuma}, {Shibuya}, {Hayashi}, {Nagao}, {Jiang}, {Malkan}, {Egami},
  {Shimasaku}, {Motohara}, \& {Ishizaki}}]{Toshikawa12}
{Toshikawa}, J., {Kashikawa}, N., {Ota}, K., {et~al.} 2012, \apj, 750, 137

\bibitem[{{Trainor} \& {Steidel}(2012)}]{Trainor12}
{Trainor}, R.~F., \& {Steidel}, C.~C. 2012, \apj, 752, 39

\bibitem[{{Utsumi} {et~al.}(2010){Utsumi}, {Goto}, {Kashikawa}, {Miyazaki},
  {Komiyama}, {Furusawa}, \& {Overzier}}]{Utsumi10}
{Utsumi}, Y., {Goto}, T., {Kashikawa}, N., {et~al.} 2010, \apj, 721, 1680

\bibitem[{{Venemans} {et~al.}(2007){Venemans}, {R{\"o}ttgering}, {Miley}, {van
  Breugel}, {de Breuck}, {Kurk}, {Pentericci}, {Stanford}, {Overzier}, {Croft},
  \& {Ford}}]{Venemans07}
{Venemans}, B.~P., {R{\"o}ttgering}, H.~J.~A., {Miley}, G.~K., {et~al.} 2007,
  \aap, 461, 823

\bibitem[{{Vestergaard}(2002)}]{Vestergaard02}
{Vestergaard}, M. 2002, \apj, 571, 733

\bibitem[{{Vikhlinin} {et~al.}(2009){Vikhlinin}, {Kravtsov}, {Burenin},
  {Ebeling}, {Forman}, {Hornstrup}, {Jones}, {Murray}, {Nagai}, {Quintana}, \&
  {Voevodkin}}]{Vikhlinin09}
{Vikhlinin}, A., {Kravtsov}, A.~V., {Burenin}, R.~A., {et~al.} 2009, \apj, 692,
  1060

\bibitem[{{White} {et~al.}(2012){White}, {Myers}, {Ross}, {Schlegel},
  {Hennawi}, {Shen}, {McGreer}, {Strauss}, {Bolton}, {Bovy}, {Fan},
  {Miralda-Escude}, {Palanque-Delabrouille}, {Paris}, {Petitjean}, {Schneider},
  {Viel}, {Weinberg}, {Yeche}, {Zehavi}, {Pan}, {Snedden}, {Bizyaev},
  {Brewington}, {Brinkmann}, {Malanushenko}, {Malanushenko}, {Oravetz},
  {Simmons}, {Sheldon}, \& {Weaver}}]{White12}
{White}, M., {Myers}, A.~D., {Ross}, N.~P., {et~al.} 2012, \mnras, 424, 933

\bibitem[{{Willott} {et~al.}(2005){Willott}, {Percival}, {McLure}, {Crampton},
  {Hutchings}, {Jarvis}, {Sawicki}, \& {Simard}}]{Willott05}
{Willott}, C.~J., {Percival}, W.~J., {McLure}, R.~J., {et~al.} 2005, \apj, 626,
  657

\bibitem[{{Worseck} \& {Prochaska}(2011)}]{Worseck11}
{Worseck}, G., \& {Prochaska}, J.~X. 2011, \apj, 728, 23

\bibitem[{{Wyithe} \& {Loeb}(2002)}]{Wyithe02}
{Wyithe}, J.~S.~B., \& {Loeb}, A. 2002, \apj, 581, 886

\bibitem[{{York} {et~al.}(2000){York}, {Adelman}, {Anderson}, {Anderson},
  {Annis}, {Bahcall}, {Bakken}, {Barkhouser}, {Bastian}, {Berman}, {Boroski},
  {Bracker}, {Briegel}, {Briggs}, {Brinkmann}, {Brunner}, {Burles}, {Carey},
  {Carr}, {Castander}, {Chen}, {Colestock}, {Connolly}, {Crocker}, {Csabai},
  {Czarapata}, {Davis}, {Doi}, {Dombeck}, {Eisenstein}, {Ellman}, {Elms},
  {Evans}, {Fan}, {Federwitz}, {Fiscelli}, {Friedman}, {Frieman}, {Fukugita},
  {Gillespie}, {Gunn}, {Gurbani}, {de Haas}, {Haldeman}, {Harris}, {Hayes},
  {Heckman}, {Hennessy}, {Hindsley}, {Holm}, {Holmgren}, {Huang}, {Hull},
  {Husby}, {Ichikawa}, {Ichikawa}, {Ivezi{\'c}}, {Kent}, {Kim}, {Kinney},
  {Klaene}, {Kleinman}, {Kleinman}, {Knapp}, {Korienek}, {Kron}, {Kunszt},
  {Lamb}, {Lee}, {Leger}, {Limmongkol}, {Lindenmeyer}, {Long}, {Loomis},
  {Loveday}, {Lucinio}, {Lupton}, {MacKinnon}, {Mannery}, {Mantsch}, {Margon},
  {McGehee}, {McKay}, {Meiksin}, {Merelli}, {Monet}, {Munn}, {Narayanan},
  {Nash}, {Neilsen}, {Neswold}, {Newberg}, {Nichol}, {Nicinski}, {Nonino},
  {Okada}, {Okamura}, {Ostriker}, {Owen}, {Pauls}, {Peoples}, {Peterson},
  {Petravick}, {Pier}, {Pope}, {Pordes}, {Prosapio}, {Rechenmacher}, {Quinn},
  {Richards}, {Richmond}, {Rivetta}, {Rockosi}, {Ruthmansdorfer}, {Sandford},
  {Schlegel}, {Schneider}, {Sekiguchi}, {Sergey}, {Shimasaku}, {Siegmund},
  {Smee}, {Smith}, {Snedden}, {Stone}, {Stoughton}, {Strauss}, {Stubbs},
  {SubbaRao}, {Szalay}, {Szapudi}, {Szokoly}, {Thakar}, {Tremonti}, {Tucker},
  {Uomoto}, {Vanden Berk}, {Vogeley}, {Waddell}, {Wang}, {Watanabe},
  {Weinberg}, {Yanny}, {Yasuda}, \& {SDSS Collaboration}}]{York00}
{York}, D.~G., {Adelman}, J., {Anderson}, Jr., J.~E., {et~al.} 2000, \aj, 120,
  1579

\bibitem[{{Zheng} {et~al.}(2006){Zheng}, {Overzier}, {Bouwens}, {White},
  {Ford}, {Ben{\'{\i}}tez}, {Blakeslee}, {Bradley}, {Jee}, {Martel}, {Mei},
  {Zirm}, {Illingworth}, {Clampin}, {Hartig}, {Ardila}, {Bartko}, {Broadhurst},
  {Brown}, {Burrows}, {Cheng}, {Cross}, {Demarco}, {Feldman}, {Franx},
  {Golimowski}, {Goto}, {Gronwall}, {Holden}, {Homeier}, {Infante}, {Kimble},
  {Krist}, {Lesser}, {Menanteau}, {Meurer}, {Miley}, {Motta}, {Postman},
  {Rosati}, {Sirianni}, {Sparks}, {Tran}, \& {Tsvetanov}}]{Zheng06}
{Zheng}, W., {Overzier}, R.~A., {Bouwens}, R.~J., {et~al.} 2006, \apj, 640, 574

\end{thebibliography}

\end{document}